\documentclass[aos,preprint]{imsart}

\usepackage[english]{babel}
\usepackage{amsmath}
\usepackage{amssymb}
\usepackage{ulem}
\usepackage{comment}
\usepackage{url}
\usepackage{hyperref}
\usepackage{graphicx}
\usepackage[right = 2cm, top=2cm, bottom=3cm,left=2cm]{geometry}
\usepackage{mathdots}
\usepackage{amsthm}
\usepackage{bbm}
\usepackage{enumitem}
\usepackage{float}

\usepackage[section]{placeins} 

\pubyear{2024}
\arxiv{arXiv:0000.0000}

\usepackage{framed}

\usepackage[T1]{fontenc}
\usepackage[utf8]{inputenc}
\usepackage{csquotes}

\makeatletter
\let\c@author\relax
\makeatother
\usepackage[backend=biber]{biblatex}
\begin{document}

\newcommand{\Rho}{P}
\newcommand{\IN}{\mathbb{N}}
\newcommand{\IQ}{\mathbb{Q}}
\newcommand{\IZ}{\mathbb{Z}}
\newcommand{\IR}{\mathbb{R}}
\newcommand{\IC}{\mathbb{C}}
\newcommand{\Ima}{\mbox{Im}}
\newcommand{\dif}{\ \mbox{d}}
\newcommand{\TR}{\text{TR}}
\newcommand{\cov}{\mbox{cov}}

\newcommand{\Lp}{\mathcal{L}}
\newcommand{\sI}{\mathcal{I}}
\newcommand{\sA}{\mathcal{A}}
\newcommand{\sB}{\mathcal{B}}
\newcommand{\sP}{\mathcal{P}}
\newcommand{\sE}{\mathcal{E}}
\newcommand{\sF}{\mathcal{F}}
\newcommand{\sG}{\mathcal{G}}
\newcommand{\sH}{\mathcal{H}}
\newcommand{\sT}{\mathcal{T}}
\newcommand{\sV}{\mathcal{V}}
\newcommand{\sL}{\mathcal{L}}
\newcommand{\SUP}{\text{SUP }}

\renewcommand{\Re}{\mbox{Re }}
\renewcommand{\Im}{\mbox{Im }}

\newcommand{\reff}[1]{(\ref{#1})}

\newcommand{\IP}{\mathbb{P}}
\newcommand{\IE}{\mathbb{E}}
\newcommand{\Ii}{\mathbbm{1}}
\newcommand{\supp}{\mbox{supp}}
\newcommand{\Hess}{\mbox{Hess}}
\newcommand{\Var}{\mbox{Var}}
\newcommand{\sX}{\mathcal{X}}
\newcommand{\Kov}{\mbox{Kov}}
\newcommand{\Cov}{\mbox{Cov}}
\newcommand{\tr}{\mbox{tr}}
\newcommand{\gdw}{\Leftrightarrow}
\newcommand{\pto}{\overset{\IP}{\to}}
\newcommand{\fsto}{\overset{f.s.}{\to}}
\newcommand{\dto}{\overset{d}{\to}}
\newcommand{\lto}{\overset{L^2}{\to}}
\newcommand{\sD}{\mathcal{D}}
\newcommand{\iid}{\overset{\mbox{iid}}{\sim}}
\renewcommand{\l}{\ell}
\renewcommand{\i}{|}

\renewcommand{\supp}{\text{supp}}

\newcommand{\err}{\mbox{err}}
\newcommand{\bias}{\mbox{bias}}

\newcommand{\norm}[1]{\left\lVert#1\right\rVert}

\newtheorem{theorem}{Theorem}[section]
\newtheorem{corollary}[theorem]{Corollary}
\newtheorem{definition}[theorem]{Definition}
\newtheorem{proposition}[theorem]{Proposition}
\newtheorem{lemma}[theorem]{Lemma}
\newtheorem{remark}[theorem]{Remark}
\newtheorem{exampleremark}[theorem]{Examples/Remarks}
\newtheorem{example}[theorem]{Example}
\newtheorem{assumption}[theorem]{Assumption}

\begin{frontmatter}

\title{Financial market geometry: The tube oscillator}
\runtitle{The tube oscillator}

\begin{aug}

  \author{\fnms{Dragoljub} \snm{Katic}\ead[label=e1]{d.katic@blacklace.finance}}
	\and
  \author{\fnms{Stefan}  \snm{Richter}\ead[label=e2]{stefan.richter@iwr.uni-heidelberg.de}}

  \runauthor{D. Katic and S. Richter}

  \affiliation{Blacklace Finance and Heidelberg University}

\end{aug}

\begin{abstract}:
	Based on geometrical considerations, we propose a new oscillator for technical market analysis, the tube oscillator. This oscillator measures the trending behavior of a fixed market instrument based on its past history. It is shown in an empirical analysis of the German DAX and the Forex EUR/USD exchange rate that a simple trading strategy based on this oscillator and fixed threshold leads to consistent positive monthly returns of average magnitude of 2\% or more.
	
	The oscillator is derived from a broader understanding of the geometric behavior of prices throughout a fixed period, which we term financial market geometry. The remarkable profit results of the presented technique show that 1) prices of financial market instruments have a strong underlying deterministic component which can be detected and quantified with a matching approach and 2) financial market geometry is capable of providing such detectors.
\end{abstract}


\end{frontmatter}

\bigskip
\noindent
\begin{center}
\begin{minipage}[t]{.45\textwidth}
	\small
	Blacklace Finance\\
 Josefst\"{a}dter Str. 43\\
  1080 Vienna, Austria\\
  Tel. +43 660 5353 588\\
  Email: \url{d.katic@blacklace.finance}
\end{minipage}
\noindent
\begin{minipage}[t]{.45\textwidth}
	\small
Institut f\"{u}r Angewandte Mathematik\\
  Universit\"{a}t Heidelberg\\
  Im Neuenheimer Feld 205\\
  69120 Heidelberg, Germany\\
  Email: \url{stefan.richter@iwr.uni-heidelberg.de}
\end{minipage}
\end{center}
\vspace{3mm}



\section{Introduction}

For stock market traders, technical analysis is one of the most important tools to generate trading decisions. A seminal work in the field was the book of J. Murphy in 1986 \cite{murphy1986technical}, who accumulated lots of experience in the financial markets to provide a comprehensive summary. Based on this work, several further techniques were invented and the new name 'chart analysis' emerged; for an overview, consider \cite{edward1996technical}, \cite{lo2000foundations}, 
\cite{kirkpatrick2010technical} or the more recent work \cite{edwards2018technical}.

However, there exist several works which have investigated methods of technical analysis from a mathematical point of view or based on empirical studies. They challenge its efficacy and reliability as a method for predicting market movements. In his book \cite{malkiel1999random}, Malkiel argues that market prices follow a stochastic random walk, meaning basically that they are unpredictable in the short term. The same statement was derived by Fama \cite{fama}. He criticizes technical analysis as a futile endeavor, asserting that past price movements do not provide any useful information for predicting future price movements. There exist several empirical works which examine the profitability of technical trading rules in the foreign exchange market and find that they do not generate consistent profits after transaction costs or are outperformed by simple rational traders, e.g. \cite{curcio1997technical}.

Another tool to analyze the market and price movements is stochastic finance. With the subfield of financial time series analysis, including e.g. generalized autoregressive conditional heteroscedastic models (GARCH, \cite{bollerslev1986generalized}), model-based approaches were invented to forecast future prices given the past with uncertainty quantification. The most prominent model for stock prices in continuous stochastic finance is the Black-Scholes model \cite{black1973pricing}, which formalizes the evolution of a price as an exponential random walk. While this model was improved and adjusted in many different ways, the core idea remained the same, describing the evolution of a price as a sequence of random movements. The popularity of the Black-Scholes model and the connected formulas which are used to calculate option prices and risk measurements is maybe the strongest evidence against technical analysis, whose essence is that price movements in the short-run are not random. On the other hand, the models from stochastic finance are designed in a way that they indeed allow for some deterministic movements, like the 'drift' term $\mu$ in the Black-Scholes model.

\emph{In this paper, we introduce a new subdiscipline of technical analysis called \emph{financial market geometry}. Contrary to existing chart techniques, we use a larger time frame (e. g. a whole day or more days) to determine the behavior of the price. To calculate signals, geometric properties of the chart are used. We are able to show empirically that even very basic methods constructed from our theory are able to provide positive revenue without any exhaustive parameter tuning. As an example, we introduce the so called tube oscillator.}

It shall be noted that existing works about geometric approaches like \cite{PIOTROWSKI2007228} or \cite{evertsz1995fractal} are completely different: the first working in information theory and the second analyzing financial time series in the setting of stochastic finance. 

To classify our results in current research, it should be noted that also other promising approaches exist to forecast prices. Recently, neural networks were used for nonparametric forecasting price movements \cite{lawrence1997using},\cite{di2016artificial}. It is clear that the rapid evolution of artificial intelligence allows for increasingly complex strategies, while big data sources allow to include vast amounts of information into those strategies. This may not only be the past prices, but also political decisions all over the world, regarding the country or company which one is interested in, etc. However, the past data at hand for a specific forecasting problem like the Dow Jones price is limited, in particular if one acknowledges that data from 20 years ago cannot anymore resemble mechanisms from today due to advanced technical evolution and globalization. Moreover, neural network algorithms from artificial intelligence still lack of a sound statistical description, making them hardly interpretable \cite{9380482}. This raises challenges in designing the correct architecture and selecting relevant information to avoid overfitting on past happenings and wrong forecasts. The best results with neural networks are still obtained if the underlying mechanism is well predictable also with more simple methods. A prominent example is forecasting of electricity prices \cite{SINGHAL2011550} which can be tackled also with much more simple regression models \cite{CHE20101911}. However, stock prices are well known to be hardly predictable, in particular in the short-run. This makes it unlikely that complex models are able to capture the important dynamics. Furthermore, it was often seen that simple forecasting models often outperform more complex ones \cite{green2015simple}. The problem of overfitting complex models is even more present in the case that there is high volatility, which obviously is the case for most stock prices. Overall, it seems that complex models will not be able to give meaningful forecasts without a lot of expert knowledge and some fixed guidelines provided beforehand.

\emph{In opposite to the complex methods above, financial market geometry does only use the chart of the current financial market instrument. This drastically reduces complexity of the corresponding algorithms. Due to their graphical motivation, the methods are also easy to understand and interpretable. This makes them a rational tool to develop trading decisions.}

In Section \ref{sec_framework}, the general framework of financial market geometry is explained. In Section \ref{sec_tube}, the tube oscillator is introduced as a specific method. Section \ref{sec_tube_algorithm} provides an explicit algorithm for computing the tube oscillator and discusses a simple trading strategy based on this oscillator and thresholding. Section \ref{sec_empirical} provides empirical evidence for the quality of the simple trading strategy and shows that remarkable, consistently positive returns can be achieved on instruments like German DAX 40 or Forex EUR/USD. In Section \ref{sec_conclusion}, a conclusion is drawn.

\section{The framework of financial market geometry}
\label{sec_framework}

The setting of financial market geometry is based on a description of all quantities in terms of the time $t$. In this article, it is assumed for simplicity that the time $t$ is measured in seconds, starting from some reference point $t = 0$ in the past. The \emph{last price} of the stock of interest is denoted by a real-valued sequence $(S_t)_{t\in\IN_0}$, where $\IN_0 = \{0,1,2,...\}$ are the natural numbers.

Time is separated in \emph{periods}. One may think for example about stock market days (other periods like weeks or months are also possible, but here we stick to days). The sequence describing the start of each period is denoted by $(P_i)_{i\in\IN_0}$. For the $i$-th stock market day, $P_i$ is the number of seconds after the reference point to 0:00 a.m. of that day. As an example, in timezone GMT-4 and using Unix timestamps starting at January 1st, 1970, the stock market day April 11, 2024 would be described by $P_i = 1.712.782.800$. 

\begin{figure}[h!]
	\centering
	\caption{Graphical illustration of the framework}
	\includegraphics[width=13cm]{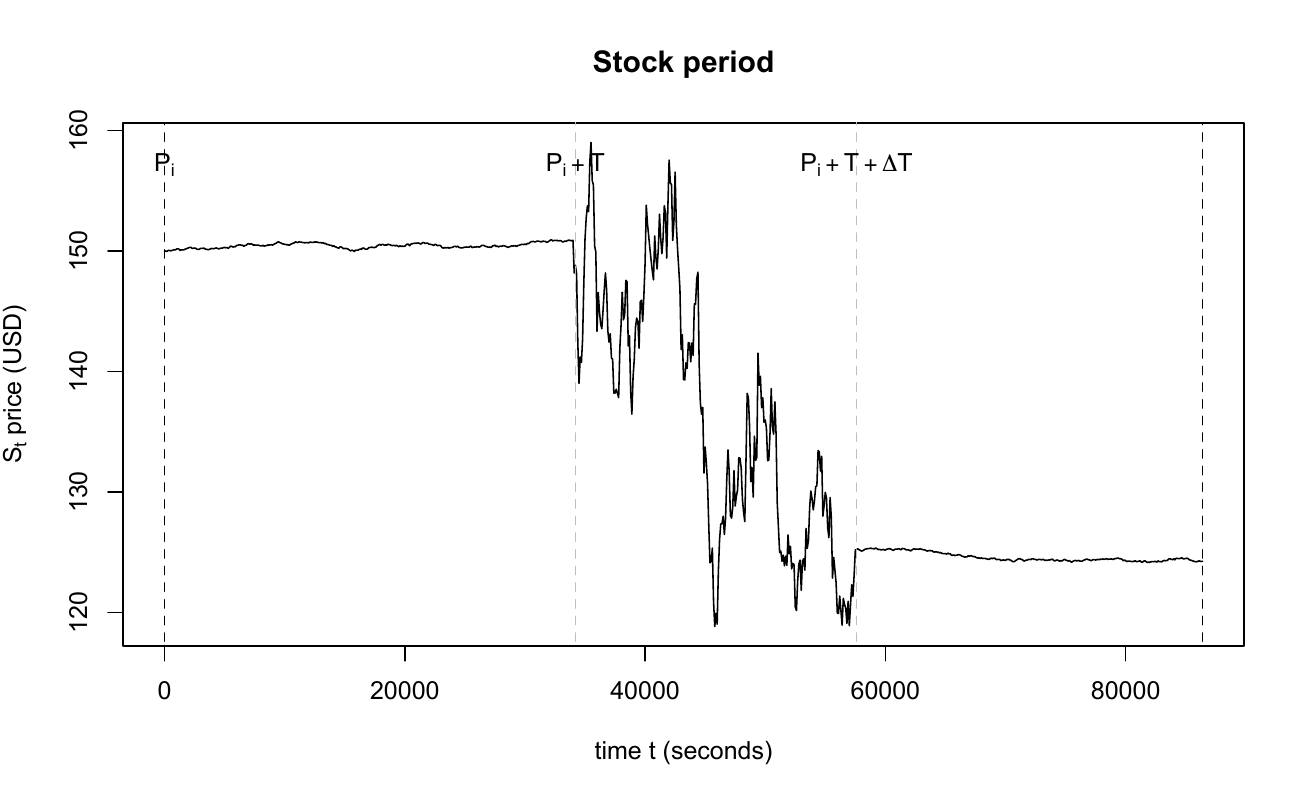}
	\label{fig1}
\end{figure}

Inside a period, there is a specific time of interest which is assumed to be same for all periods. Regarding stock market days, this would be the time of the opening times of the stock exchange where the stock $S$ is traded. For instance, the New York Stock Exchange is open from 9:30 a.m. to 4:00 p.m. (UTC/GMT-4). This can be described by the starting time $T$ and the duration $\Delta T$,
\[
	T = 3600 \cdot 9,5 = 34.200 \quad \text{(9:30 a.m.)}, \quad\quad \Delta T = 3600\cdot 6,5 = 23.400\quad \text{(4:00 p.m.)}
\]
In this way, the time $t$ is separated in \emph{zones of interest},
\[
	\sT_i = P_i + [T, T + \Delta T], \quad i\in\IN_0.
\]
For an illustration, see Figure \ref{fig1}.

\begin{figure}[h!]
	\centering
	\caption{Illustration of a line $\ell_{(t_0,s_0),m}(t)$ defined in \reff{definition_line} with $t_0 = P_i + T$, $s_0 = 140$ and $m = 0.0003$.}
	\includegraphics[width=13cm]{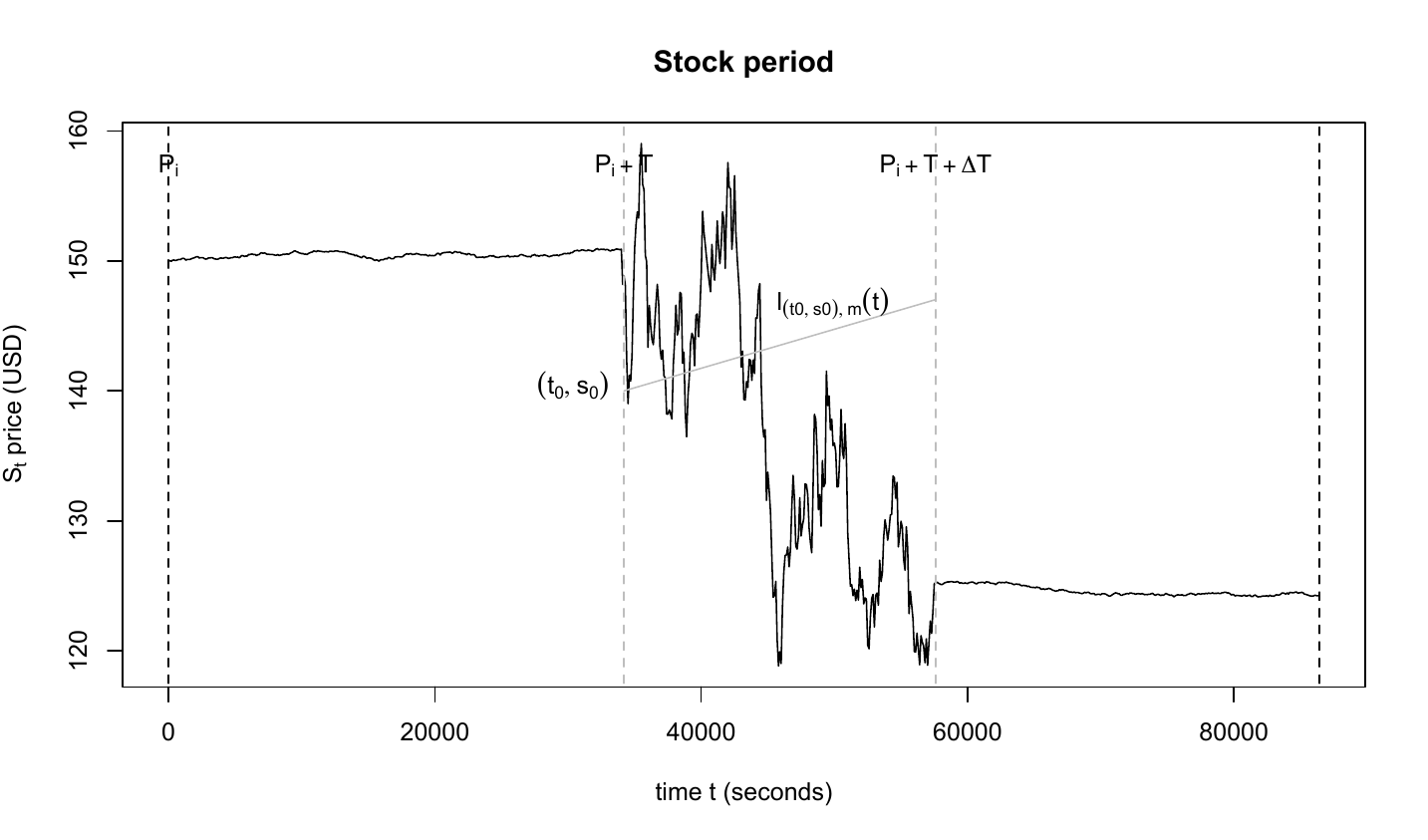}
	\label{fig2}
\end{figure}

Now, characteristics of the stock price $S_t$ in each period can be summarized. For $i\in\IN_0$, let
\begin{equation}
	M_i = \max_{t\in \sT_i} S_t, \quad\quad m_i = \min_{t\in \sT_i}S_t\label{def_max_val}
\end{equation}
be the maximum and minimum last price of the stock in period $i$. Furthermore, the closing price of the period $i$ is
\[
	C_i = S_{P_i + T + \Delta T}.
\]
The famous \emph{Intraday Pivot Point} in period $i$ then can be obtained via
\[
	PP_i = \frac{\text{High+Low+Close}}{3} = \frac{M_i + m_i + C_i}{3}.
\]
Financial market geometry has the aim to describe \emph{global} characteristics of $S_t$ in a specific time period $\sT_i$ given the knowledge of the past. An important part is the identification of meaningful supporting and resistance lines over the whole period. Breakthoughs through lines specified can be an indicator In the framework specified, lines are parametrized with a starting point $(t_0,s_0)$ and slope $m$, meaning that the line is given by
\begin{equation}
	\ell_{(t_0,s_0), m}(t) = s_0 + m\cdot (t - t_0)\label{definition_line}
\end{equation}
This is illustrated in Figure \ref{fig2}.

\section{The tube oscillator}
\label{sec_tube}

\subsection{Motivation and definition}

The \emph{tube oscillator} is derived as an example of the use of global characteristics to react to local occurencies. Additionally, it uses the central idea of technical analysis that local trends tend to persist for a certain period of time. The aim of the tube oscillator therefore is to detect if the price of a stock goes up or down \emph{in the shortrun.} It is designed for short trading periods, typically not exceeding 30 minutes.

There exist already a large variety of oscillators that have the same goal, for instance the William's percent range \%R or the stochastic oscillator (cf. \cite[Table II]{oszillatoren}). Those are built in a very basic way by considering the actual price and comparing it with past prices, making them very sensitive but also rough since they do not include lots of movement information.

The tube oscillator instead is, roughly spoken, a moving average for the current slope of the price. Unlike many smoothing approaches like kernel-based estimators, it is much more stable also for small bandwidth since its calculation uses a global grid of resistance and supporting lines.

\begin{figure}[h!]
	\centering
	\caption{Illustration of the idea of crossing lines as an indicator for the direction of the price, illustrated are lines $\ell_{(t_0,s_0),m}(t)$ with negative slope  $m = -0,002$.}
	\includegraphics[width=13cm]{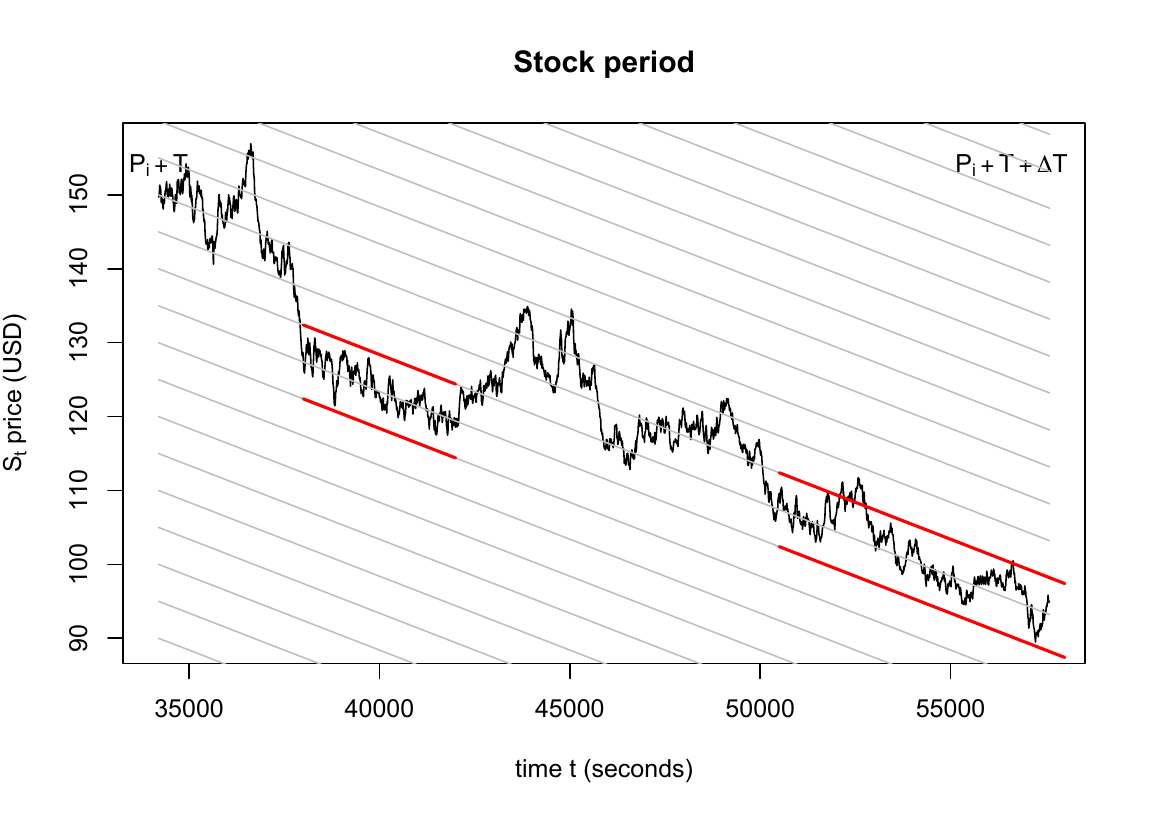}
	\label{fig3}
\end{figure}

The idea is that one tries to detect if the price curve $(t,S_t)$ currently moves inside a \emph{tube} and to calculate the slope of this motion. This is done by checking how many lines in a grid are crossed. If the curve $(t,S_t)$ crosses only few lines with negativ slope, it is more likely that there is a motion downwards, cf. Figure \ref{fig3}.

For a given straight line $\ell_{(t_0,s_0), m}$, point $(t,s)$ is \emph{above} the line if and only if
\begin{eqnarray*}
	&&\ell_{(t_0,s_0), m}(t) < s\\
	&\Longleftrightarrow& \ell_{(t_0,s_0), m}(t) - s < 0\\
	&\Longleftrightarrow& \text{sign}\big(\ell_{(t_0,s_0), m}(t) - s\big) = -1,
\end{eqnarray*}
where $\text{sign}(\cdot)$ denotes the signum function which is 1 if the argument is positive and $-1$ if the argument is negative. Therefore, crossings from time point $t$ to time point $t+1$ can be measured with
\begin{equation}
	D_{t+1}^{(t_0,s_0),m} := \frac{1}{2}\Big[\text{sign}\big(\ell_{(t_0,s_0), m}(t+1) - S_{t+1}\big) - \text{sign}\big(\ell_{(t_0,s_0), m}(t) - S_t\big)\Big],\label{def_single_slope_information}
\end{equation}
so that, graphically,
\[
	D_{t+1}^{(t_0,s_0),m} = \begin{cases}
		1, & \text{$S_t$ crosses the line from above to below}\\
		0, & \text{no crossing happens},\\
		-1, & \text{$S_t$ crosses the line from below to above}.
	\end{cases}
\]
Note that the time index of $D_{t+1}$ was denoted by $t+1$ since it includes information of time point $t+1$. It is clear that for many time points $t$, one has $D_{t+1}^{(t_0,s_0),m}=0$ when just considering one single line $\ell_{(t_0,s_0), m}$. However, the information can be accumulated by summing up the results for different lines. For a chosen slope $m$,
\[
	\text{fixed } t_0 \quad\text{and}\quad \text{chosen points } s_j, j\in \{1,...,N_s\},
\]
one defines the \emph{crossing information for slope $m$} as
\begin{equation}
	D_{t}^{m} = \sum_{j=1}^{N_s}D_t^{(t_0,s_j),m}.\label{def_crossinginformation}
\end{equation}
Here it is important that the points $s_j$ are chosen in such a way that the corresponding lines cover the whole potential window of the price in the period. In Figure \ref{fig3}, for instance, the lines were chosen with $t_0 = P_i + T$ the starting point of the period and $s_j = 85 + 5\cdot j$, $j\in \{1,...,N_s = 20\}$. The resulting $D_{t+1}^{m}$ typically still has values in $\{-1,0,1\}$, since a larger value only is possible if the price curve cross two lines from $t$ to $t+1$. This obviously also depends on how close the chosen points $s_j$ are together.

The crossing information can be accumulated along $t$ by including and weighting the information of past time points. In the most simple way, this is done by specifying a \emph{bandwidth} $\delta T \in \IN$ and performing an average over the last $\delta T$ time points as
\begin{equation}
	D_{t}^{m,\delta T} = \frac{1}{\delta T}\sum_{i=0}^{\delta T-1}D_{t-i}^{m},\label{easy_weighting}
\end{equation}
which we call \emph{accumulated crossing information for slope $m$}. More refined weightings are possible, for instance by discounting past information according to some $\gamma \in (0,1)$ (close to 1, e.g. $\gamma = 0.99$) via
\begin{equation}
	D_{t}^{m,\delta T,\gamma} = \frac{1}{\delta T}\sum_{i=0}^{\delta T-1}\gamma^{i} \cdot D_{t-i}^{m}.\label{discounted_weighting}
\end{equation}
In the presentation, we stick to the more simple version in equation \reff{easy_weighting}.

\begin{figure}[h!]
	\centering
	\caption{Accumulated crossing information for fixed slope $m = -0,002$ based on lines with starting points $s_j = 85 + 5\cdot j$, $j\in \{1,...,N_s = 20\}$ (top) and $s_j = 85 + 1\cdot j$, $j\in \{1,...,N_s = 100\}$ (bottom).}
	\begin{tabular}{c}
		\includegraphics[width=13cm]{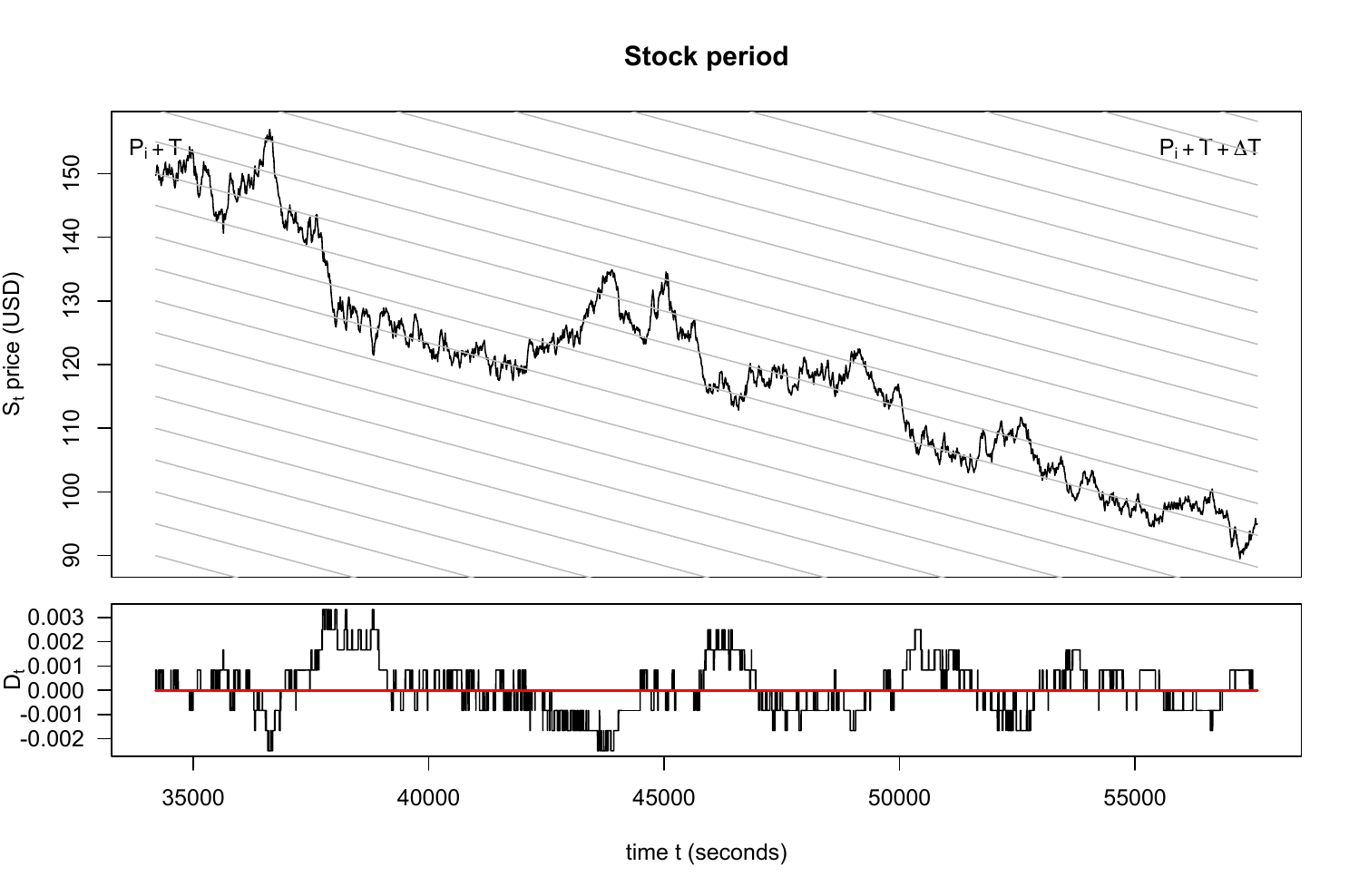}\\
		\includegraphics[width=13cm]{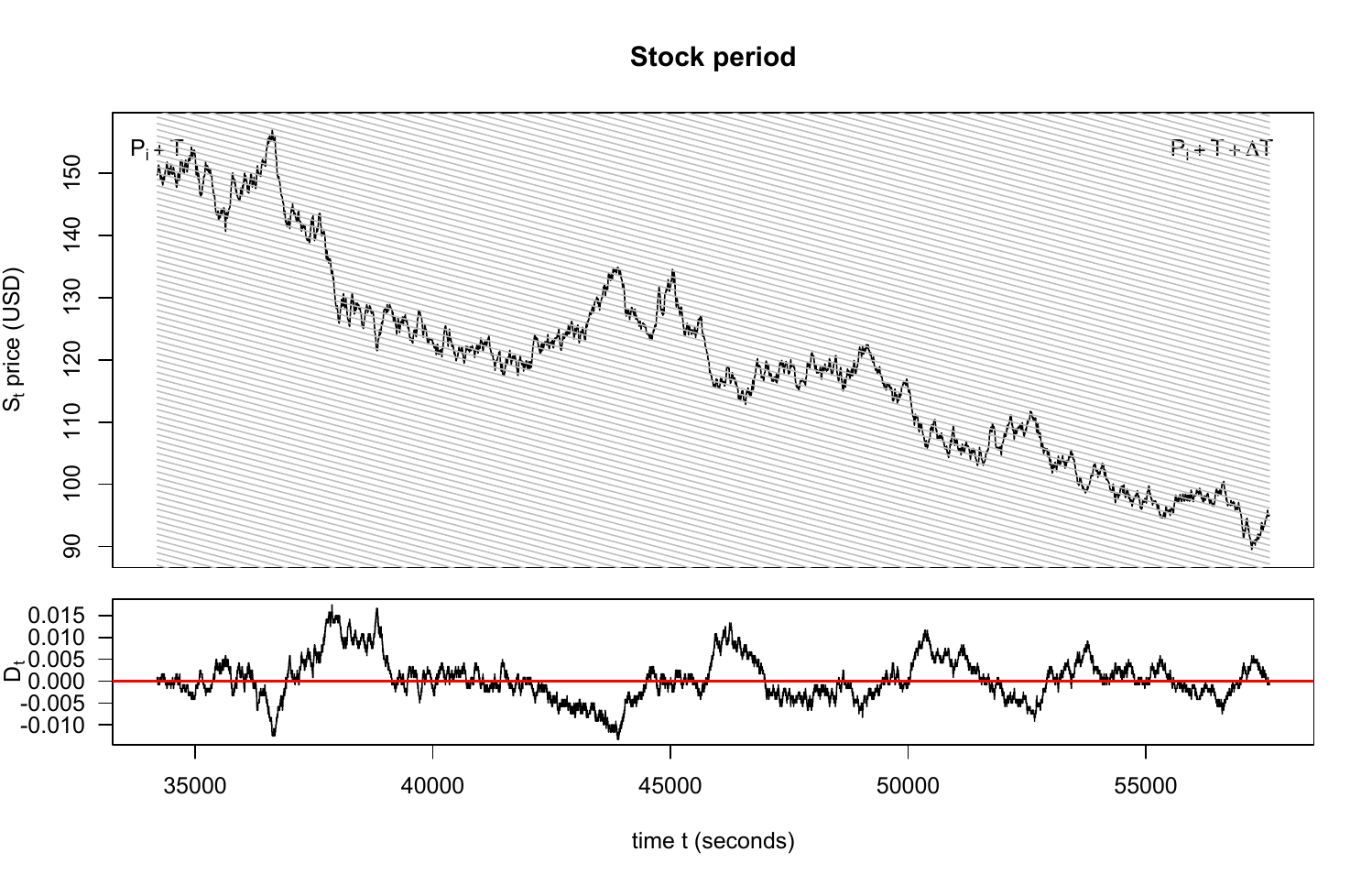}
	\end{tabular}
	\label{fig4}
\end{figure}

The resulting $D_{t}^{m,\delta T}$ with $\delta T = 1200$ (which corresponds to 1200 seconds, that is, 20 minutes) for the price from Figure \ref{fig3} is shown in Figure \ref{fig4}. For instance, at time $t \approx 38000$ relatively large values of $D_{t}^{m,\delta T}$ are observed, which is due to the fact that in the last 1200 seconds before this time point, several crossings from 'above to below' did happen. 

\begin{figure}[h!]
	\centering
	\caption{One specific selection of slopes based on equation \reff{generation_slopes} with $m_{basic}=0,003$ and $f_k = \tan(\frac{\pi}{2}\cdot \frac{k}{10})$.}
	\includegraphics[width=10cm]{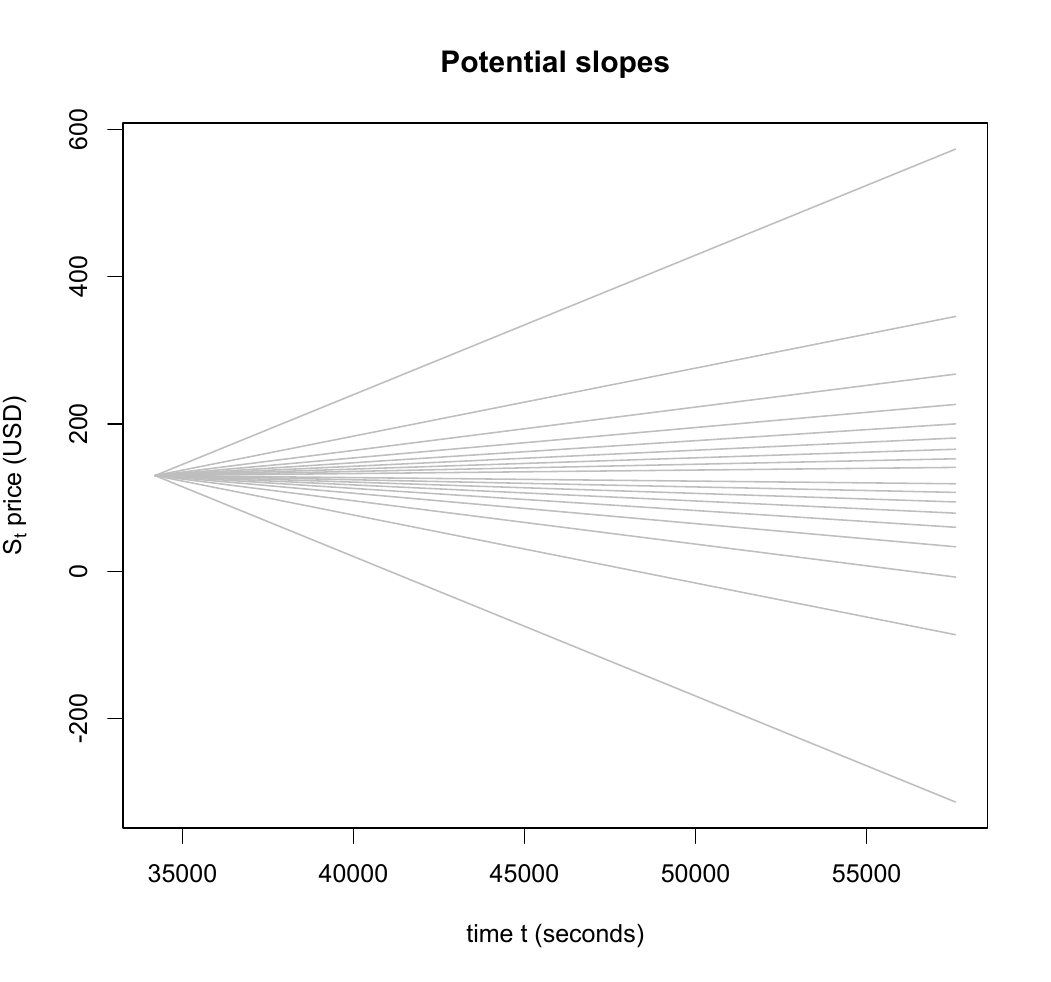}
	\label{fig5}
\end{figure}

The power of the machinery comes now from choosing a more dense grid of points $s_j$ (cf. Figure \ref{fig4} below) combined with different slopes. For instance, one may select a \emph{basic slope} $m_{basic} > 0$, and dependent on this an array of factors,
\[
	f_k > 0, \quad k \in \{1,...,N_f\}
\]
which produces all slopes by multiplying the positive and negative version of $m_{basic}$,
\begin{equation}
	m_{2k} = m_{basic}\cdot f_k, \quad\quad m_{2k+1} = -m_{basic}\cdot f_k, \quad k\in \{1,...,N_f\}.\label{generation_slopes}
\end{equation}
An example is given in Figure \ref{fig5}, where it was chosen $m_{basic} = 0.003$ and $f_k = \tan(\frac{\pi}{2}\cdot \frac{k}{10})$, $k\in \{1,...,N_f = 9\}$.

\begin{figure}[h!]
	\centering
	\caption{The tube oscillator based on slopes from \reff{generation_slopes} with $m_{basic}=0,003$ and $f_k = \tan(\frac{\pi}{2}\cdot \frac{k}{10})$, as well as $s_j = 85 + 1\cdot j$, $j\in \{1,...,N_s = 100\}$ as starting points for three different bandwidths $\delta T \in \{300,600,1200\}$.}
	\includegraphics[width=13cm]{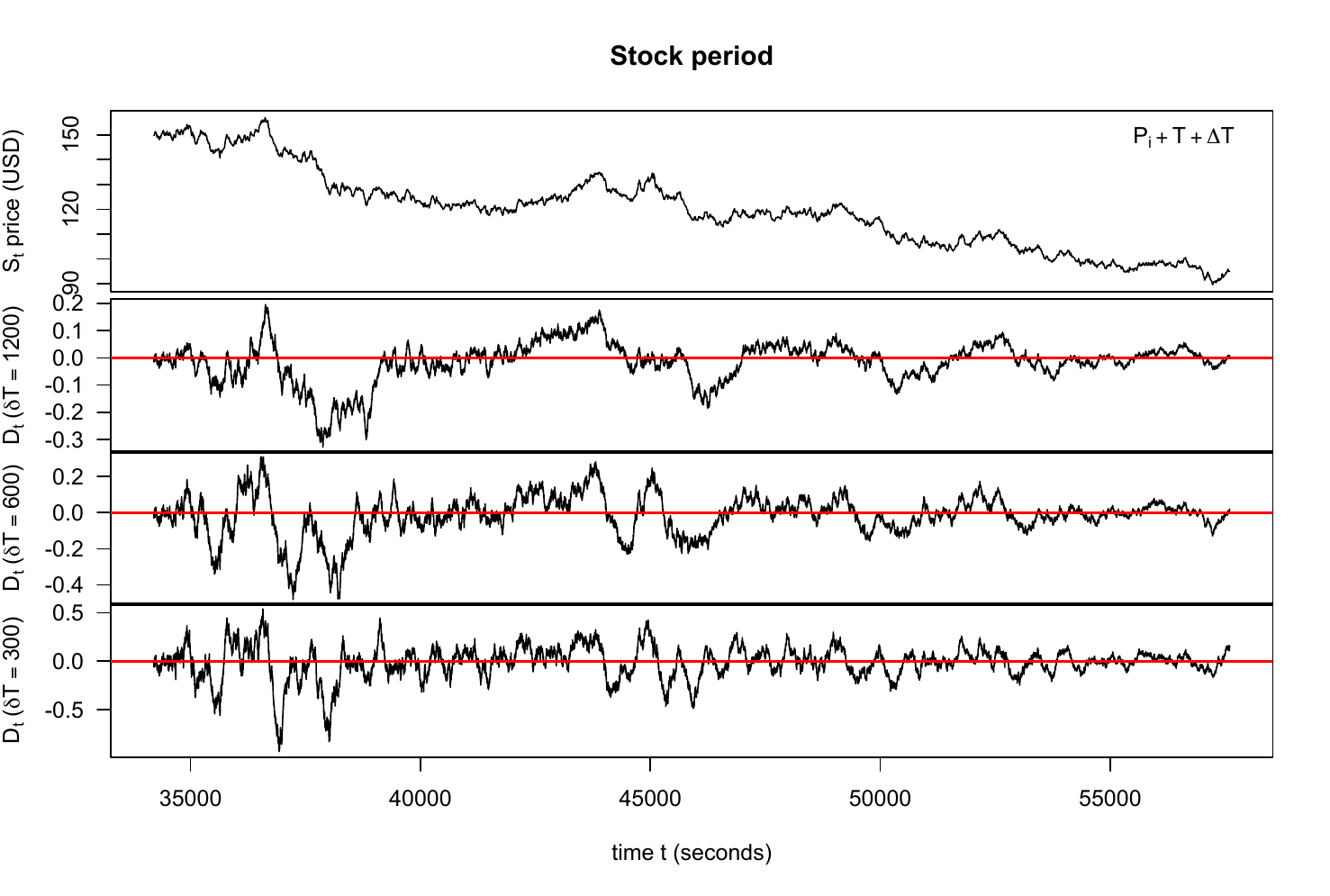}
	\label{fig6}
\end{figure}

The final \emph{tube oscillator} is now obtained by \emph{averaging and inverting} the information of the different $D_{t}^{m_k,\delta T}$ as
\begin{equation}
	O_t^{\delta T} = -\frac{1}{2N_f}\sum_{k=1}^{2N_f}D_{t}^{m_k,\delta T}.\label{def_oscillator}
\end{equation}
The inversion has to take place so that $O_t$ is \emph{positive} if there is a positive slope of the price, and \emph{negative} if there is a negative slope. An example for the slopes above with $O_t^{\delta T}$ for $\delta T \in \{300,600,1200\}$ is provided in Figure \ref{fig6}.

It is nicely seen that the oscillator $O_t^{\delta T}$ is capable of detecting the slope of the price. The version with $\delta T = 300$ (which refers to an averaging of 300 seconds = 5 minutes in the past) is able to detect every slope change after at most 5 minutes.\\

\emph{Normalization}

In contrast to other oscillators like William's \%R or the stochastic oscillator whose values are normalized to an interval $[0,1]$, the tube oscillator $O_t^{\delta T}$ is \emph{not normalized}. The reason is that compared to the aforementioned oscillators, the formula does not consist of ratio where the nominator is smaller than the denominator. Instead, the tube oscillator measures the slope in an absolute way.

The strongest influence on the size of $O_t^{\delta T}$ is the number, or more specifically, the denseness of the points $s_j, j=1,...,N_s$ used for constructing the lines. The more dense the points are, the more often it may occur that a price movement not only crosses one but several lines, which substantially increases the crossing information $D_t^m$ in equation \reff{def_crossinginformation}. Compared to that, the influence of the number of slopes $N_f$ used is marginal due to the averaging in the construction (cf. \reff{def_oscillator}). The same holds for the  bandwidth $\delta T$ in \reff{easy_weighting}. Therefore, one can expect that for a fixed time discretization and a fixed stock, it is possible to \emph{calibrate} the values in a stable way as long as the stock does not vary in its general behavior too much over time. This is also verified in Section \ref{sec_empirical}. 

Finally, it should be mentioned that the missing normalization can also be seen as an advantage. The normalization of, e.g, the stochastic oscillator or William's \%R is obtained by forming the ratio of the actual price (minus lowest past price) and highest past price (minus lowest past price). If the price observed right now is the highest price, William's \%R will be constantly 100\%, providing only few information about the nature of the price movement. Contrary to that, the tube oscillator is an average over the number of crossings in a fixed grid of lines. Thus, it cannot grow arbitrarily high and there is indeed some kind of normalization if the maximum of possible crossings is achieved. However, for real data this 'cap' is not achieved and the tube oscillator permanently provides information.

\section{Tube oscillator computation and trading}
\label{sec_tube_algorithm}

\subsection{Tube oscillator computation}
Summarizing the construction, the tube oscillator is computed based on the equations \reff{def_single_slope_information}, \reff{def_crossinginformation},  \reff{easy_weighting} and \reff{def_oscillator} as follows:

\begin{definition}[Tube oscillator computation]\noindent\\
	\label{def_oscillator_definition}\textbf{Input:}
	\begin{itemize}
		\item a set of starting points $s_j > 0$, $j \in \{1,...,N_s\}$ for the lines of the grid,
		\item a set of slope factors $f_k \in \IR$, $k \in \{1,...,N_f\}$, a basic slope $m_{basic} > 0$,
		\item a bandwidth $\delta T \in \IN$.
		\item time $t\ge P_i + T$ where the oscillator shall be computed
	\end{itemize}
	\textbf{Output:} Oscillator value $O_{t}^{\delta T}$ at time $t$.\\
	\textbf{Algorithm:}
	\begin{itemize}
		\item[1)] Put $t_0 = P_i + T$.
		\item[2)] Define $m_{2k} = m_{basic}\cdot f_k$, $m_{2k+1} = -m_{basic}\cdot f_k$ ($k\in \{1,...,N_f\}$).
		\item[3)] For $t \ge t_0$, compute
		\begin{itemize}
			\item[3.1)] For each $k\in \{1,...,2N_f\}$,
			\begin{itemize}
				\item[3.1.1)] For each $j \in \{1,...,N_s\}$, compute
				\begin{eqnarray*}
				D_{t}^{(t_0,s_j),m_k} &:=& \frac{1}{2}\Big[\text{sign}\big(s_j + m_k\cdot (t - t_0) - S_{t}\big)\\
				&&\quad\quad - \text{sign}\big(s_j + m_k\cdot (t - 1 - t_0) - S_{t-1}\big)\Big].
				\end{eqnarray*}
			\end{itemize}
			\item[3.2)] Compute
			\begin{eqnarray*}
				D_{t}^{m_k} &:=& \sum_{j=1}^{N_s}D_t^{(t_0,s_j),m_k},\\
				D_{t}^{m_k,\delta T} &:=& \frac{1}{\delta T}\sum_{i=0}^{\delta T-1}D_{t-i}^{(t_0,s_j),m_k}.
			\end{eqnarray*}
		\end{itemize}
		\item[4)] Compute the tube oscillator
		\[
				O_t^{\delta T} := -\frac{1}{2N_f}\sum_{k=1}^{2N_f}D_{t}^{m_k,\delta T}.
		\]
	\end{itemize}
\end{definition}

Suitable values of $s_j$ can be chosen as follows:
\begin{itemize}
	\item Depending on how fast the oscillator shall update to new behavior, $\delta T$ can be chosen as $\delta T = 300$ (seconds) or larger.
	\item A good choice for the basic slope is the slope observed in the last period $P_{i-1}$, given by
	\[
		m_{basic} = \frac{M_{i-1} - m_{i-1}}{\Delta T},
	\]
	where $\Delta T$ is the length of the whole period and $M_{i-1}$, $m_{i-1}$ are the maximum and minimum price observed in that period, cf. \reff{def_max_val}. It is also possible to choose a fixed slope which is obtained by inspecting the general variation expected over one day of the stock price.
	\item The slope factors $f_k$ can be, for instance, chosen as
	\[
		f_k = \tan(\frac{\pi}{2}\cdot \frac{k}{10}), \quad k\in \{1,...,N_f = 9\}.
	\]
	\item The set of starting points has to be chosen in a way that the resulting grid covers well and fine enough the price curve $(t,S_t$). This can be done by an inspection of past periods. A good starting point is to use the difference of maximum and minimum of the past period, $\Delta S := M_{i-1} - m_{i-1}$, and choose
	\[
		s_j = S_{t_0} - 2 \Delta S + \frac{j}{N_s}\cdot 4\Delta S, \quad j = 1,...,N_s, \quad\quad N_s \approx 300.
	\]
\end{itemize}

It is important to note three basic positive facts:
\begin{itemize}
	\item First, the oscillator is an average over values of the past, but much more stable than usual kernel-based weighting schemes since the approach is based on a global grid of supporting and resistance lines.
	\item Second, even though it has a relatively complex construction, it is easy to understand from a graphical point of view.
	\item Third, it can be computed fast with basic operations, therefore an online calculation scheme can be constructed such that the already existing oscillator value $O_t^{\delta T}$ can be updated based on a new price observation at time $t+1$.
\end{itemize}

\subsection{Trading with the Tube Oscillator}

The most simple way to do trading solely based on the tube oscillator is as follows: Since a high (positive) value indicates an increasing price and a strongly negative value indicates a decreasing price, one shall choose four parameters:

\begin{figure}[h!]
	\centering
	\caption{The arrangement of the thresholds chosen.}
	\includegraphics[width=8cm]{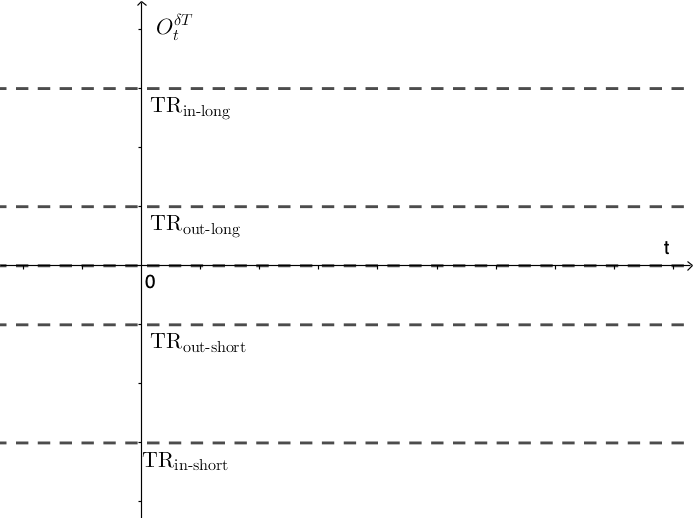}
	\label{fig_tresholds}
\end{figure}

\begin{itemize}
	\item The thresholds for opening and closing a long position, $\TR_{\text{in-long}} > \TR_{\text{out-long}} > 0$. If the tube oscillator is greater than $\TR_{\text{in-long}}$,
	\[
		O_t^{\delta T} > \TR_{\text{in-long}},
	\]
	a long position on the stock shall be opened. If the value of the oscillator drops below $TR_{\text{out-long}}$,
	\[
		O_t^{\delta T} < \TR_{\text{out-long}},
	\]
	the corresponding position shall be closed.
	\item The thresholds for opening and closing a short position, $\TR_{\text{in-short}} < \TR_{\text{out-short}} < 0$. Vice versa to the situation for longs, $O_t^{\delta T} < \TR_{\text{in-short}}$ now signalizes opening a short position while $O_t^{\delta T} > \TR_{\text{out-short}}$ indicates to close this position. If one does not expect any specific positive or negative movements of the price $S_t$, it seems reasonable to set the thresholds for short positions as the negative values fixed for long positions,
\[
	\TR_{\text{in-short}} := -\TR_{\text{in-long}},\quad\quad \TR_{\text{out-short}} := -\TR_{\text{out-short}}.
\]
This reduces the number of parameters to choose to only two.
\end{itemize}
For a graphical visualization see Figure \ref{fig_tresholds}. The trading according to those thresholds can be formalized as follows.

\begin{definition}[Trading with the tube oscillator]\label{tradingstrategy}\noindent\\
	\label{def_algorithm_trading}\textbf{Input:}
	\begin{itemize}
		\item the tube oscillator $O_t^{\delta T}$, $t \in \sT_i$
		\item Thresholds for long positions, $\TR_{\text{in-long}} > \TR_{\text{out-long}} > 0$,
		\item (Thresholds for short positions, $\TR_{\text{in-short}} < \TR_{\text{out-short}} < 0$),
		\item Trade volume $v$ for each trade.
	\end{itemize}
	\textbf{Output:} Trading strategy.\\
	\textbf{Algorithm:}
	\begin{itemize}
		\item[1)] Put $t_0 = P_i + T$.
		\item[2)] For $t \ge t_0$,
		\begin{itemize}
			\item[2.1)] If no position is open:
			\begin{itemize}
				\item[2.1.1)] If $O_t^{\delta T} > \TR_{\text{in-long}}$, then open a long position with trade volume $v$.
				\item[2.1.2)] If $O_t^{\delta T} < \TR_{\text{in-short}}$, then open a short position with trade volume $v$.
			\end{itemize}
			\item[2.2)] If a position is open:
			\begin{itemize}
				\item[2.2.1)] If a long position is open and $O_t^{\delta T} < \TR_{\text{out-long}}$, close the long position.
				\item[2.2.2)] If a short position is open and $O_t^{\delta T} > \TR_{\text{out-short}}$, then close the short position.
				\item[2.2.3)] If the end of the period $\sT_i$ is reached, the position is closed.
			\end{itemize}
		\end{itemize}
	\end{itemize}
\end{definition}

It shall be noted that this trading algorithm \emph{only} depends on the values of the tube oscillator. The algorithm can be extended by implementing stop-loss or more sophisticated mechanisms to control the closing of the positions.

It is clear that the choice of the thresholds is very important. Roughly spoken, the higher $\TR_{\text{in-long}}$ (or lower $\TR_{\text{in-short}}$, respectively), the stronger the signal of the oscillator has to be to open a position. Therefore, higher 'in'-thresholds lead to less trades. Conversely, the lower $\TR_{\text{out-long}}$ (or the higher $\TR_{\text{out-short}}$, respectively), the longer the position will be kept open, increasing the possibility for losses but also making the procedure more robust for small deviances from the general movement. Based on the original idea, the 'out'-thresholds should not be chosen too near to 0. The reason is that the tube oscillator only detects price movements on a very short time scale; but the longer the position is open, the less probable it becomes that the same movement persists. In other words: By choosing the 'out'-thresholds near to 0 (or even, incorrectly specified, $\TR_{\text{out-long}} < 0$ or $\TR_{\text{out-short}} > 0$), the position, originally opened on a mathematical founded basis, is left to the randomness of the market.

The thresholds have to be calibrated in accordance with the tube oscillator values. It is important to note that only \emph{one} calibration as long as the very overall behavior or magnitude of the price does not change. In Section \ref{sec_empirical} it is seen that over a period of 5 years, no change of the thresholds was necessary to provide satisfying results.

\section{Empirical verification}
\label{sec_empirical}

In this section it is shown that trading based on the tube oscillator can lead to remarkable profits. Since the oscillator indicates short range movements, it is necessary to use instruments which are (highly) liquid on the market, allowing for buying and selling in seconds. Two examples are the German DAX 40 and the Forex EUR/USD exchange rate. Data in tick resolution for Ask and Bid price from January 2019 to May 2024 were obtained from \cite{dukascopy}. All computations were performed with scripts in the statistics program R, version 4.3.1. In a first step, the tick data was summarized to \emph{data with resolution in seconds} by assigning to each second of the day the last prices observed. Note that this manipulation mimics a trader who gets the prices each second. It does not introduce any knowledge from the future.

\subsection{Oscillator and trading parameters}

The trading is performed solely based on the algorithm from Definition \ref{def_algorithm_trading}, with the convention that the thresholds for long and short positions are the same in absolute value,
\[
	\TR_{\text{in-short}} := -\TR_{\text{in-long}},\quad\quad \TR_{\text{out-short}} := -\TR_{\text{out-short}}.
\]
As a trading period serves one day. Regarding the trading environment, the following conventions are made:
\begin{itemize}
	\item Opening a short position (at bid price) or a long position (at ask price) can be performed instantly (or with maximum delay of one second). The same holds for closing a short position (at ask price) or a long position (at bid price). In particular, the empirical results take into account the loss of one bid-ask-spread for each trade.
	\item Besides the bid-ask-spread, there are no additional costs for opening or closing a position.
\end{itemize}
These two conventions are widely used in testing trading algorithms (e.g. in the MetaTrader). Note that the impact of these conventions on the overall results is small, since additional costs from brokers are negligible by scaling the investments. Furthermore, as seen below in the empirical results, the time a position is hold is in average longer than 100 seconds, and a delay of one or even some seconds does not influence the profit much.\\

The oscillator was computed according to Definition \ref{def_oscillator_definition}. The parameters of the procedure were not changed over the whole period from 2019 to 2024, and were chosen as given in Table \ref{table_parameters}. In particular, the starting points $s_j$ of the lines were chosen equidistantly around the first price of the day's period. The tube oscillator was multiplied with a constant Multiplicator of 20 or 50, respectively, such that its peaks are roughly around $\pm 0.5$, that is, instead of $O_t^{\delta T}$ it was used
\[
	\tilde O_t^{\delta T} = \text{Multiplicator} \cdot O_t^{\delta T}.
\]
\emph{This was done only for presentational purposes}; the value of the oscillator only comes into play through comparison with the thresholds; the same trading results are obtained with the original tube oscillator by dividing the thresholds by the Multiplicator. The detailed parameters used are presented in Table \ref{table_parameters}. To analyze the stability of the trading strategy, in total four configurations were investigated: The bandwidth $\delta T$ of the oscillator was either chosen to be 300 or 600 seconds in the past, and the thresholds used to indicate the trading actions were either chosen to be 0.4/0.1 (which provides a more active strategy) or 0.8/0.2 (which provides a more conservative strategy).

\begin{table}[h!]
	\centering
	\begin{tabular}{rr|c|c|}
		&& DAX & EUR/USD\\
		\hline
		Daily period && 9 a.m. - 6 p.m. MESZ & 1 p.m. - 22 p.m. MESZ\\
		\hline
		Basic slope $m_{basic}$ && \#\#\# & \#\#\#\\
		\hline
		Slopes $f_k$ && \#\#\# & \#\#\#\\
		\hline
		$N_s$ && \#\#\# & \#\#\#\\
		Starting points $s_j$ && \#\#\# & \#\#\#\\
		\hline
		bandwidth $\delta T$/ Multiplicator $\delta T$ & Config. 1 & \multicolumn{2}{|c|}{600s / 50}\\
		& Config. 2 & \multicolumn{2}{|c|}{300s / 20}\\
		\hline
		\hline
		$\TR_{\text{in-long}}$ / $\TR_{\text{out-long}}$ & Config. A & \multicolumn{2}{|c|}{0.8 / 0.2}\\
		 & Config. B&  \multicolumn{2}{|c|}{0.4 / 0.1}\\
		\hline
	\end{tabular}
	\caption{The parameters chosen for computation of the tube oscillator the German DAX 40 and the Forex EUR/USD. \textbf{In this preliminary version, the parameters are not disclosed}. The time $t_0$ represents the first time point of the period (9 a.m for DAX, 1 p.m. for EUR/USD), such that $S_{t_0}$ is the first price observed during the trading day. The thresholds were chosen symmetric around 0, that is, $\TR_{\text{in-long}}:= -\TR_{\text{in-short}}$ and $\TR_{\text{out-long}} := -\TR_{\text{out-short}}$.}
	\label{table_parameters}
\end{table}

Note that the procedures for DAX and EUR/USD only differ regarding the basic slope $m_{basic}$ and the grid of the starting points. This adjustment is necessary due to the different magnitudes of the units. The units of the DAX have an average order of magnitude around $\approx 10^4$ while EUR/USD is an exchange rate with an order of magnitude $\approx 1$. Additionally, besides the trading volume $v$, the values in Table \ref{table_parameters} describe the oscillator computation and the trading procedure completely.\\

The values in Table \ref{table_parameters} used for this backtesting were \emph{not optimized} and \emph{not} cherry-picked. The chosen basic slopes $m_{basic}$ roughly lead to slopes which cover the typical variation of the prices during a day. 
The bandwidths $\delta T \in \{300s, 600s\}$ were chosen rather arbitrarily. Only the threshold values and the Multiplicator were adjusted visually along 3 trading days to calibrate the oscillator. The fact that the same values are used for DAX and EUR/USD, and their simple appearance (0.4 and not 0.374892...) should underline that the procedure was not fitted to the specific prices.\\

The \emph{trading strategy} used is the simple trading strategy described in Definition \ref{tradingstrategy}. No adjustments were made, in particular no stop loss was introduced and no leverage instruments were used. Outside the daily period, the oscillator was set to 0, forcing a stop of open positions at the end of the daily period according to the trading strategy. To analyze the evolution of the strategies in the long run, a starting balance of 10'000 EUR was used and each trade was performed with a 100\% investment of the current balance.\\

\subsection{Results}

All results are presented for both DAX and EUR/USD and all four configurations of the bandwidth and the thresholds (cf. Table \ref{table_parameters} for those configurations). As main indicator of profitability we have chosen the monthly returns and the Sharpe Ratio. More detailed, the monthly Sharpe Ratio and the extended yearly Sharpe Ratio were computed. The monthly Sharpe Ratio is given as $\text{SR}_{monthly} = \frac{\overline{r_t} - \overline{rf_t}}{\sigma_{r_t - rf_t}}$, where $r_t$ are the monthly returns and $rf_t$ the risk-free rates. The extrapolated annual Sharpe Ratio is obtained as $\text{SR}_{yearly} = \sqrt{12}\cdot \text{SR}_{monthly} \approx 3.464\cdot \text{SR}_{monthly}$. One-Month U.S. treasury bond yields as risk-free rates were taken from the Federal Reserve Bank of St. Louis \cite{federalbank}. They are provided as annual rates on a daily basis. To obtain one-month rates, they were compounded linearly according to $r_{one-month} = \frac{r_{one-year}}{12}$. The daily data was then summarized with averaging to a 1-month treasury bond yield for each month, producing a risk-free rate $rf_t$ for each month.\\

\begin{figure}[h!]
	\centering
	\caption{EUR/USD, Bandwidth 300s, In/Out: 0.4/0.1: Behavior of the trading strategy from 2019/01 to 2024/05. The ask price is in the upper part of the plot (together with the starting points and slopes used for the oscillator in grey), while below the computed oscillator is shown. In the oscillator plots, blue points mark signals for going-in longs and orange points mark signals for going-in shorts. The trades performed are depicted in the price plot as blue (longs) and orange (shorts) lines. The text at the lines shows the profit of the trade, positive numbers are green and indicate a winning trade, negative numbers are red and indicate a loosing trade. Due to the small range of the price, the profits were multiplied by 1000 only for visual purposes. The price and also the oscillator were only plotted at 1000 time points during the day. Note that a long is bought for the ask price but sold for the bid price, therefore the line ends do not always correspond to the ask price in the price plot.}
	\begin{tabular}{c}
		\includegraphics[width=13cm]{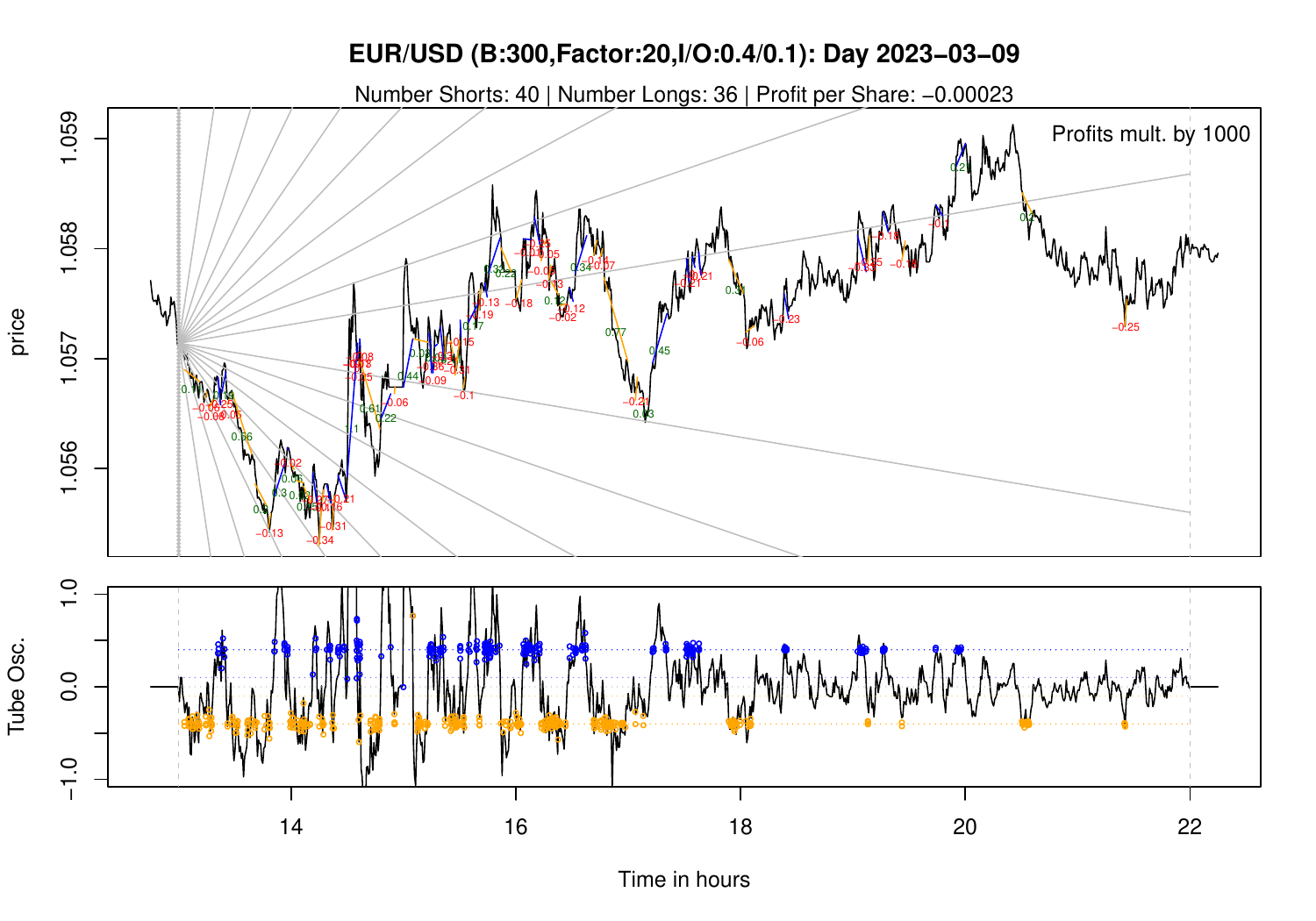}\\
		\includegraphics[width=13cm]{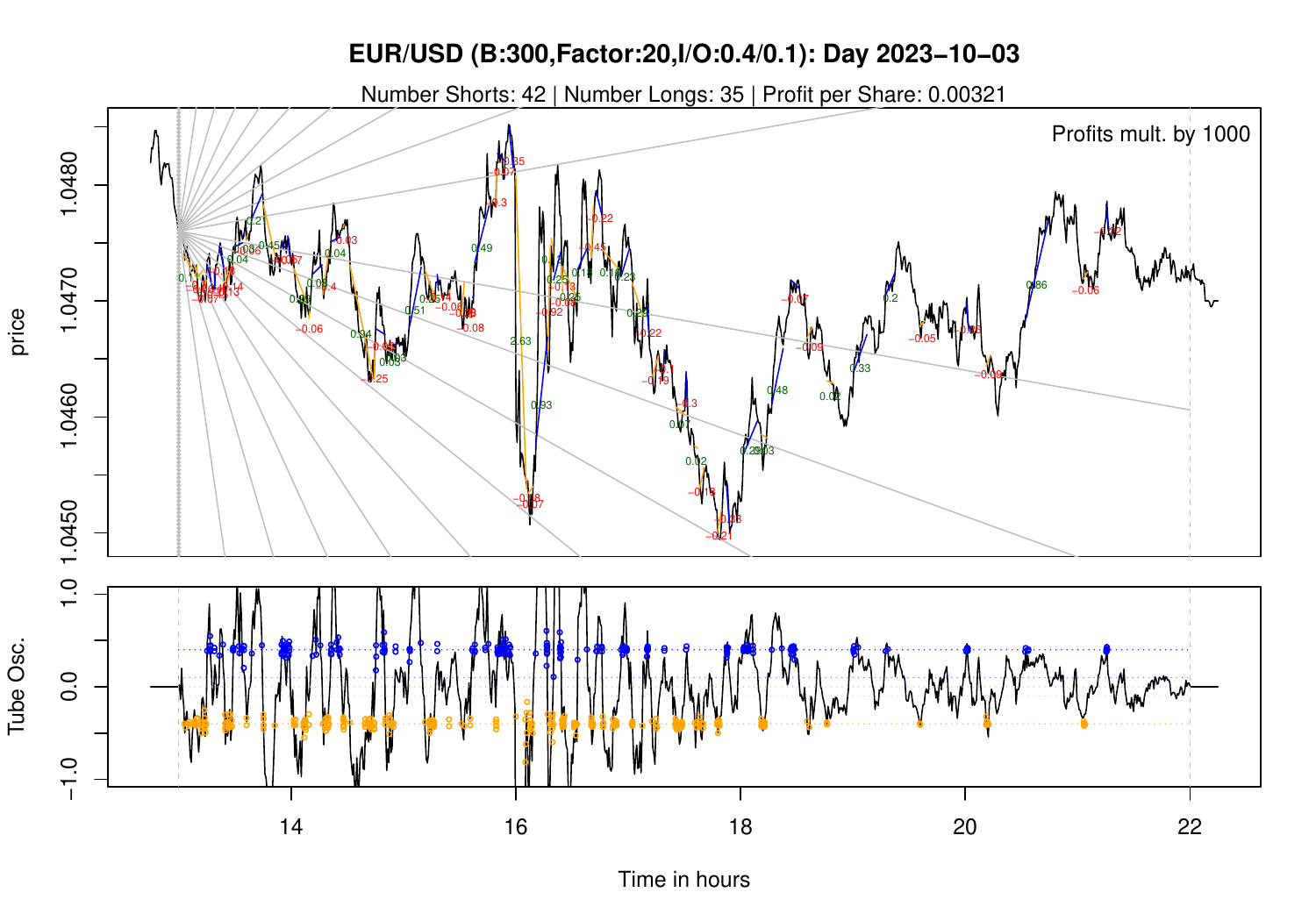}\\
	\end{tabular}
	\label{fig_EURUSD_300_OSC20_IN04_OUT01_example}
\end{figure}

The results of the trading strategy for the two instruments EUR/USD and DAX are now discussed separately. First, Figure \ref{fig_EURUSD_300_OSC20_IN04_OUT01_example} shows the behavior of the trading strategy on two days with the oscillator configuration bandwidth 300s and I/O 0.4/0.1.

Tables \ref{tab2_EURUSD_results} summarizes the results for the trading strategies applied to EUR/USD. Note that one 'share' is considered as the instrument price, which ranges between 0.96 and 1.2 in the whole trading time from 2019/01 to 2024/05. The best configuration regarding total profit (bandwidth 300s, I/O 0.4/0.1) lead to average monthly returns of $4.659\%$ (standard deviation: $3.334\%$) and median monthly returns of $4.422\%$ (MAD: $2.438\%$). The corresponding monthly Sharpe Ratio was 1.352 (extrapolated yearly Sharpe Ratio: 4.683). Over the whole trading time from 2019/01 to 2024/05, the initial capital of 10'000 EUR could be increased to 187'016 EUR (initial capital plus profit) in 86'774 trades, which is an overall return on investment of
\[
	\frac{187'016 - 10'000}{10'000} \approx 1770 \%.
\]

\begin{table}[h!]
	\centering
	\caption{EUR/USD results: Monthly and (extrapolated) yearly Sharpe Ratio SR, as well as monthly return characteristics of the trading strategy over the time horizon from 2019/01 until 2024/05. $\delta T$ is the bandwidth of the oscillator used, Mult. $O_{t}^{\delta T}$ refers to the Multiplicator used to enlarge the oscillator, I/O are the thresholds used for starting and stopping a position. Average refers to the mean (with standard deviation in brackets), while median refers to the statistical median (with mean absolute deviation from the median in brackets).}
	\begin{tabular}{|r|r|r|r|r|r|r|r|}
		$\delta T$ & Mult. $O_{t}^{\delta T}$ & I/O & $SR_{monthly}$ & $SR_{yearly}$ & Average Return/month & Median Return/month & Total Profit\\
		\hline
			600s & 50 & 0.4/0.1 & 1.362 & 4.718 & 3.989 (2.822)\% & 3.045 (2.169)\% & 114230.10 \\
		\hline
			600s & 50 & 0.8/0.2 & 1.249 & 4.326 & 2.369 (1.788)\% & 1.970 (1.389)\% & 35367.15 \\
		\hline
			300s & 20 & 0.4/0.1 & 1.352 & 4.683 & 4.659 (3.334)\% & 4.422 (2.438)\% & 177016.30 \\
		\hline
			300s & 20 & 0.8/0.2 & 1.276 & 4.422 & 2.453 (1.821)\% & 2.056 (1.404)\% & 37837.91 \\
		\hline
	\end{tabular}
	\label{tab2_EURUSD_results}
\end{table}

However, all other configurations provided results in the same order of magnitude: Even the 'worst' configuration (bandwidth 600s, I/O 0.8/0.2) had average monthly returns of $2.369\%$ (standard deviation: $1.788\%$) and median monthly returns of $1.970\%$ (MAD: $1.389\%$) with a Sharpe Ratio of 1.249 (extrapolated yearly Sharpe Ratio: 4.326).\\

\begin{figure}[h!]
	\centering
	\caption{EUR/USD, Bandwidth 300s, In/Out: 0.4/0.1: Characteristics of the trading strategy from 2019/01 to 2024/05.}
	\begin{tabular}{cc}
		\includegraphics[width=6.5cm]{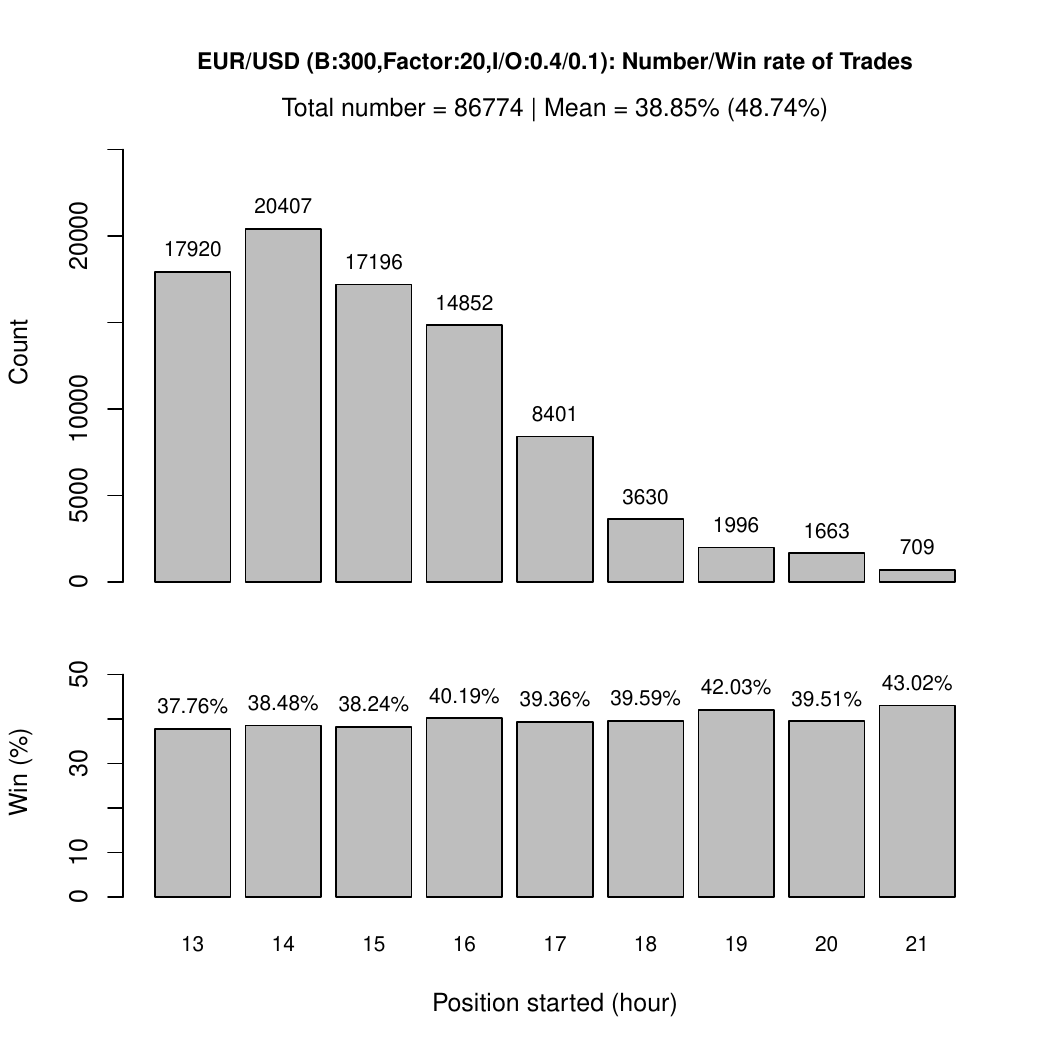} & \includegraphics[width=6.5cm]{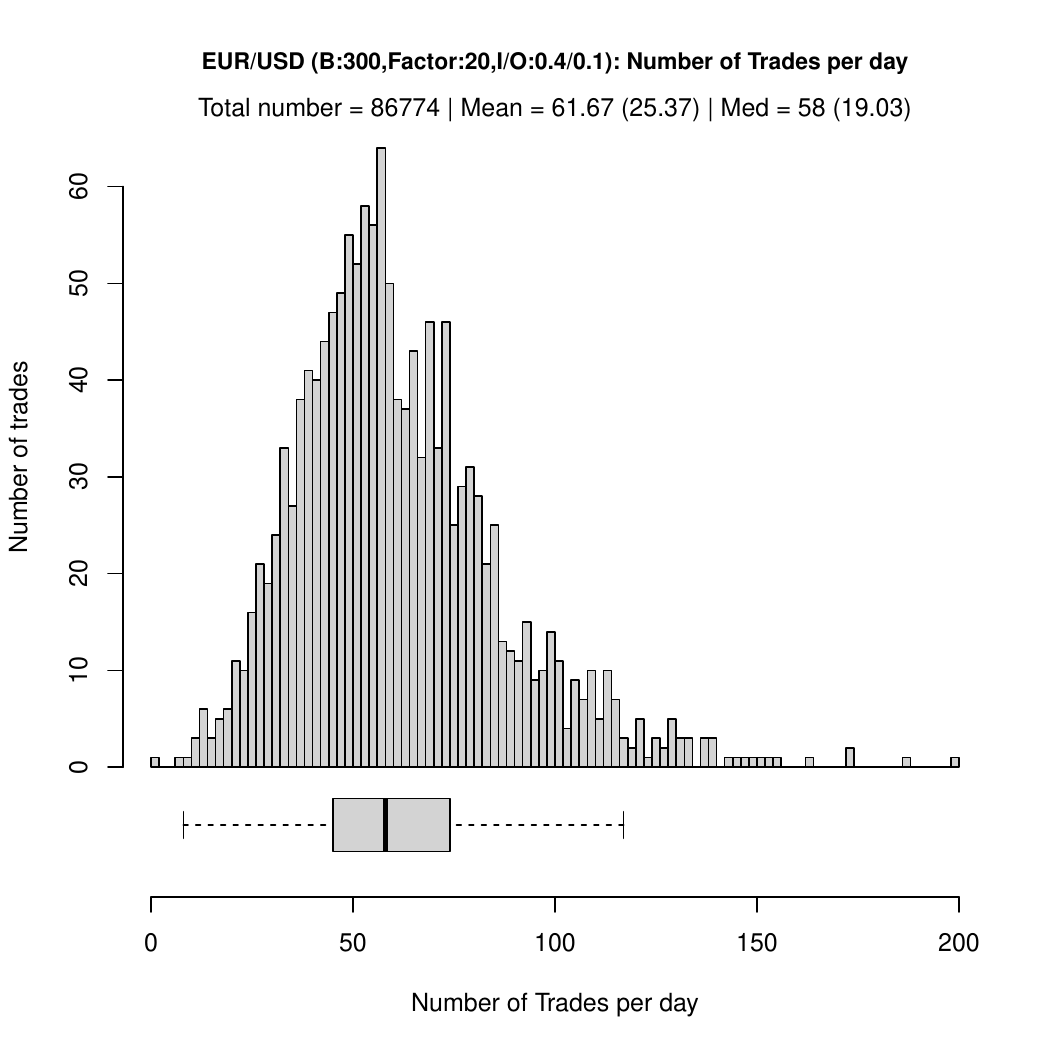}\\
		\includegraphics[width=6.5cm]{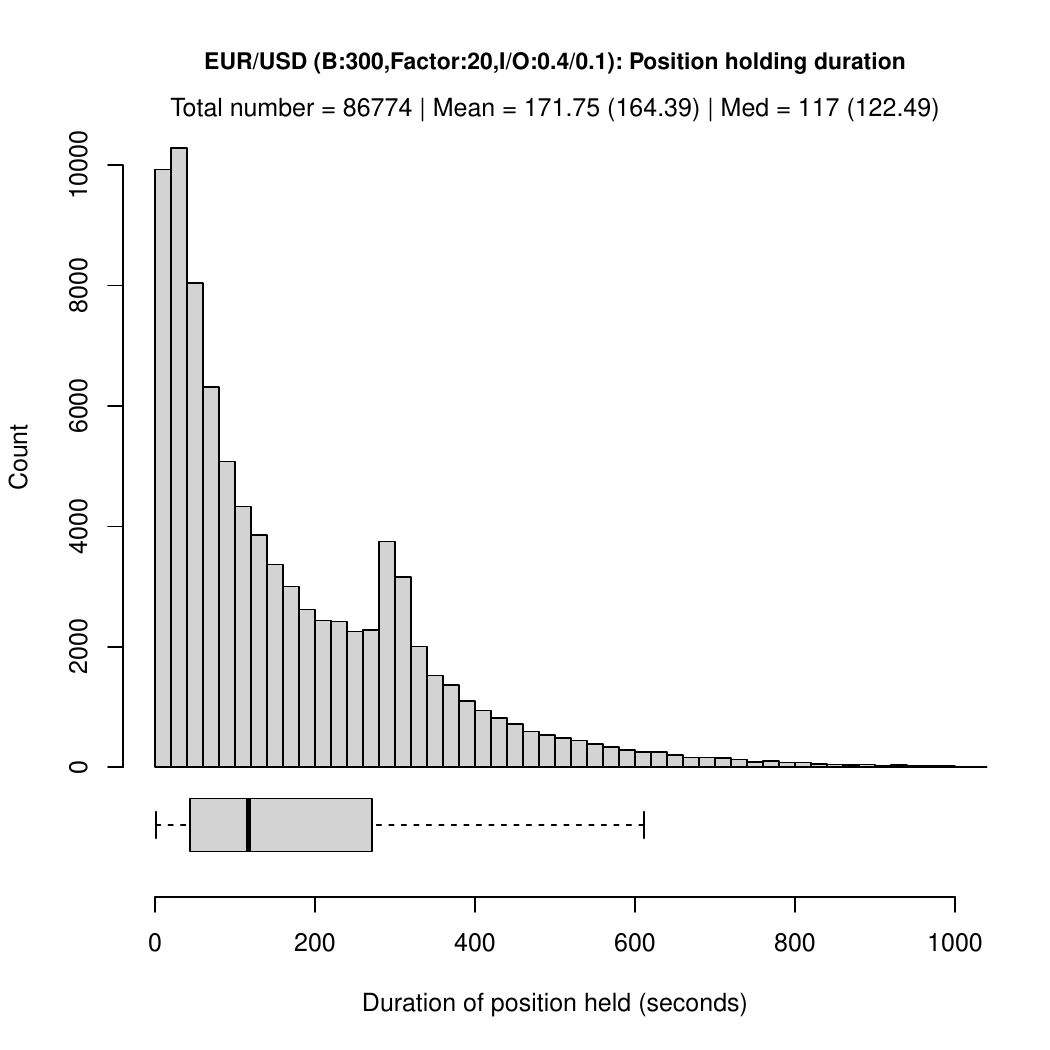} & \includegraphics[width=6.5cm]{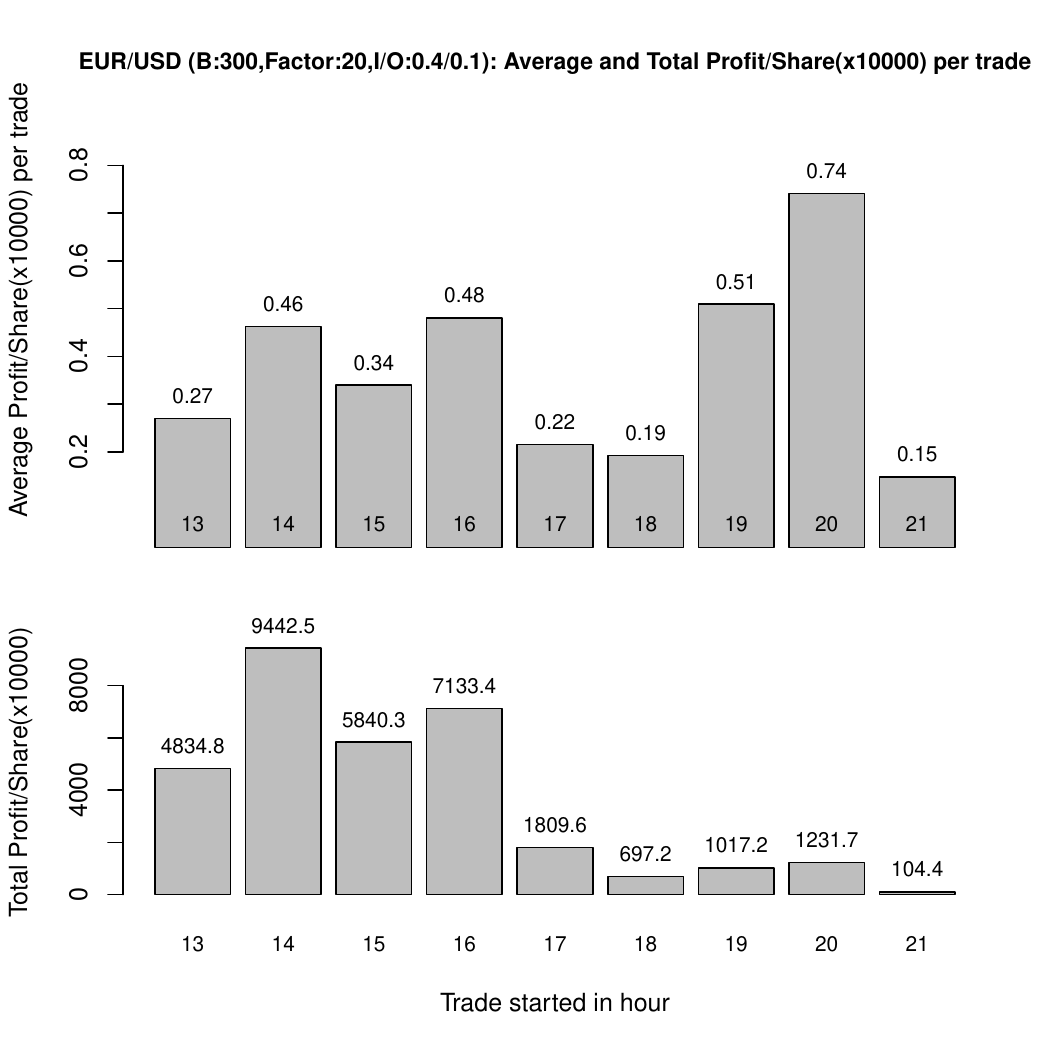}\\
		\includegraphics[width=6.5cm]{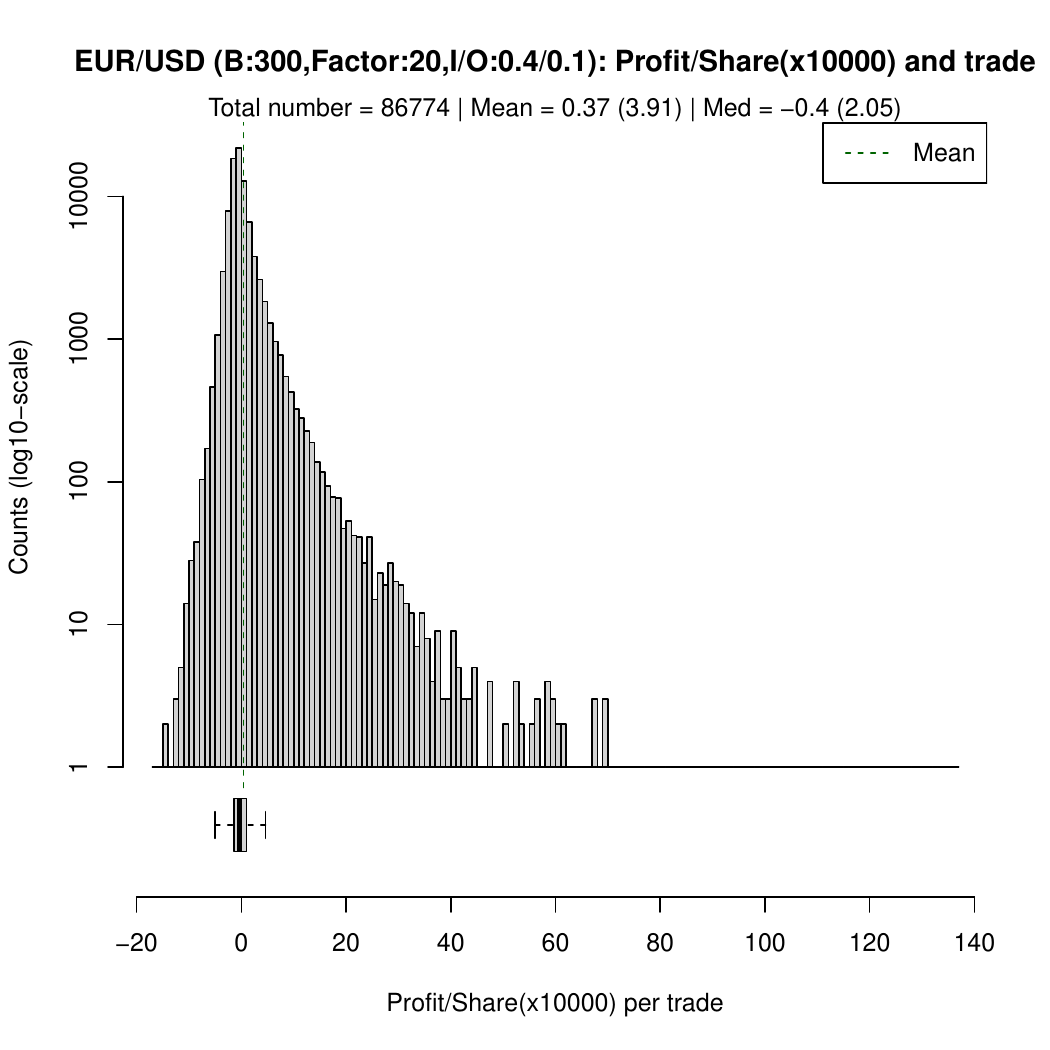} & \includegraphics[width=6.5cm]{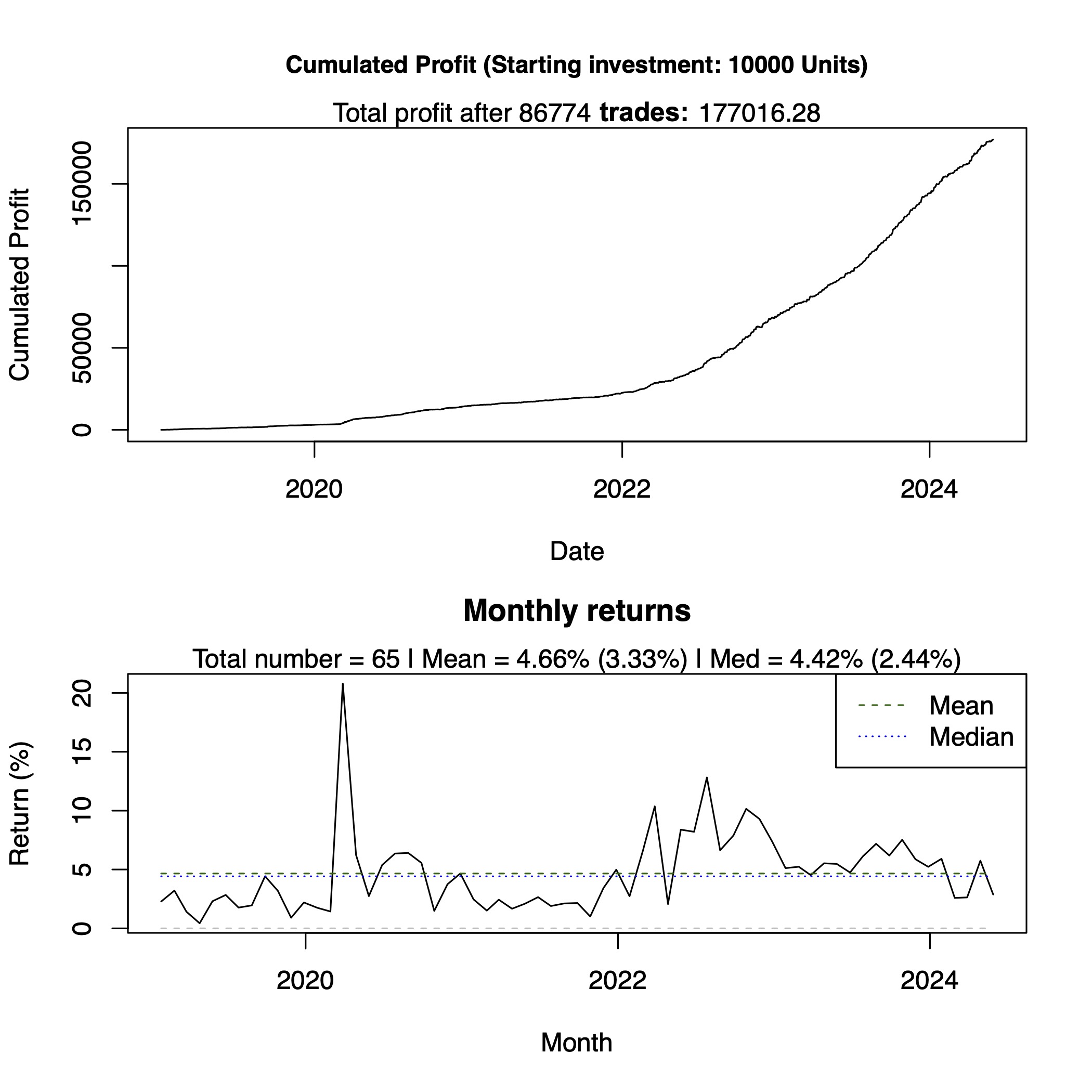}\\
	\end{tabular}
	\label{fig_EURUSD_300_OSC20_IN04_OUT01}
\end{figure}

Figure \ref{fig_EURUSD_300_OSC20_IN04_OUT01} provides a more detailed analysis of the trading strategy with configuration bandwidth 300s, I/O 0.4/0.1. In the bottom-left, a log-histogram of the profit per share for all trades is provided (that is, each count corresponds to one performed trade). The average profit/share is $0.37\cdot 10^{-5}$ (in particular, positive), while the median profit/share is $-0.4\cdot 10^{-5}$ (negative). \emph{The trading strategy therefore tends to have more loosing than winning trades, but the winning trades provide higher revenue.}  In the bottom-right of Figure \ref{fig_EURUSD_300_OSC20_IN04_OUT01} it is seen that the monthly returns are \emph{consistently positive} and besides one positive outlier in early 2020, \emph{have a relatively small variation}. In the top-right and the middle-left of Figure \ref{fig_EURUSD_300_OSC20_IN04_OUT01} it is seen that in average, 61.67 trades are performed per day (standard deviation: 25.37), with an average position holding time of 171.75 seconds (standard deviation: 164.39). While the number of trades is distributed nearly symmetrically around the average, the position holding duration is right-skewed with lots of positions gave up in less than 30 seconds. Most of those very short trades are loss trades where soon after the going-in signaled by the oscillator, a going-out signal was retrieved. As it is seen on top-left and middle-right of Figure \ref{fig_EURUSD_300_OSC20_IN04_OUT01}, most of the trades were performed between 1 p.m. and 5 p.m., which is also the period with the most profitable profit/share. As it could be seen in Figure \ref{fig_EURUSD_300_OSC20_IN04_OUT01_example}, the oscillator tends to be smaller at the end of the day, leading to less signals for going-in. The mathematical reason is that the lines on which the oscillator is based do not cover the price of the end of the day anymore. This could easily be fixed by choosing a larger range of starting points.\\

The other three configurations of the oscillator lead to overall similar results; the corresponding visualizations of the behavior and the characteristics as shown in Figure \ref{fig_EURUSD_300_OSC20_IN04_OUT01_example} and Figure \ref{fig_EURUSD_300_OSC20_IN04_OUT01} are postponed to the Appendix, sections \ref{EURUSD_examples} and \ref{EURUSD_characteristics}. In Table \ref{tab1_EURUSD_results}, the main trade characteristics for all four configurations are summarized. As expected, the configurations with In/Out-thresholds 0.8/0.2 are more conservative as the configurations with In/Out-thresholds 0.4/0.1, resulting in less trades and longer position hold durations (for example, for a bandwidth of 300s the I/O:0.8/0.2 configuration has in average 21.80 trades per day with position holding duration 202.61 seconds, while the I/O:0.4/0.1 configuration has in average 61.67 trades per day with position holding duration 171.75 seconds). The win rates and also the profit/share per trade of the more conservative configurations are slightly higher (for example, for a bandwidth of 300s the I/O:0.8/0.2 configuration has a win rate of 41.13\% with profit/share $0.562\cdot 10^{-5}$, while the I/O:0.4/0.1 configuration has a win rate of 38.85\% with profit/share $0.370\cdot 10^{-5}$). The results show that the simple in-out strategy based on the oscillator is, at least to some extend, robust against changes of the bandwidth and the thresholds regarding the profitability.

\begin{table}[h!]
	\centering
	\caption{EUR/USD results: Characteristics of the trading strategy over the time horizon from 2019/01 until 2024/05. $\delta T$ is the bandwidth of the oscillator used, Mult. $O_{t}^{\delta T}$ refers to the multiplicator used to enlarge the oscillator, I/O are the thresholds used for starting and stopping a position. SD refers to standard deviation, while MAD is the mean absolute deviation from the median.}
	\begin{tabular}{|r|r|r|r|r|r|}
		$\delta T$ & Mult. $O_{t}^{\delta T}$ & I/O & Characteristic & Mean (SD) & Median (MAD)\\
		\hline
			600s & 50 & 0.4/0.1 & Trades: 89297 & & \\
			 &  &  & Trade Duration (s) & 262.46 (317.24) & 124.00 (216.07) \\
			 &  &  & Profit/Share in one trade & 0.309 (4.210) $\cdot 10^{-5}$ & -0.500 (2.084) $\cdot 10^{-5}$ \\
			 & & & Win Rate (\%) & 35.91 (47.97) & \\
			 & & & Trades per day & 63.47 (22.51) & 60.00 (16.76)\\
		\hline
			600s & 50 & 0.8/0.2 & Trades: 40255 & & \\
			& & & Trade Duration (s) & 331.79 (313.00) & 229.00 (234.98) \\
			 &  &  & Profit/Share in one trade & 0.411 (5.443) $\cdot 10^{-5}$ & -0.700 (2.916) $\cdot 10^{-5}$ \\
			 & & & Win Rate (\%) & 37.72 (48.47) & \\
			 & & & Trades per day & 28.61 (13.41) & 26.00 (9.91)\\
		\hline
			300s & 20 & 0.4/0.1 & Trades: 86774 & & \\
			& & & Trade Duration (s) & 171.75 (164.39) & 117.00 (122.49) \\
			 &  &  & Profit/Share in one trade & 0.370 (3.909) $\cdot 10^{-5}$ & -0.400 (2.052) $\cdot 10^{-5}$ \\
			 & & & Win Rate (\%) & 38.85 (48.74) & \\
			 & & & Trades per day & 61.67 (25.37) & 58.00 (19.03)\\
		\hline
			300s & 20 & 0.8/0.2 & Trades: 30439 & & \\
			& & & Trade Duration (s) & 202.61 (152.86) & 174.00 (118.05) \\
			 &  &  & Profit/Share in one trade & 0.562 (5.287) $\cdot 10^{-5}$ & -0.500 (2.919) $\cdot 10^{-5}$ \\
			 & & & Win Rate (\%) & 41.13 (49.21) & \\
			 & & & Trades per day & 21.80 (13.89) & 19.00 (10.33)\\
		\hline
	\end{tabular}
	\label{tab1_EURUSD_results}
\end{table}

\FloatBarrier

\newpage

\subsubsection{German DAX 40}

The trading strategy as applied for the EUR/USD (but with different slope and starting point parameters, cf. Table \ref{table_parameters}) was applied to the German DAX 40 from 2019/01 to 2024/05. Two behavior of the strategy with configuration bandwidth 600s and I/O:0.8/0.2 for two sample days are shown in Figure \ref{fig_DAX_600_OSC50_IN04_OUT01_example}.

\begin{figure}[b!]
	\centering
	\caption{DAX, Bandwidth 600s, In/Out: 0.8/0.2: Behavior of the trading strategy from 2019/01 to 2024/05. For a detailed explanation of the plots, see Figure \ref{fig_EURUSD_300_OSC20_IN04_OUT01_example}.}
	\begin{tabular}{c}
		\includegraphics[width=13cm]{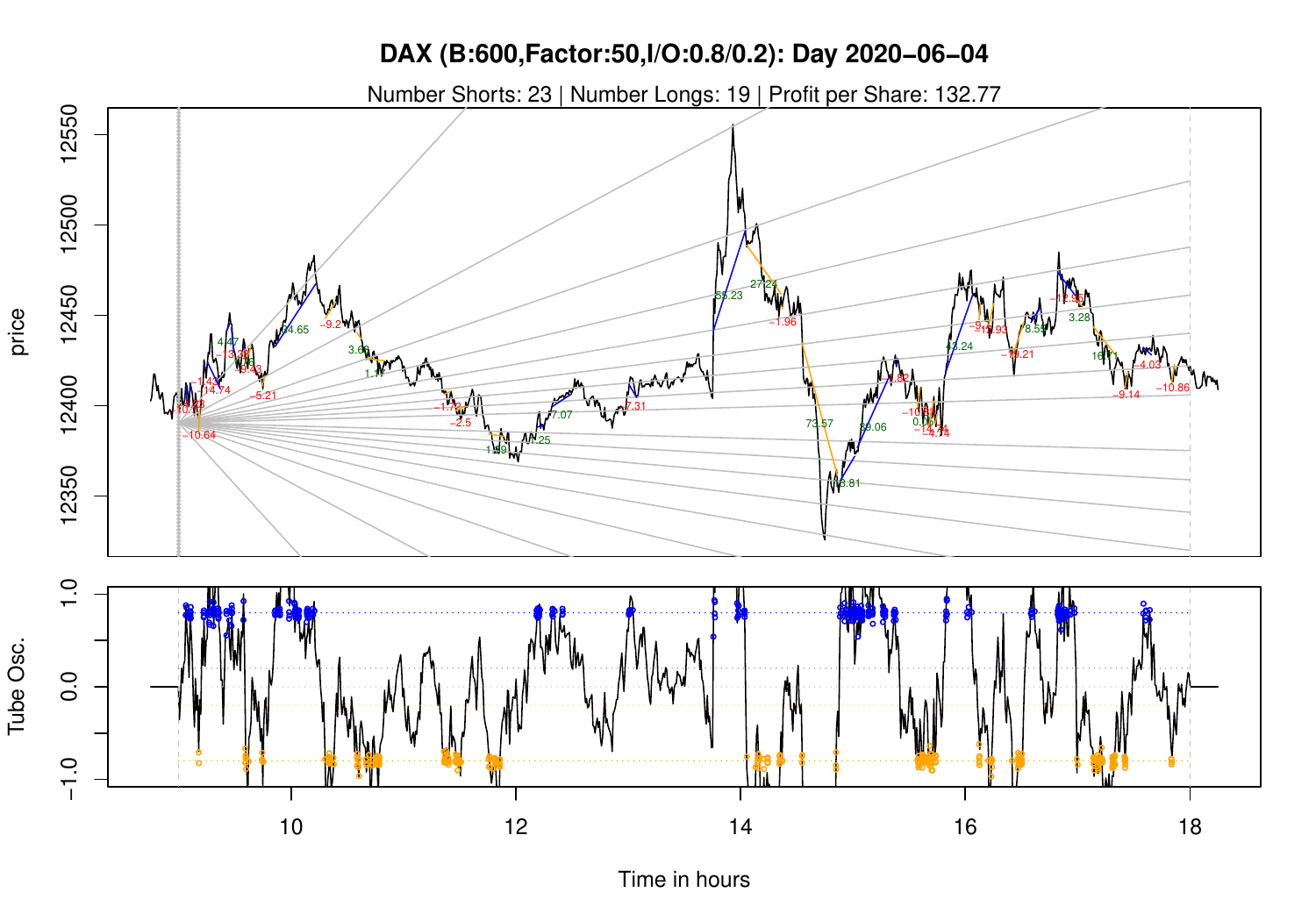}\\
		\includegraphics[width=13cm]{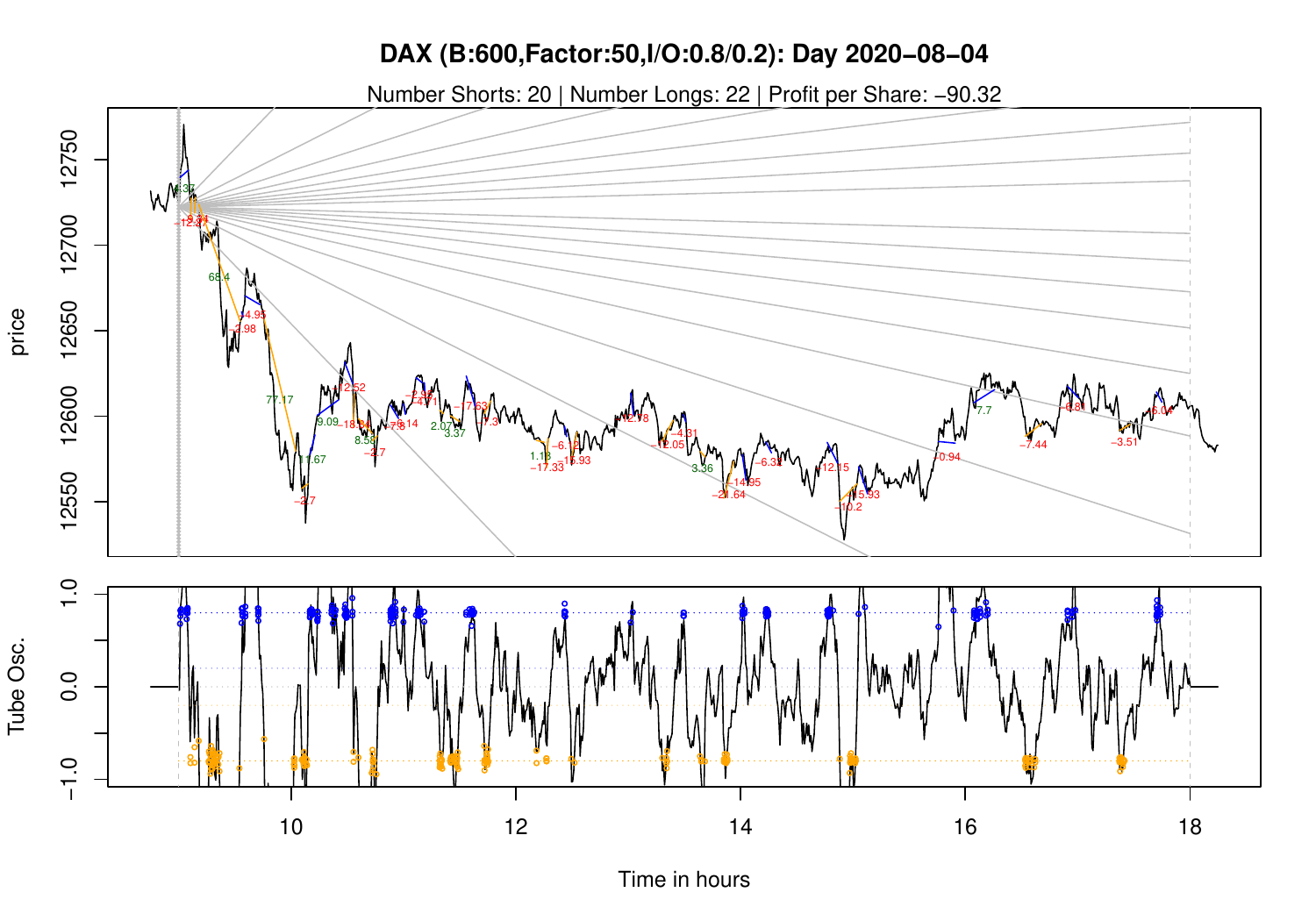}\\
	\end{tabular}
	\label{fig_DAX_600_OSC50_IN08_OUT02_example}
\end{figure}

Tables \ref{tab2_DAX_results} summarizes the results for the trading strategies applied to DAX. Here, one 'share' again has the instrument price, which ranges between 10'000 and 19'000 in the whole trading time from 2019/01 to 2024/05. The best configuration regarding Sharpe Ratio (bandwidth 600s, I/O 0.8/0.2) lead to average monthly returns of $3.658\%$ (standard deviation: $6.697\%$) and median monthly returns of $2.639\%$ (MAD: $3.546\%$). The corresponding monthly Sharpe Ratio was 0.516 (extrapolated yearly Sharpe Ratio: 1.788). Over the whole trading time from 2019/01 to 2024/05, the initial capital of 10'000 EUR could be increased to 91'923 EUR (initial capital plus profit) in 31'897 trades, which is an overall return on investment of
\[
	\frac{91'923 - 10'000}{10'000} \approx 819.23 \%.
\]

\begin{table}[h!]
	\centering
	\caption{DAX results: Monthly and (extrapolated) yearly Sharpe Ratio SR, as well as monthly return characteristics of the trading strategy over the time horizon from 2019/01 until 2024/05. $\delta T$ is the bandwidth of the oscillator used, Mult. $O_{t}^{\delta T}$ refers to the Multiplicator used to enlarge the oscillator, I/O are the thresholds used for starting and stopping a position. Average refers to the mean (with standard deviation in brackets), while median refers to the statistical median (with mean absolute deviation from the median in brackets).}
	\begin{tabular}{|r|r|r|r|r|r|r|r|}
		$\delta T$ & Mult. $O_{t}^{\delta T}$ & I/O & $SR_{monthly}$ & $SR_{yearly}$ & Average Return/month & Median Return/month & Total Profit\\
		\hline
			600s & 50 & 0.4/0.1 & 0.389 & 1.349 & 5.846 (14.519)\% & 3.019 (6.347)\% & 251'377.80 \\
		\hline
			600s & 50 & 0.8/0.2 & 0.516 & 1.788 & 3.658 (6.697)\% & 2.639 (3.546)\% & 81'923.92 \\
		\hline
			300s & 20 & 0.4/0.1 & 0.401 & 1.390 & 7.269 (17.637)\% & 3.757 (6.699)\% & 528'626.10 \\
		\hline
			300s & 20 & 0.8/0.2 & 0.449 & 1.556 & 3.592 (7.553)\% & 1.823 (3.034)\% & 76'217.42 \\
		\hline
	\end{tabular}
	\label{tab2_DAX_results}
\end{table}

\begin{figure}[h!]
	\centering
	\caption{DAX, Bandwidth 600s, In/Out: 0.8/0.2: Characteristics of the trading strategy from 2019/01 to 2024/05.}
	\begin{tabular}{cc}
		\includegraphics[width=6.5cm]{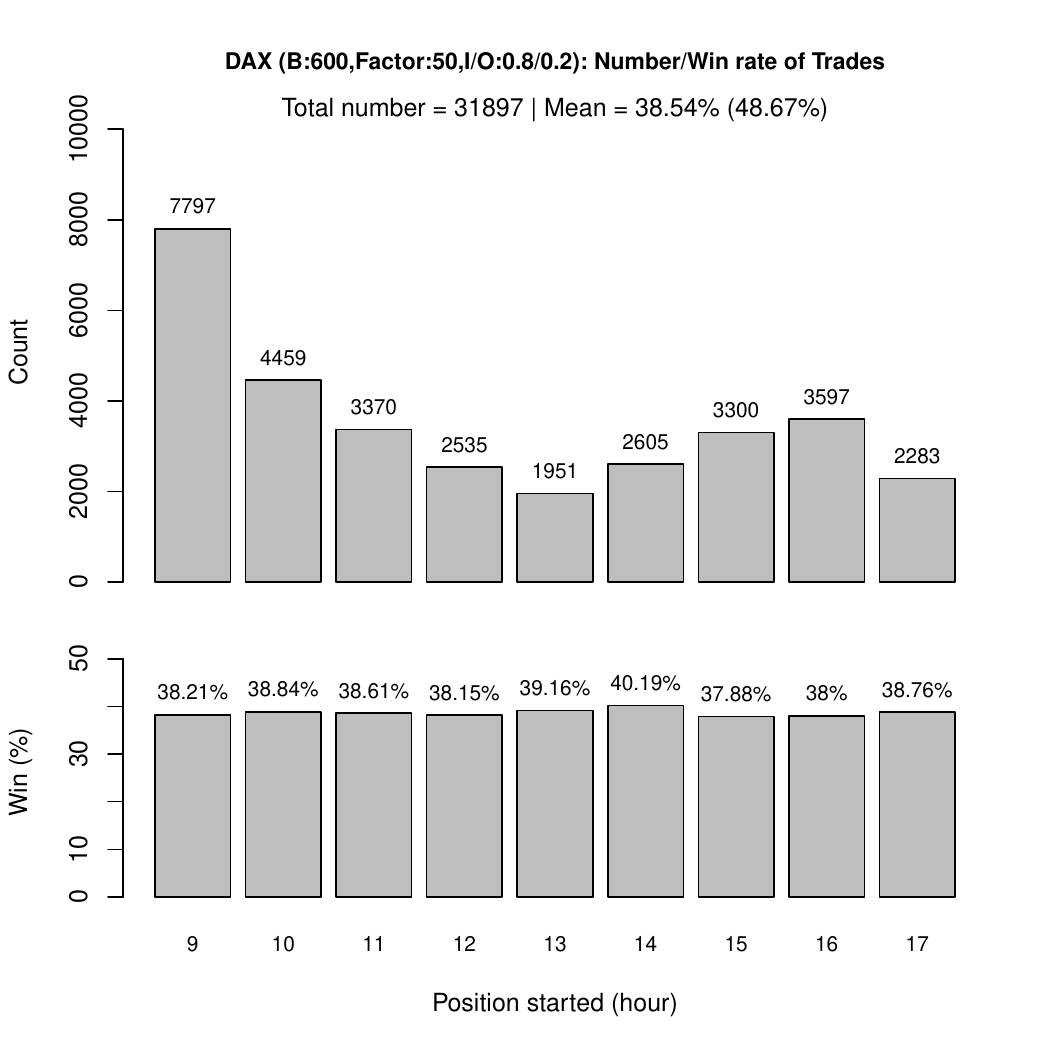} & \includegraphics[width=6.5cm]{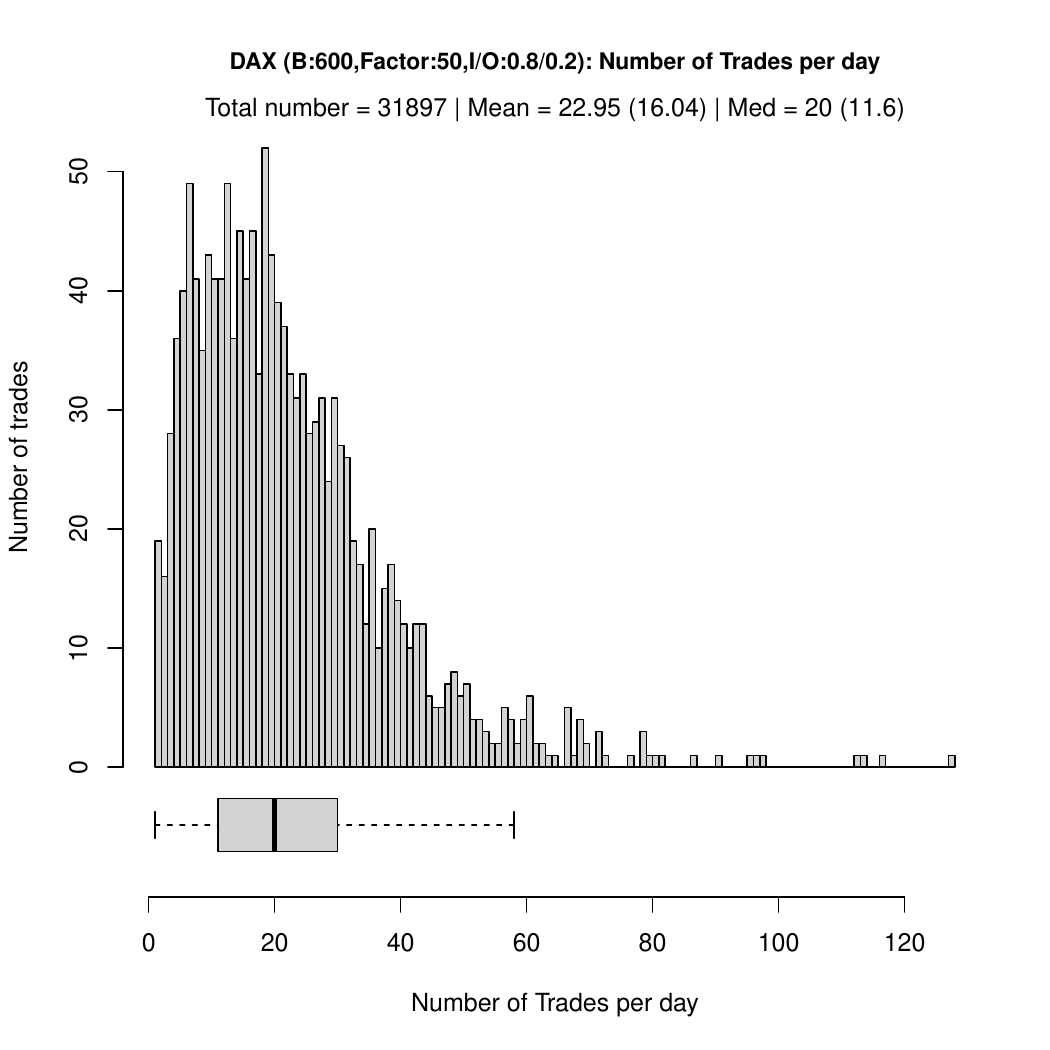}\\
		\includegraphics[width=6.5cm]{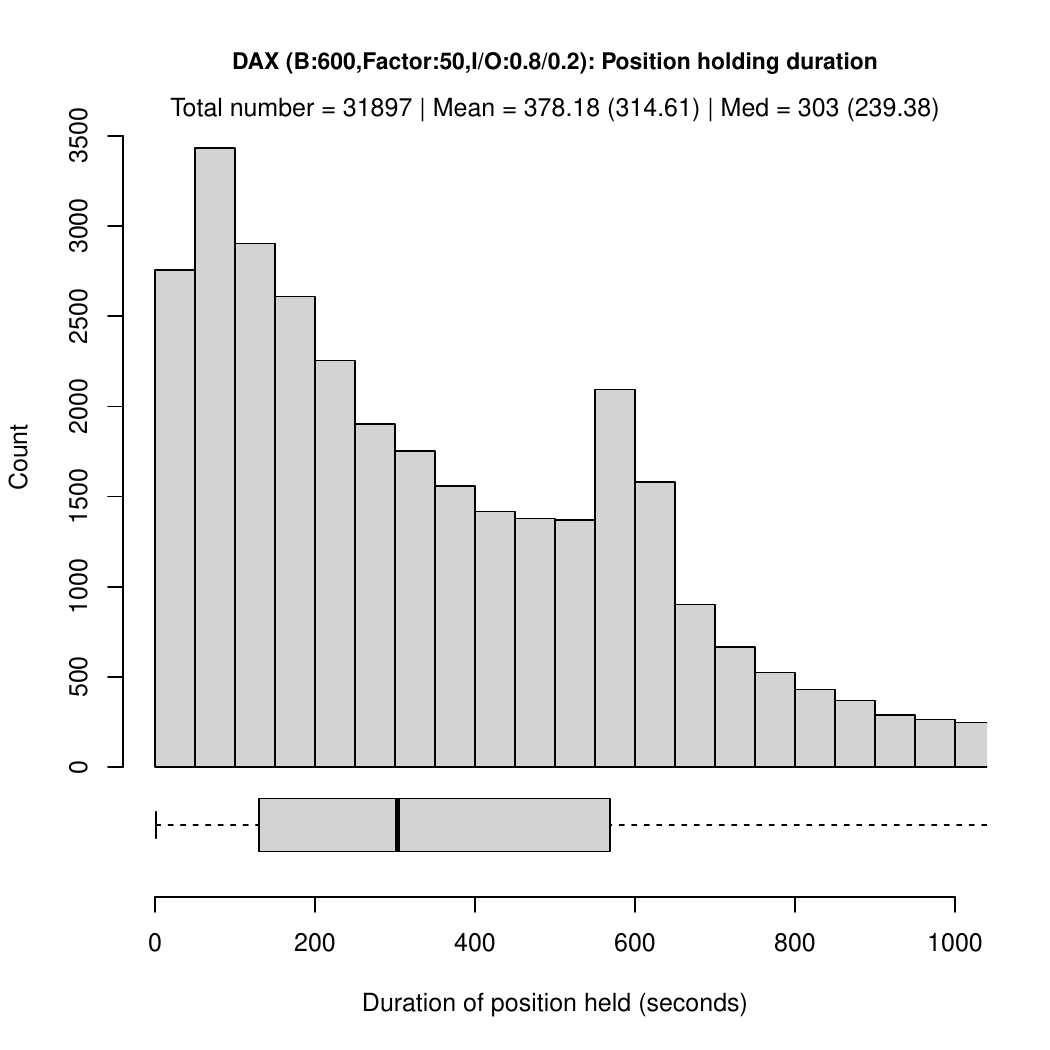} & \includegraphics[width=6.5cm]{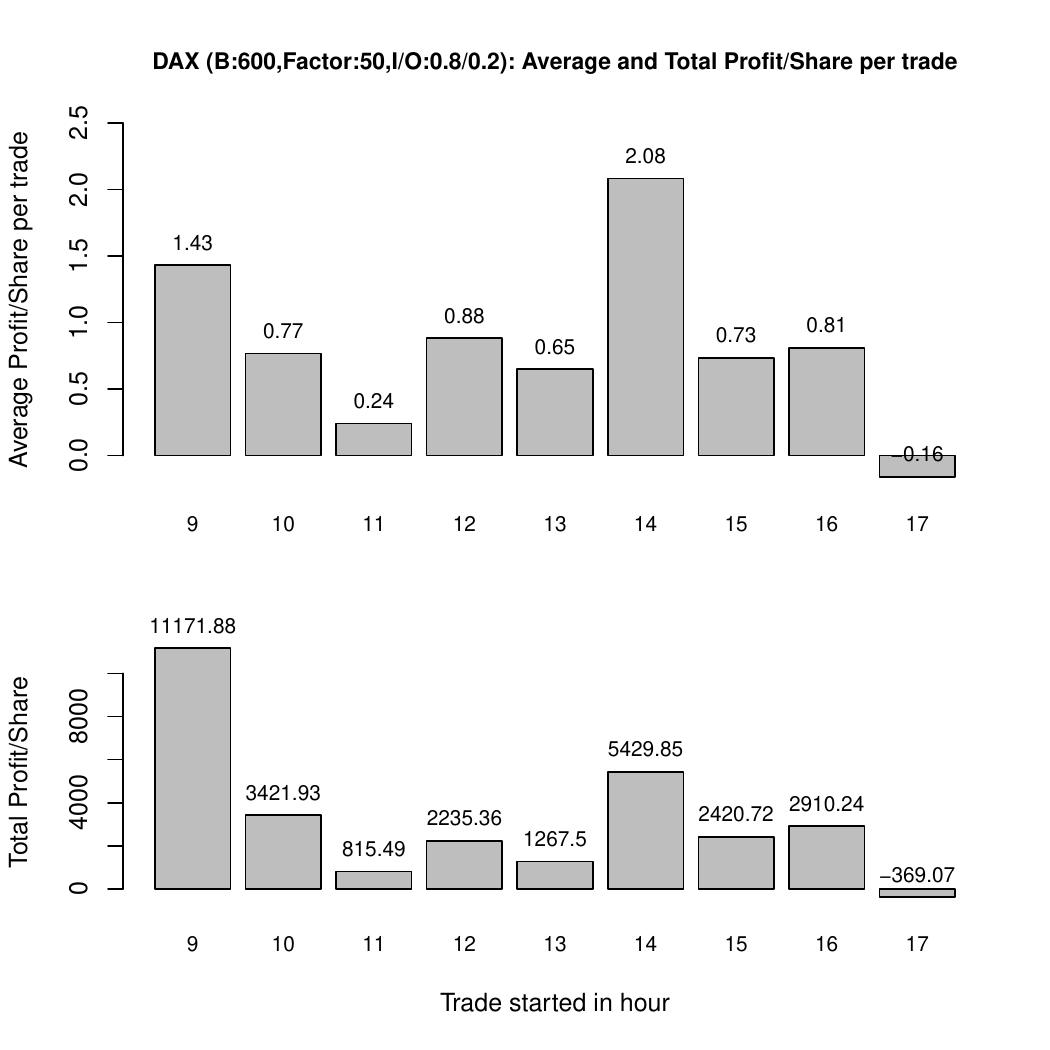}\\
		\includegraphics[width=6.5cm]{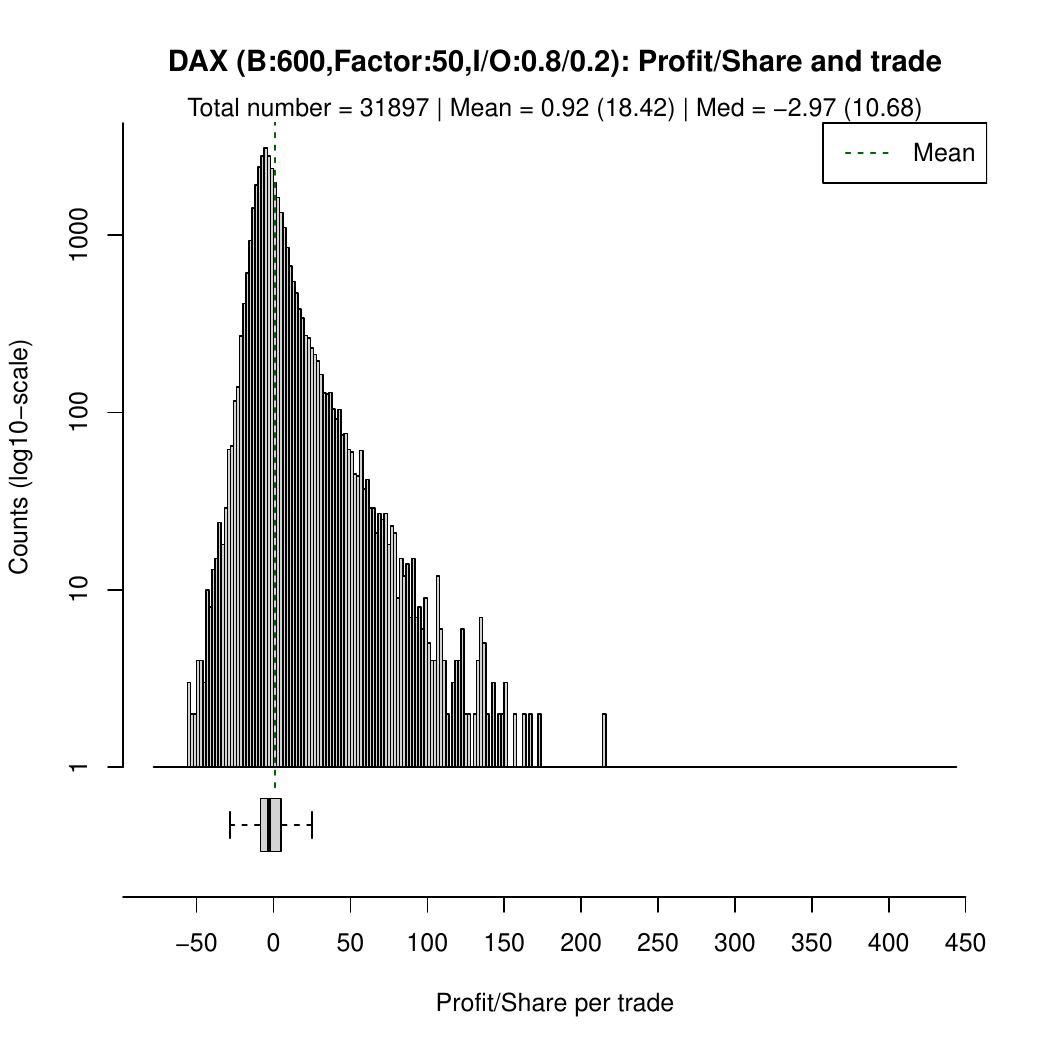} & \includegraphics[width=6.5cm]{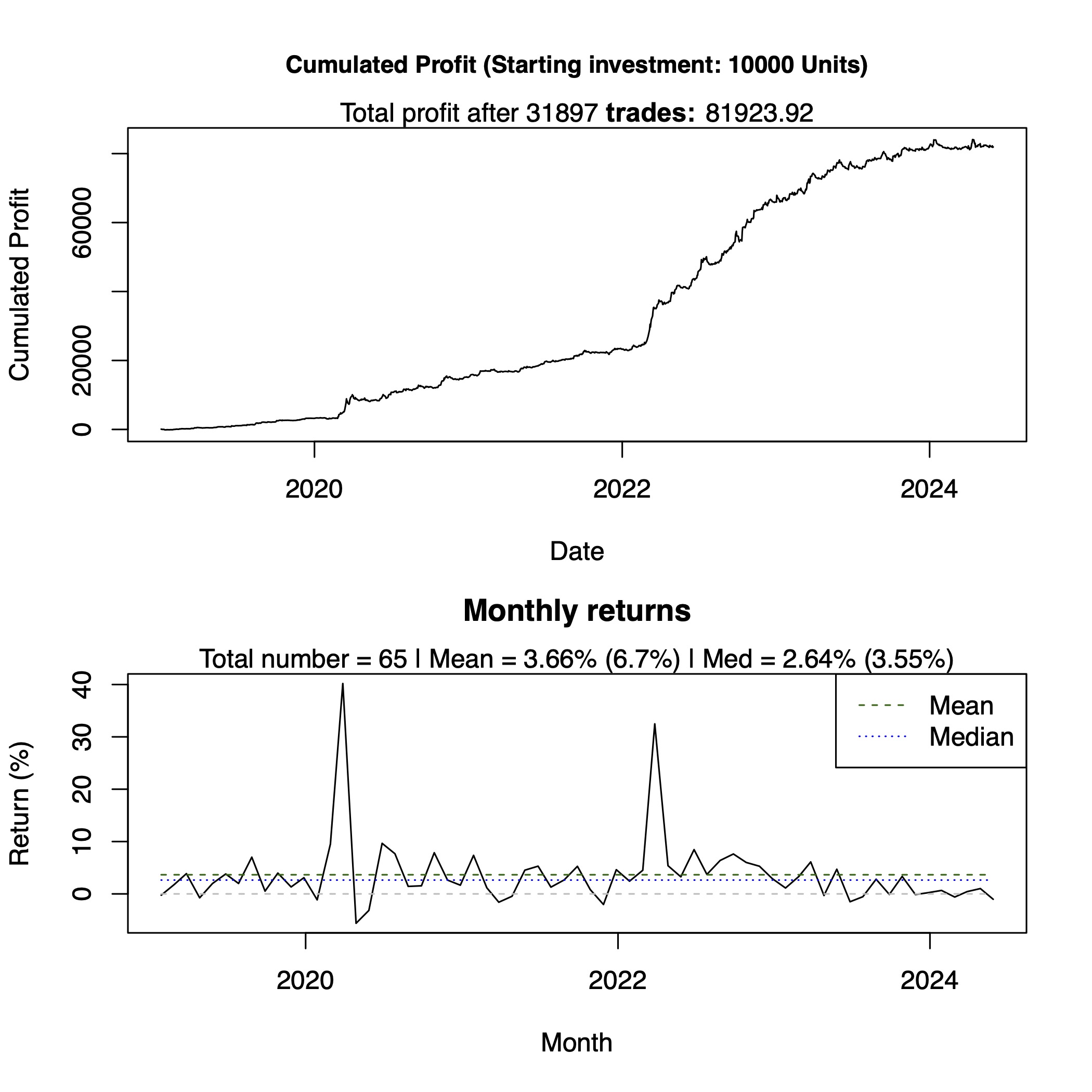}\\
	\end{tabular}
	\label{fig_DAX_600_OSC50_IN08_OUT02}
\end{figure}

As seen in Table \ref{tab2_DAX_results}, all other configurations did also perform well: In fact, configurations with smaller Sharpe Ratio resulted in much higher Total Profits (for instance, the configuration bandwidth 300s, I/O:0.4/0.1 had a monthly Shape Ratio of 0.401 but a total profit of 528'626 EUR). It can be seen later in Figure \ref{fig_DAX_600_OSC50_IN08_OUT02} that one reason for the much smaller Sharpe Ratio (compared to the results of EUR/USD) are two positive outliers in early 2020 and early 2022, which strongly increase the standard deviation (but also increase the average return). All configurations lead to average monthly returns of at least $3.592\%$ and median monthly returns of at least $1.823\%$, showing again that the method is relatively robust against changes of bandwidth and threshold parameters.

Overall, the trading strategy behaves similarly as for the EUR/USD instrument. To see this, Figure \ref{fig_DAX_600_OSC50_IN08_OUT02} provides a more detailed analysis of the trading strategy with the highest Sharpe Ratio (configuration bandwidth 600s, I/O 0.8/0.2). From the log-histogram of the profit per share for all trades it is seen that the average profit/share is positive ($0.92$), while the median profit/share is negative ($-2.97$), that is, the \emph{the trading strategy has more loosing than winning trades, but the winning trades are more profitable.} The bottom-right of Figure \ref{fig_DAX_600_OSC50_IN08_OUT02} shows the monthly returns. Compared to EUR/USD, the returns have a higher variance with positive outliers in early 2020 and 2022. The returns are not positive over the whole trading time, 15 of 65 months resulted in negative returns. However, the absolute value of the monthly returns which are negative is much smaller than the monthly returns which are positive. 

In the top-right and the middle-left of Figure \ref{fig_DAX_600_OSC50_IN08_OUT02} it is seen that in average, 22.95 trades are performed per day (standard deviation: 16.04), with an average position holding time of 378.18 seconds (standard deviation: 314.61).The distribution of the number of trades is  slightly right-skewed, while the position holding duration is right-skewed. As it is seen on top-left and middle-right of Figure \ref{fig_DAX_600_OSC50_IN08_OUT02}, most of the trades were performed between 9 a.m. and 10 a.m. The most profitable profit/share per trade was between 9 a.m. and 10 a.m. and 14 a.m. - 15 a.m. For the German DAX, 9 a.m. refers to the (highly volatile) opening time of the market which indicates that the trading strategy based on the oscillator works best in regimes with high volatility of the instrument price.

The other three configurations of the oscillator lead to overall similar characteristics, their visualization is postponed to the Appendix, sections \ref{DAX_examples} and \ref{DAX_characteristics}. Table \ref{tab1_DAX_results} summarizes the main trade characteristics for all four configurations. As already seen for the EUR/USD instrument, the configurations with In/Out-thresholds 0.8/0.2 tend to be more conservative as the configurations with In/Out-thresholds 0.4/0.1, resulting in less trades and longer position hold durations (for example, for a bandwidth of 600s the I/O:0.8/0.2 configuration has in average 22.95 trades per day with position holding duration 378.18 seconds, while the I/O:0.4/0.1 configuration has in average 59.52 trades per day with position holding duration 299.68 seconds). As for the EUR/USD, it is seen that a trading strategy with higher I/O-thresholds 0.8/0.2 results in less trades with higher win rate than using I/O-thresholds 0.4/0.1.

\begin{table}[h!]
	\centering
	\caption{DAX results: Characteristics of the trading strategy over the time horizon from 2019/01 until 2024/05. $\delta T$ is the bandwidth of the oscillator used, Mult. $O_{t}^{\delta T}$ refers to the multiplicator used to enlarge the oscillator, I/O are the thresholds used for starting and stopping a position. SD refers to standard deviation, while MAD is the mean absolute deviation from the median.}
	\begin{tabular}{|r|r|r|r|r|r|}
		$\delta T$ & Mult. $O_{t}^{\delta T}$ & I/O & Characteristic & Mean (SD) & Median (MAD)\\
		\hline
			600s & 50 & 0.4/0.1 & Trades: 83088 & & \\
			& & & Trade Duration (s) & 299.68 (327.95) & 171.00 (230.77) \\
			 &  &  & Profit/Share in one trade & 0.501 (13.860) & -2.432 (7.278) \\
			 & & & Win Rate (\%) & 35.20 (47.76) & \\
			 & & & Trades per day & 59.52 (25.02) & 56.00 (18.65)\\
		\hline
			600s & 50 & 0.8/0.2 & Trades: 31897 & & \\
			& & & Trade Duration (s) & 378.18 (314.61) & 303.00 (239.38) \\
			 &  &  & Profit/Share in one trade & 0.919 (18.420) & -2.972 (10.679) \\
			 & & & Win Rate (\%) & 38.54 (48.67) & \\
			 & & & Trades per day & 22.95 (16.04) & 20.00 (11.60)\\
		\hline
			300s & 20 & 0.4/0.1 & Trades: 73347 & & \\
			& & & Trade Duration (s) & 190.48 (161.43) & 150.00 (122.56) \\
			 &  &  & Profit/Share in one trade & 0.703 (12.712) & -1.972 (7.309) \\
			 & & & Win Rate (\%) & 38.54 (48.67) & \\
			 & & & Trades per day & 52.58 (31.94) & 47.00 (23.76)\\
		\hline
			300s & 20 & 0.8/0.2 & Trades: 18284 & & \\
			& & & Trade Duration (s) & 217.50 (145.58) & 201.00 (111.54) \\
			 &  &  & Profit/Share in one trade & 1.526 (17.952) & -2.201 (11.216) \\
			 & & & Win Rate (\%) & 42.78 (49.48) & \\
			 & & & Trades per day & 13.73 (15.45) & 9.00 (9.58)\\
		\hline
	\end{tabular}
	\label{tab1_DAX_results}
\end{table}

\FloatBarrier

\section{Conclusion}
\label{sec_conclusion}

In this paper, we have introduced financial market geometry as a subfield of technical market analysis. Unlike other approaches, financial market geometry is mainly based on the graphical behavior of the instrument price and much less on its explicit values. As an example for the detection power of financial market geometry for future price behavior, the tube oscillator was defined.

In an empirical analysis it was shown that a simple, deterministic trading strategy based on using oscillator signals for opening positions lead to consistent positive monthly returns of 2\% or more. Those results were obtained \emph{without} cherry-picking, without a fitting procedure and without leverage methods. The market instruments DAX and EUR/USD were chosen arbitrarily due to their high liquidity in the market. The oscillator as well as the trading strategy depends on several parameters which must be fitted to the market instrument to which the trading strategy is applied. Overall, the strategy can be classified as medium-frequency strategy, with position holding durations between several seconds and about 10 minutes. As shown in the empirical part, this offers flexibility in the nature of the trading behavior, from a more conservative to a more dynamic approach.

Of course, the parameters have to be chosen in an educated way to obtain profitable results. However, one main message of the empirical analysis was that remarkable results can be obtained by rule-of-thumb parameter choices and that those results are fairly robust over large domains of the parameter set. Additionally, it was found that most of the parameters are even independent of the underlying instrument (DAX or EUR/USD, respectively), as long as there is enough volatility in the price. It is likely that similar good results can be obtained by applying the trading strategy to other highly liquid market instruments.

The findings of this article have three important implications for state-of-the-art certainties about stock price evolution:
\begin{itemize}
	\item First, it shall be noted that the whole trading strategy presented does \emph{only} depend on the underlying market instrument, that is, the evolution or information of other market instruments is not used at all. Due to the good results obtained for DAX and EUR/USD with the presented strategy, this shows that a high percentage of the information necessary to provide meaningful forecasts is already included in a small proportion of the past of the instrument itself. Additionally, it raises the question if it is really necessary or 'better' to include knowledge apart from a highly liquid market instrument when forecasting its behavior, since additional information always comes with more uncertainty.
	\item Second, from a broader perspective, the results can be seen, albeit somewhat immodestly, as a 'rebirth' of technical market analysis. Financial market geometry as introduced in this paper demonstrates that a well-informed application of technical market analysis can yield consistent positive returns when trading highly liquid market instruments.  
	\item The third point addresses modeling in stochastic finance. The empirical analysis demonstrated that it is possible to forecast future prices with in average positive returns, even if the bid-ask-spread is taken into account. This indicates that even in the short run, highly liquid market instruments exhibit a strong underlying deterministic component. Note also that although the winning rate of trades in the empirical analysis ranged only between 32\% and 40\% when accounting for the bid-ask spread. Excluding this spread, it is likely that much higher winning rates above 50\% could be achieved. This raises the question if modeling one-dimensional stock price evolution mainly through noise (as for instance in the Black-Scholes model or more sophisticated approaches) is adequate, or if  more prominent short-term deterministic components should also be incorporated. 
\end{itemize}

\newpage
\printbibliography
\newpage

\appendix

\section{Additional illustrations}

\FloatBarrier
\subsection{Forex EUR/USD: Trading strategy behavior for other configurations}
\label{EURUSD_examples}

\begin{figure}[b!]
	\centering
	\caption{EUR/USD, Bandwidth 600s, In/Out: 0.4/0.1: Behavior of the trading strategy from 2019/01 to 2024/05. For a detailed explanation of the plots, see Figure \ref{fig_EURUSD_300_OSC20_IN04_OUT01_example}.}
	\begin{tabular}{c}
		\includegraphics[width=13cm]{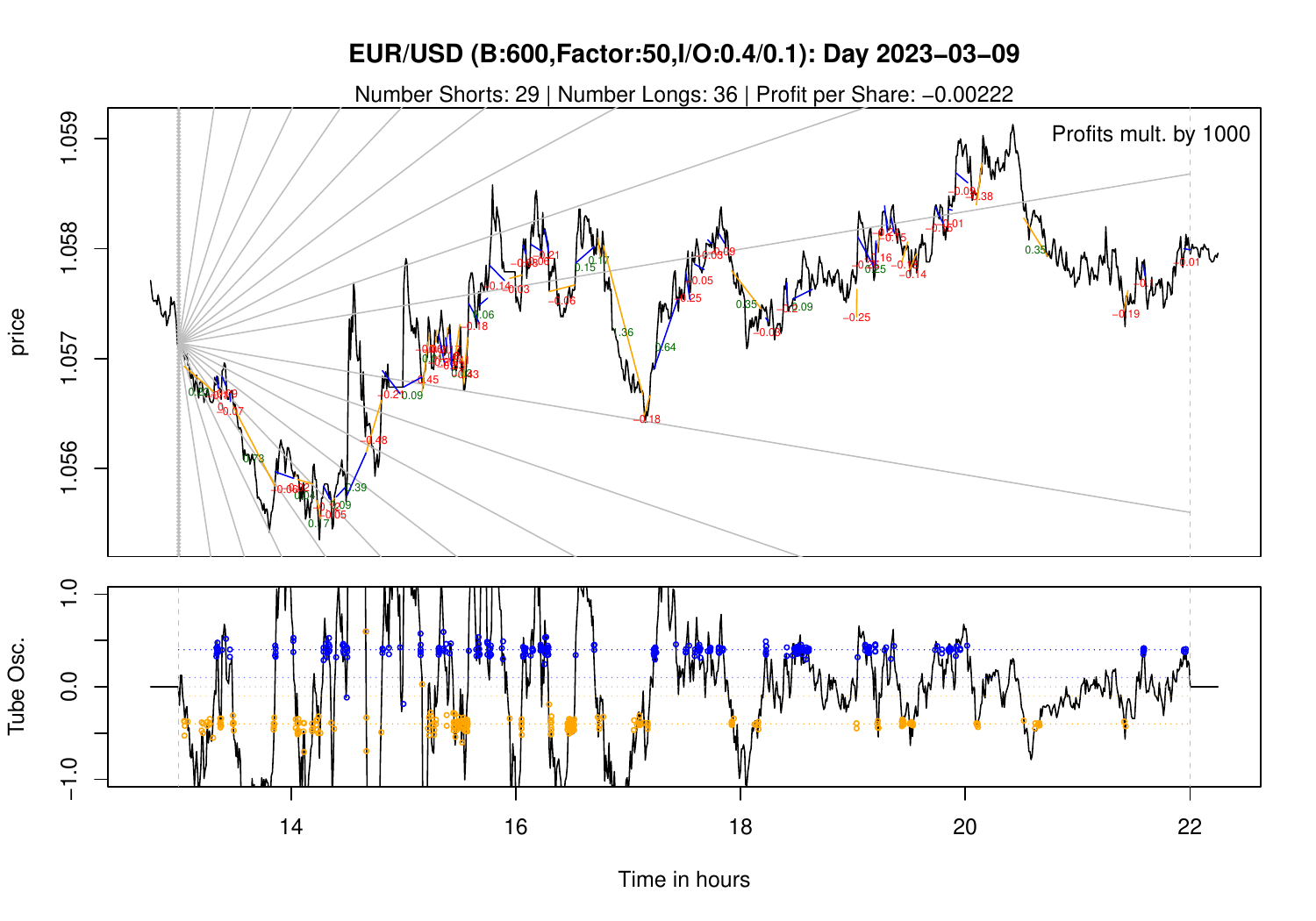}\\
		\includegraphics[width=13cm]{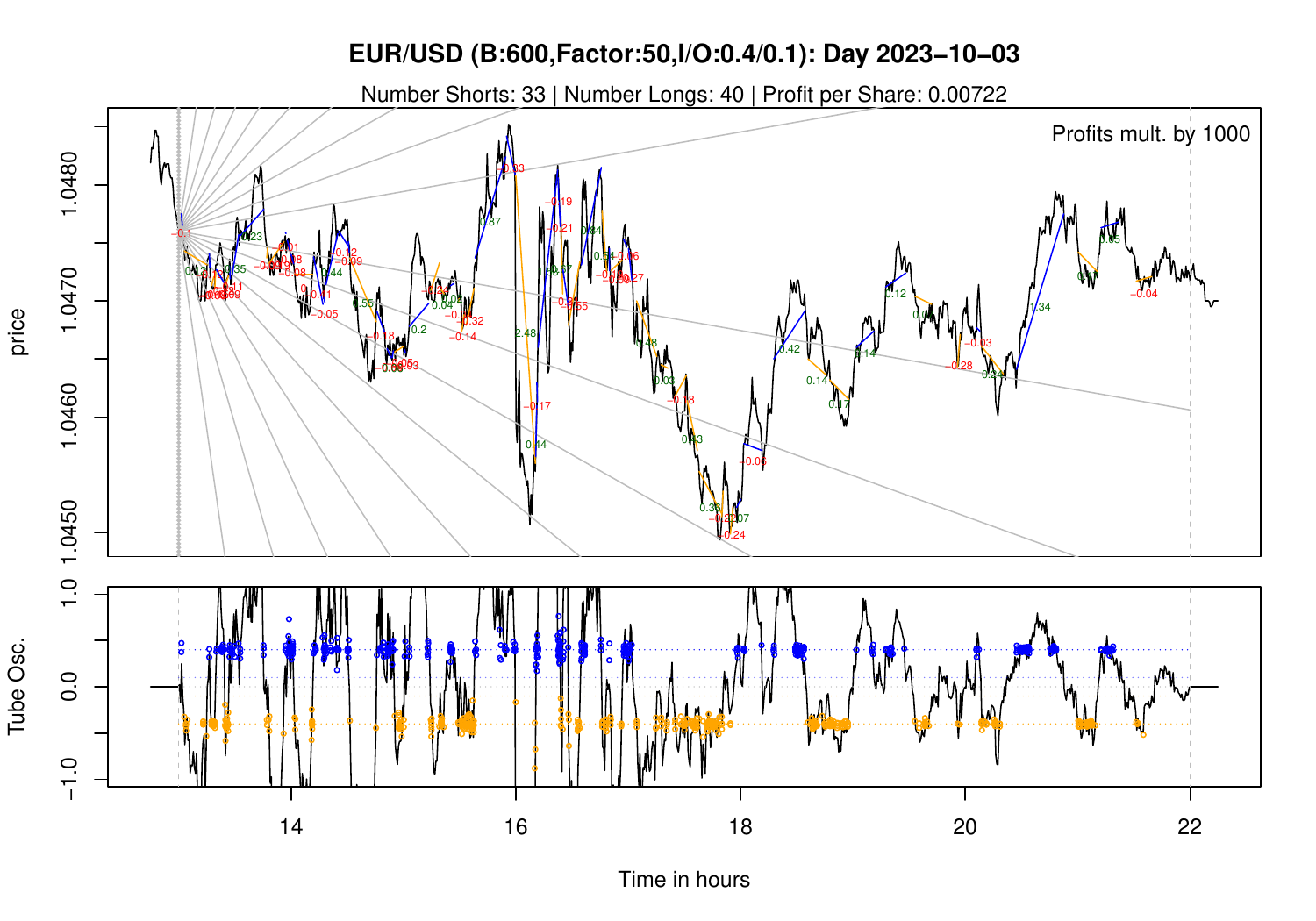}\\
	\end{tabular}
	\label{fig_EURUSD_600_OSC50_IN04_OUT01_example}
\end{figure}

\begin{figure}[h!]
	\centering
	\caption{EUR/USD, Bandwidth 600s, In/Out: 0.8/0.2: Behavior of the trading strategy from 2019/01 to 2024/05. For a detailed explanation of the plots, see Figure \ref{fig_EURUSD_300_OSC20_IN04_OUT01_example}.}
	\begin{tabular}{c}
		\includegraphics[width=13cm]{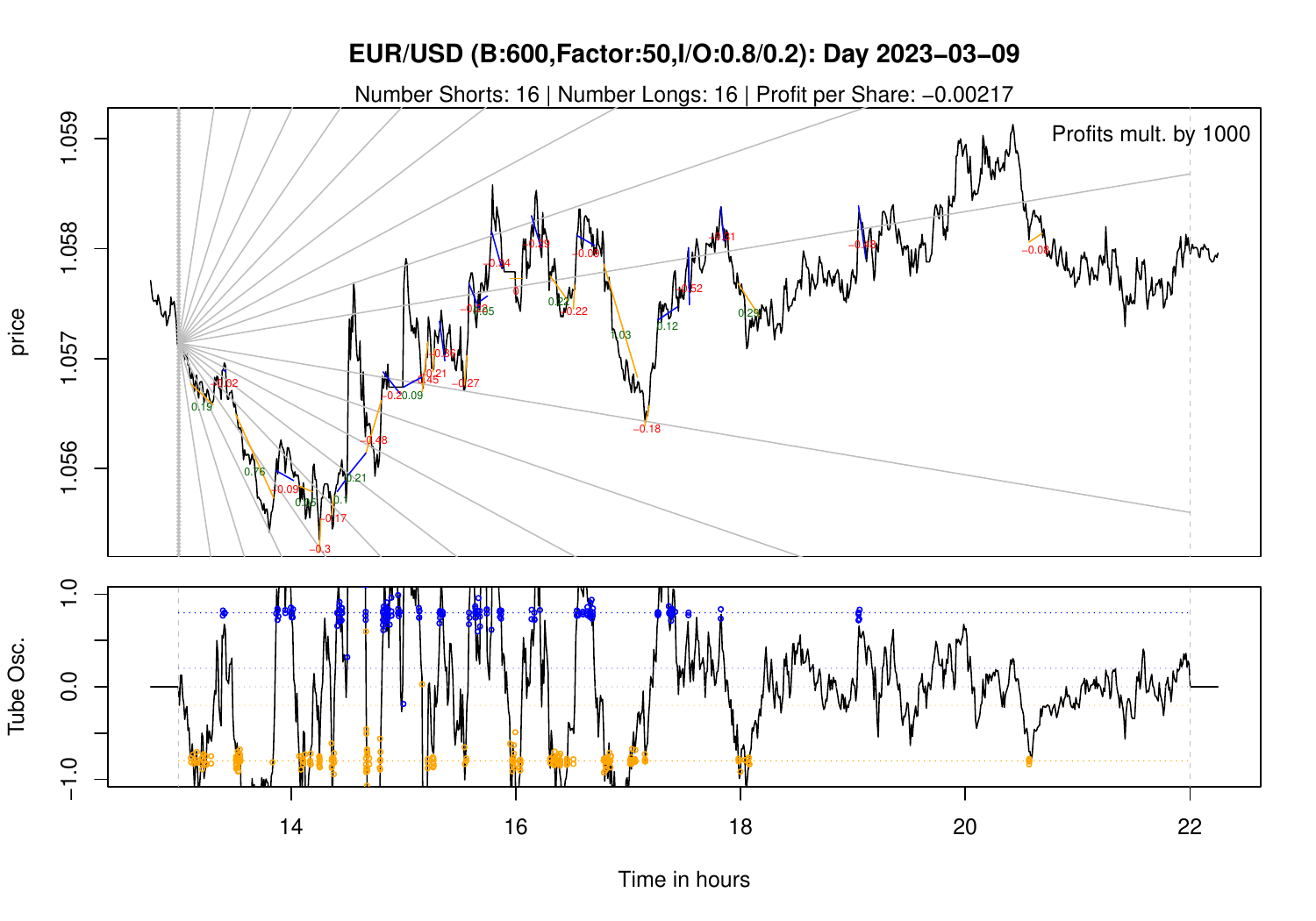}\\
		\includegraphics[width=13cm]{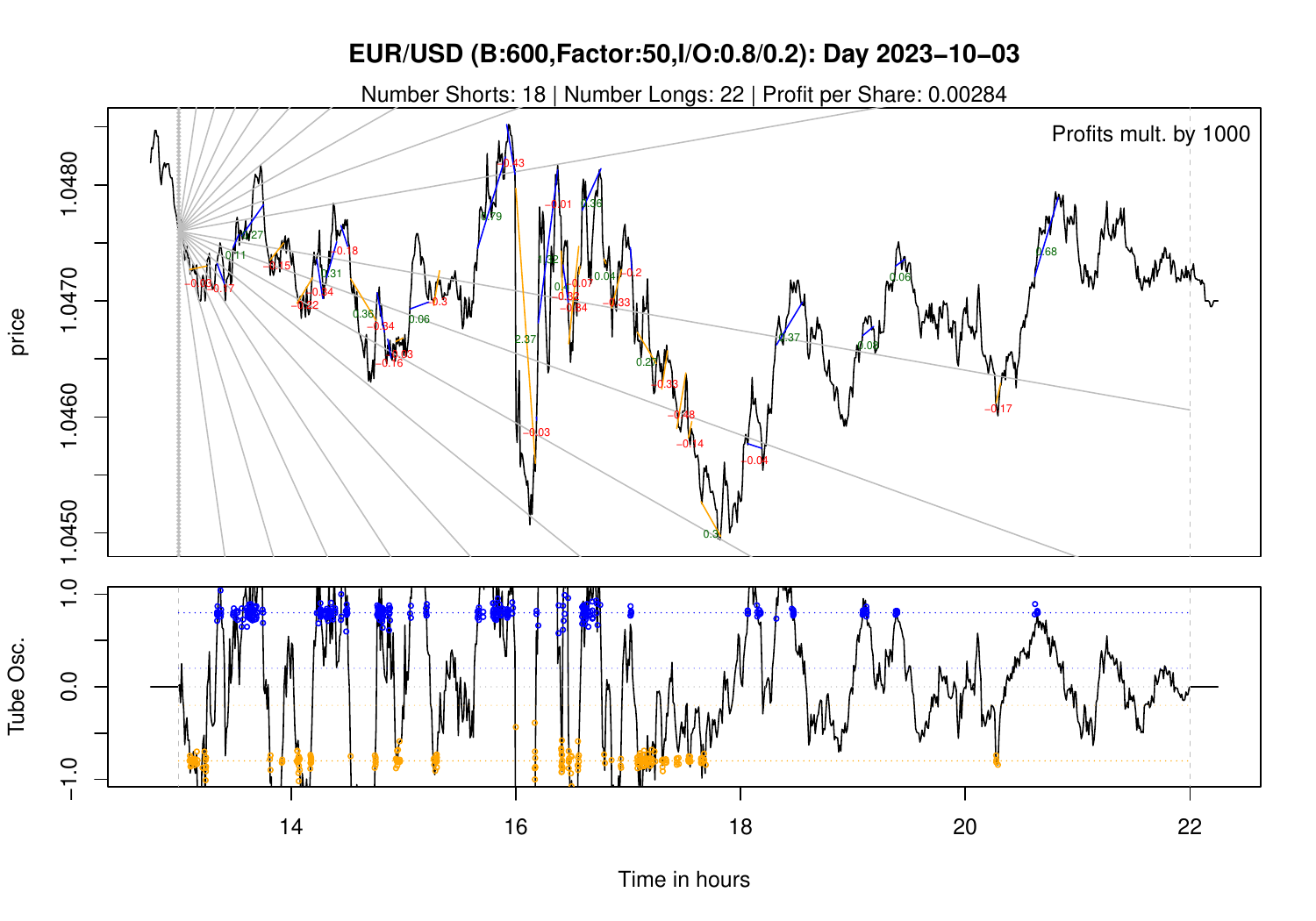}\\
	\end{tabular}
	\label{fig_EURUSD_600_OSC50_IN08_OUT02_example}
\end{figure}

\begin{figure}[h!]
	\centering
	\caption{EUR/USD, Bandwidth 300s, In/Out: 0.8/0.2: Behavior of the trading strategy from 2019/01 to 2024/05. For a detailed explanation of the plots, see Figure \ref{fig_EURUSD_300_OSC20_IN04_OUT01_example}.}
	\begin{tabular}{c}
		\includegraphics[width=13cm]{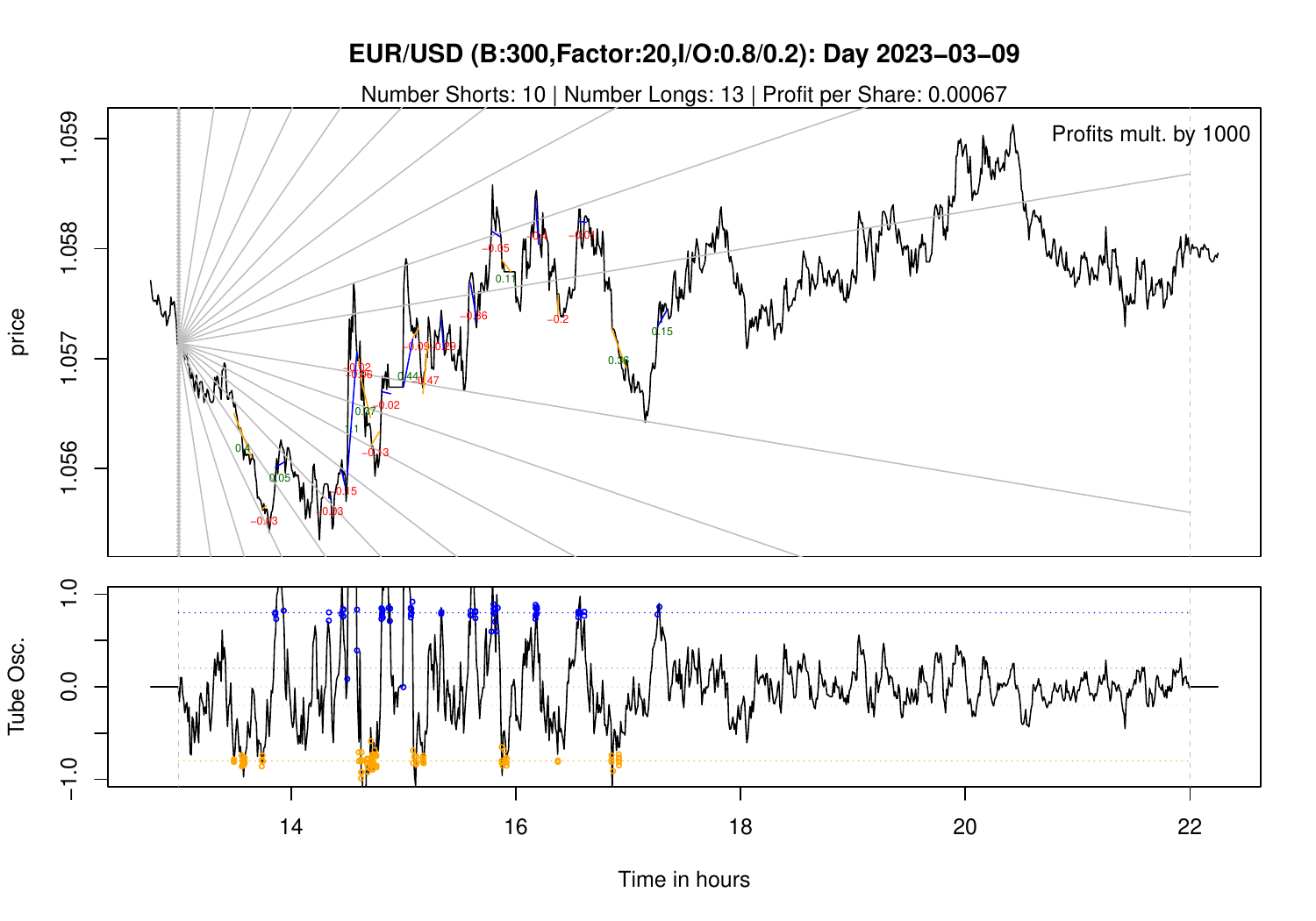}\\
		\includegraphics[width=13cm]{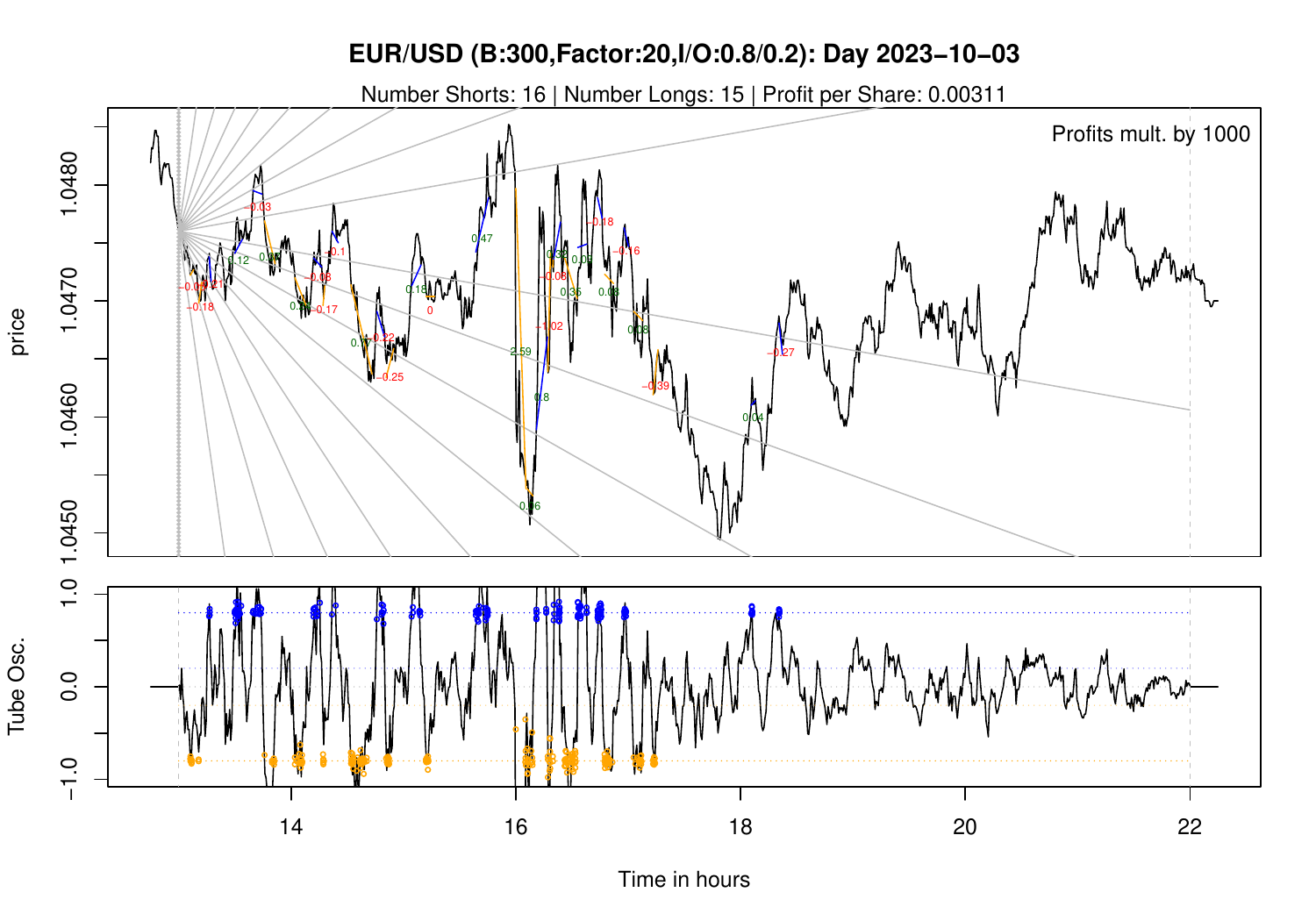}\\
	\end{tabular}
	\label{fig_EURUSD_300_OSC20_IN08_OUT02_example}
\end{figure}

\FloatBarrier

\subsection{Forex EUR/USD: Trading strategy characteristics for other configurations}
\label{EURUSD_characteristics}

\begin{figure}[b!]
	\centering
	\caption{EUR/USD, Bandwidth 600s, In/Out: 0.4/0.1: Characteristics of the trading strategy from 2019/01 to 2024/05.}
	\begin{tabular}{cc}
		\includegraphics[width=6.5cm]{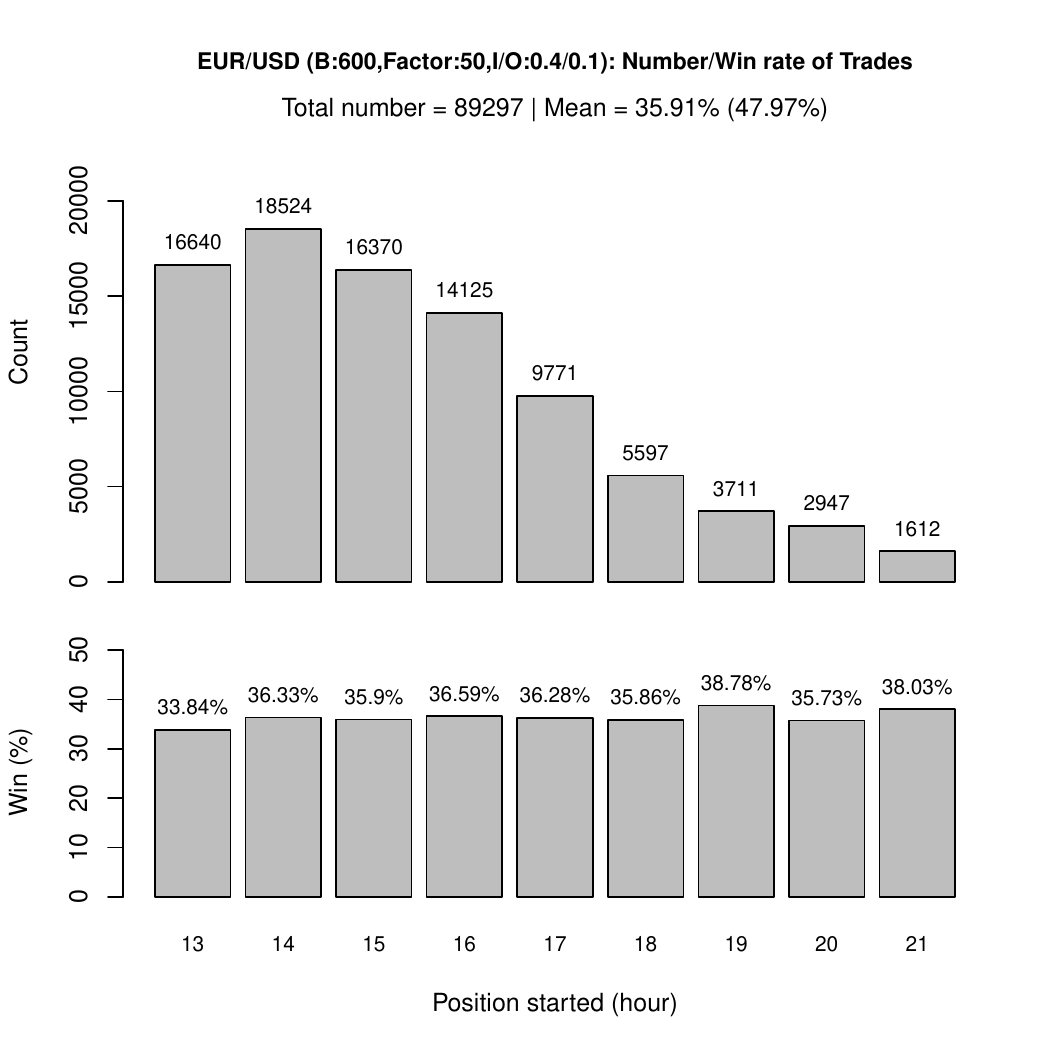} & \includegraphics[width=6.5cm]{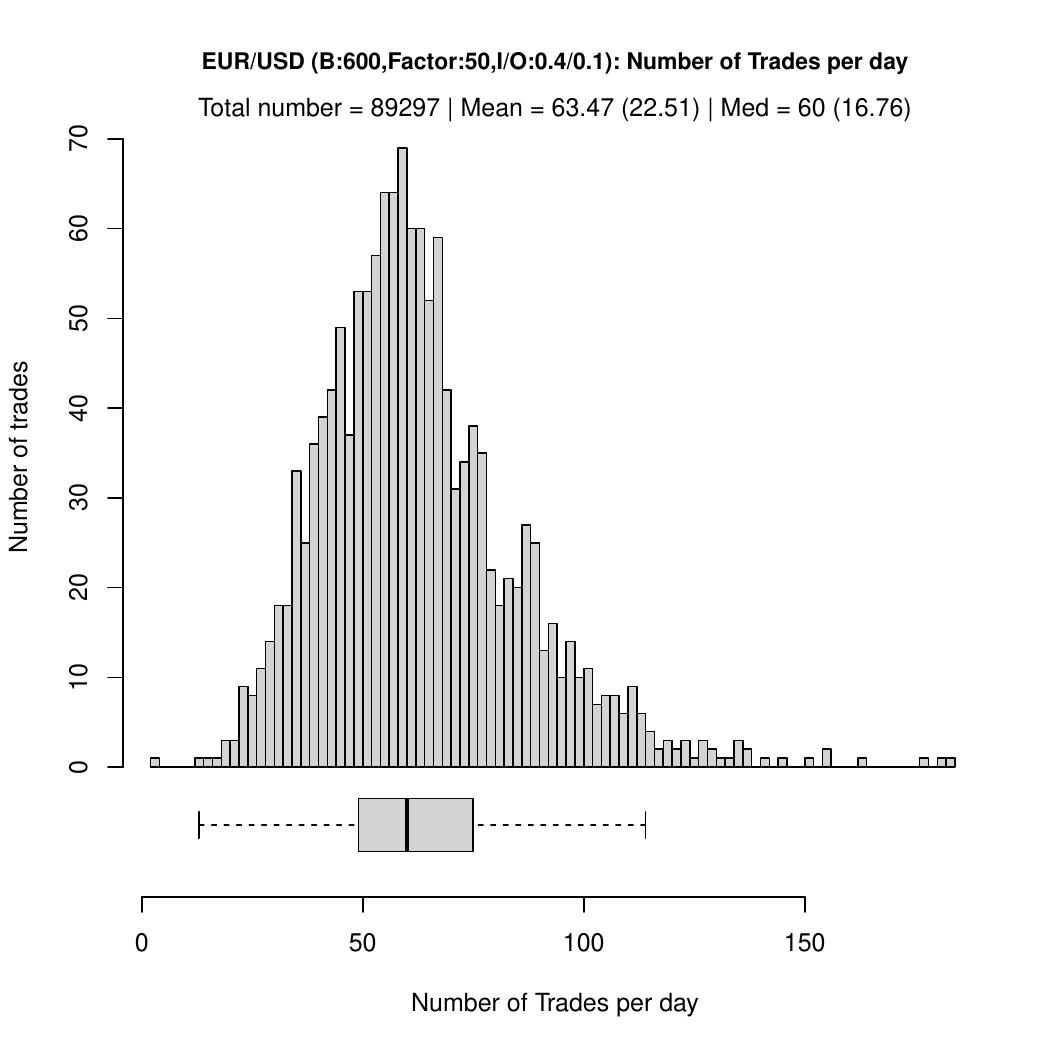}\\
		\includegraphics[width=6.5cm]{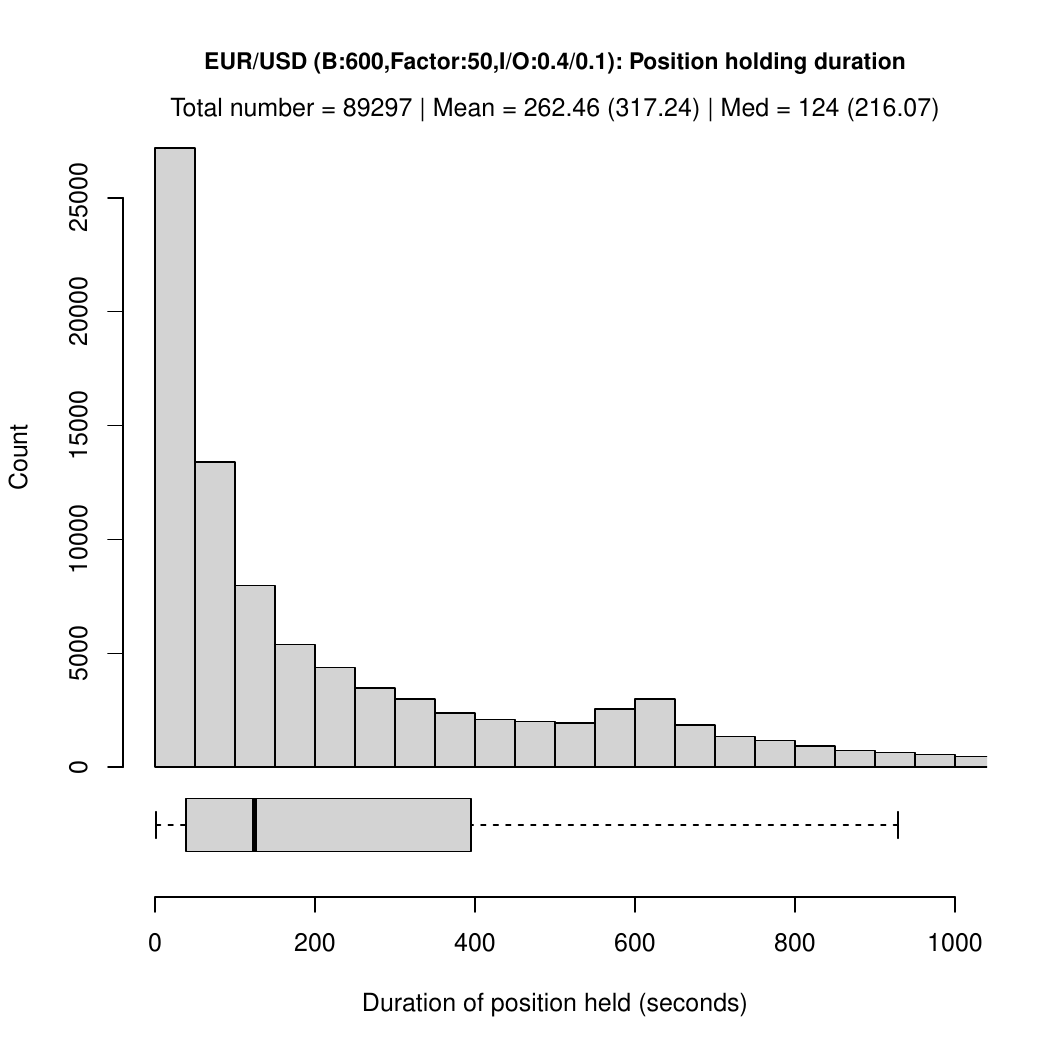} & \includegraphics[width=6.5cm]{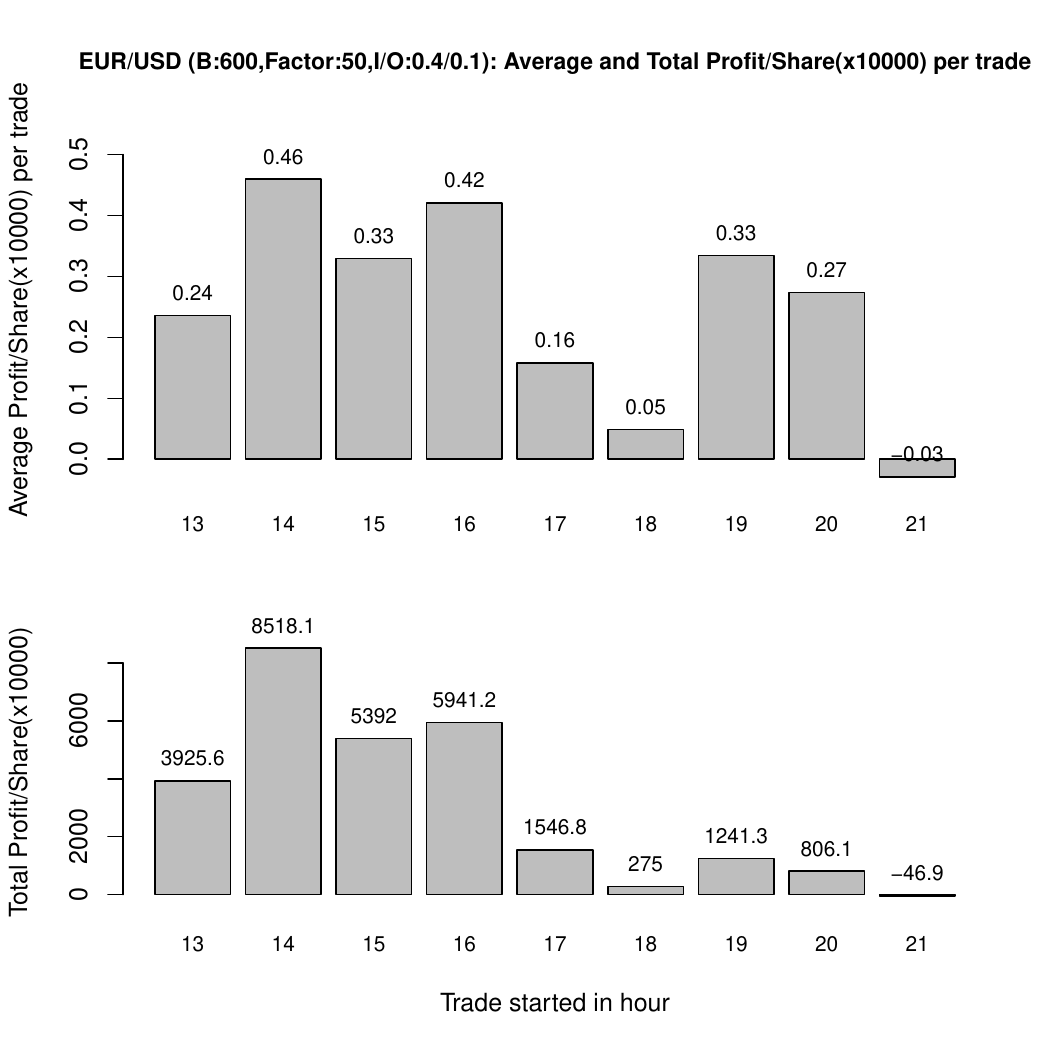}\\
		\includegraphics[width=6.5cm]{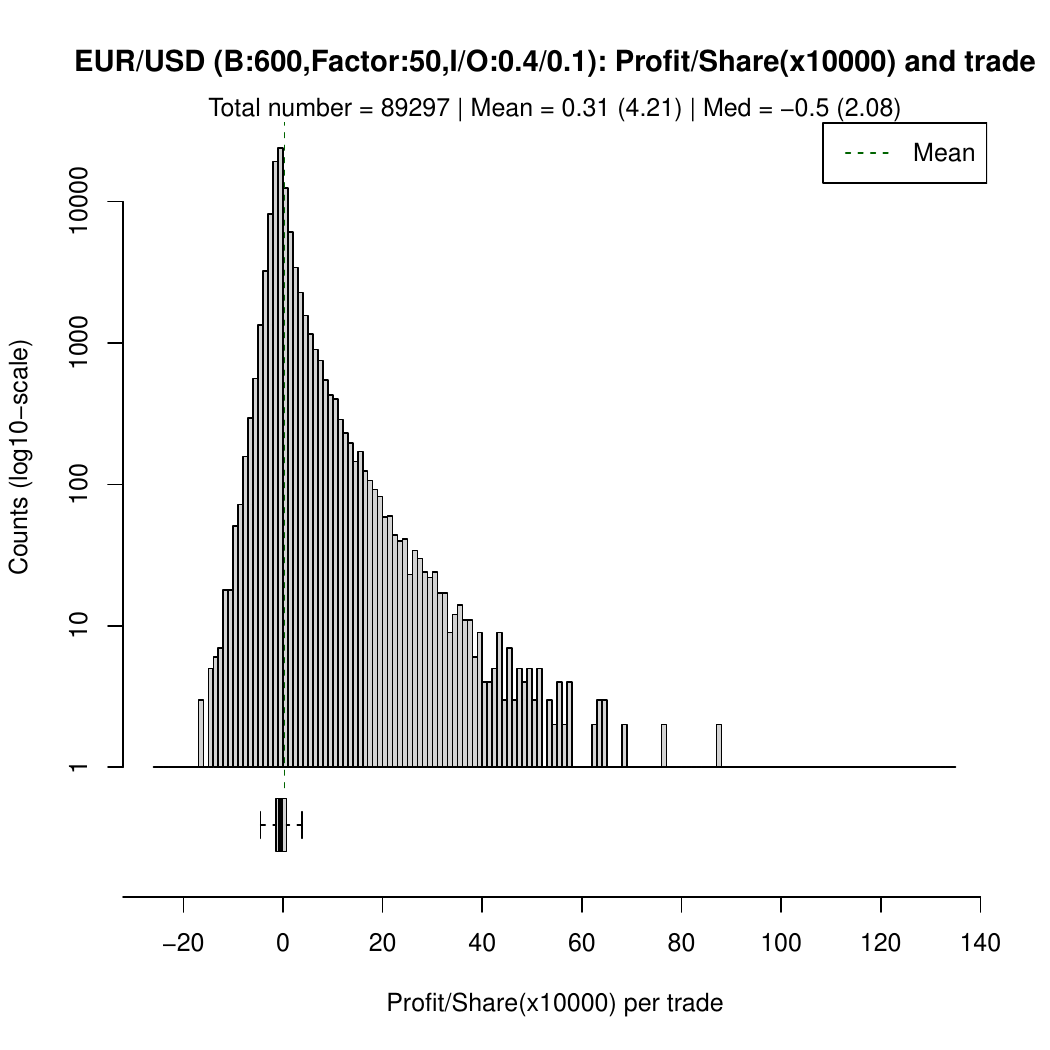} & \includegraphics[width=6.5cm]{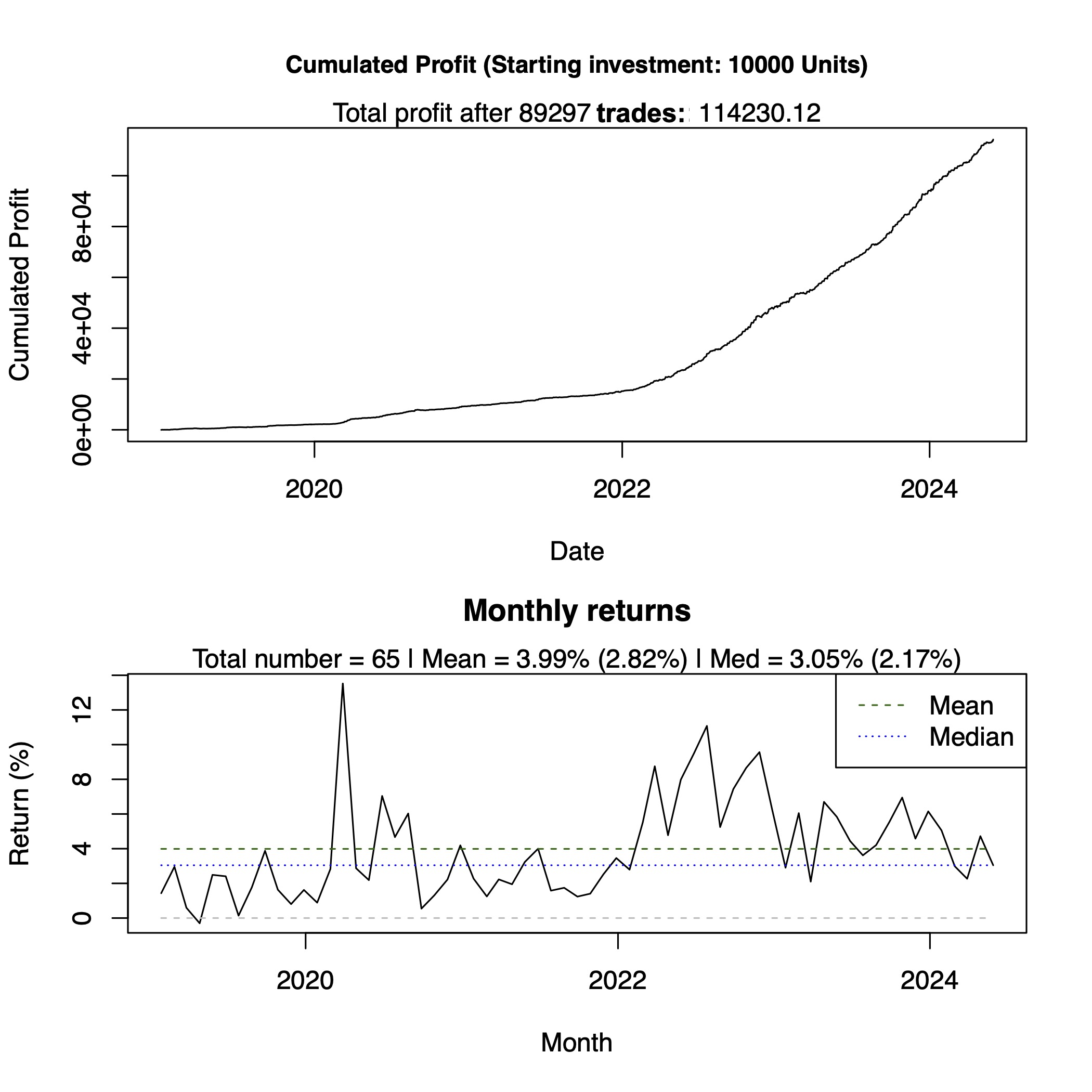}\\
	\end{tabular}
	\label{fig_EURUSD_600_OSC50_IN04_OUT01}
\end{figure}

\begin{figure}[h!]
	\centering
	\caption{EUR/USD, Bandwidth 600s, In/Out: 0.8/0.2: Characteristics of the trading strategy from 2019/01 to 2024/05.}
	\begin{tabular}{cc}
		\includegraphics[width=6.5cm]{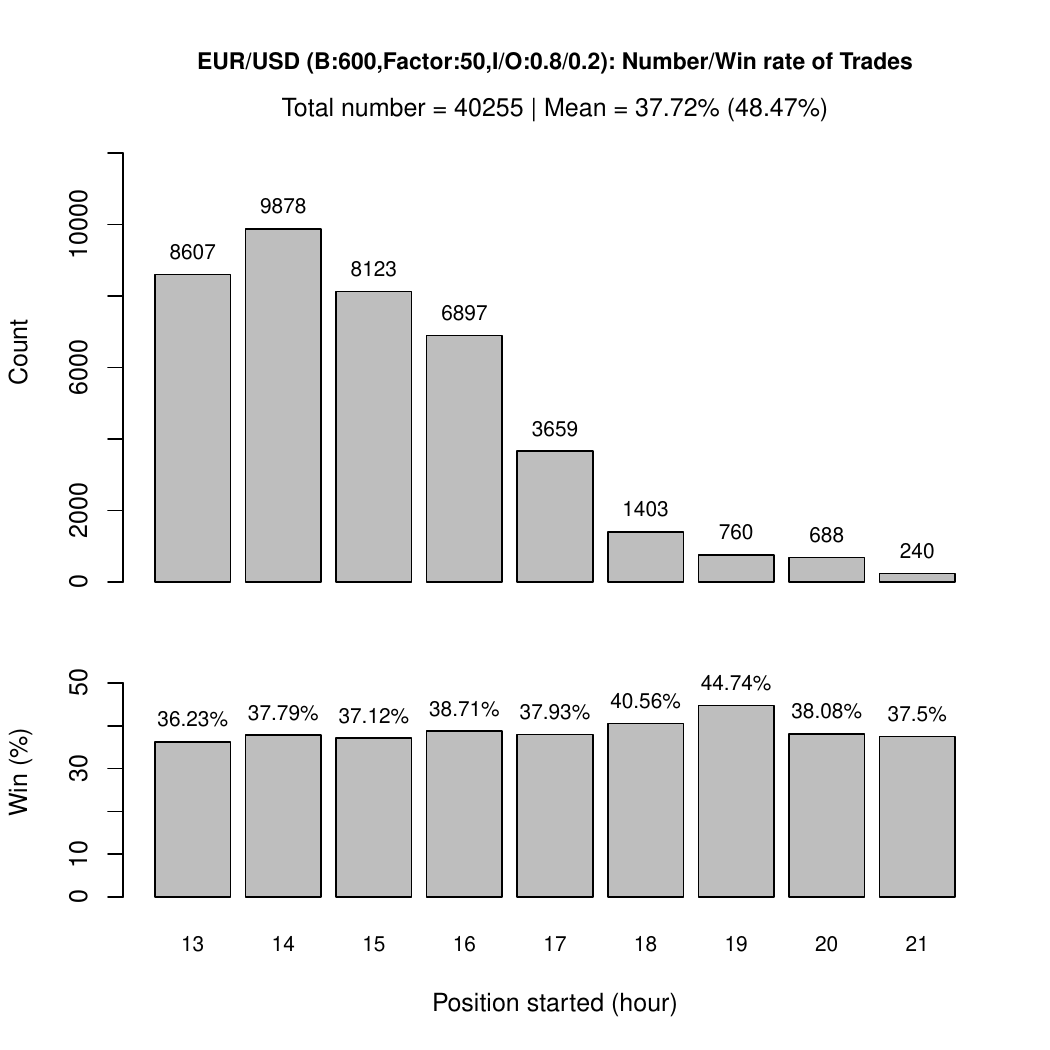} & \includegraphics[width=6.5cm]{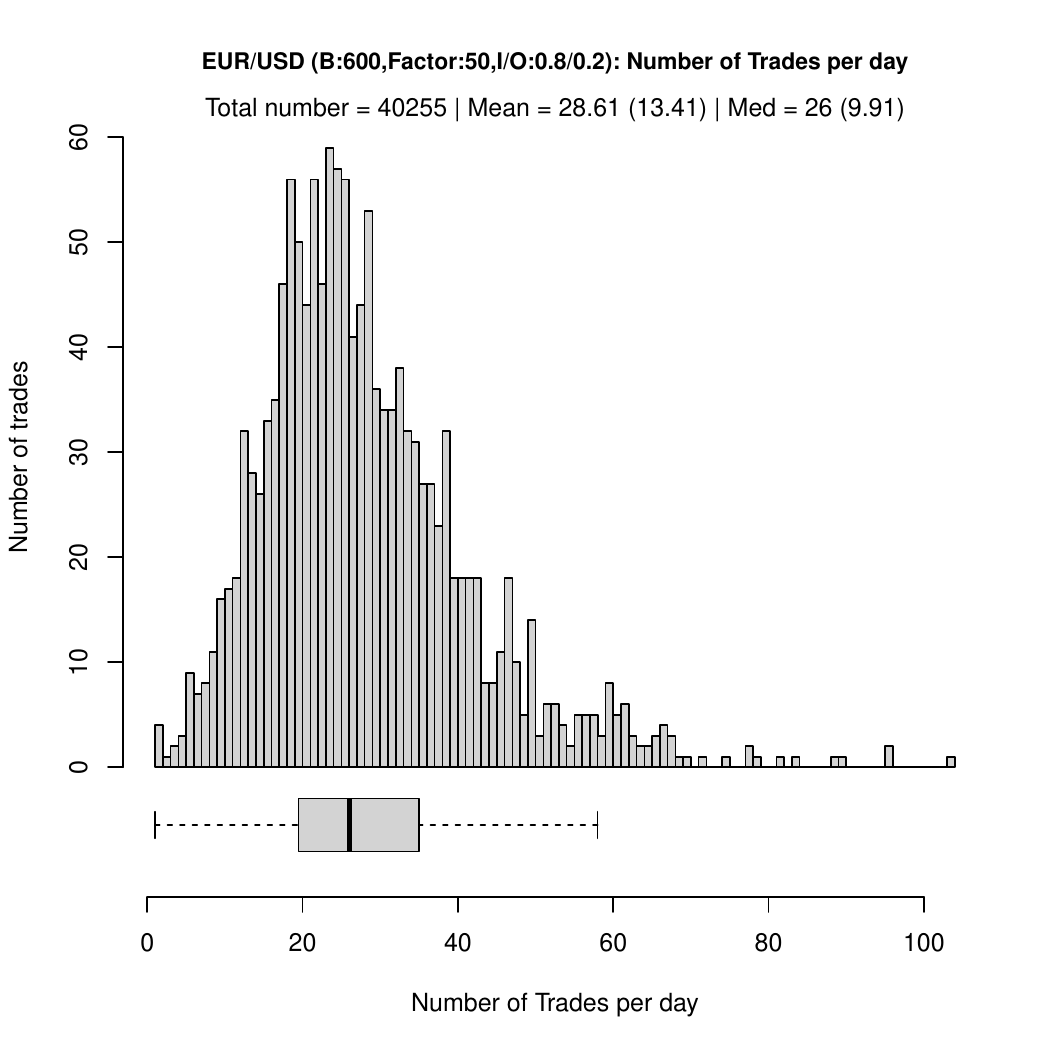}\\
		\includegraphics[width=6.5cm]{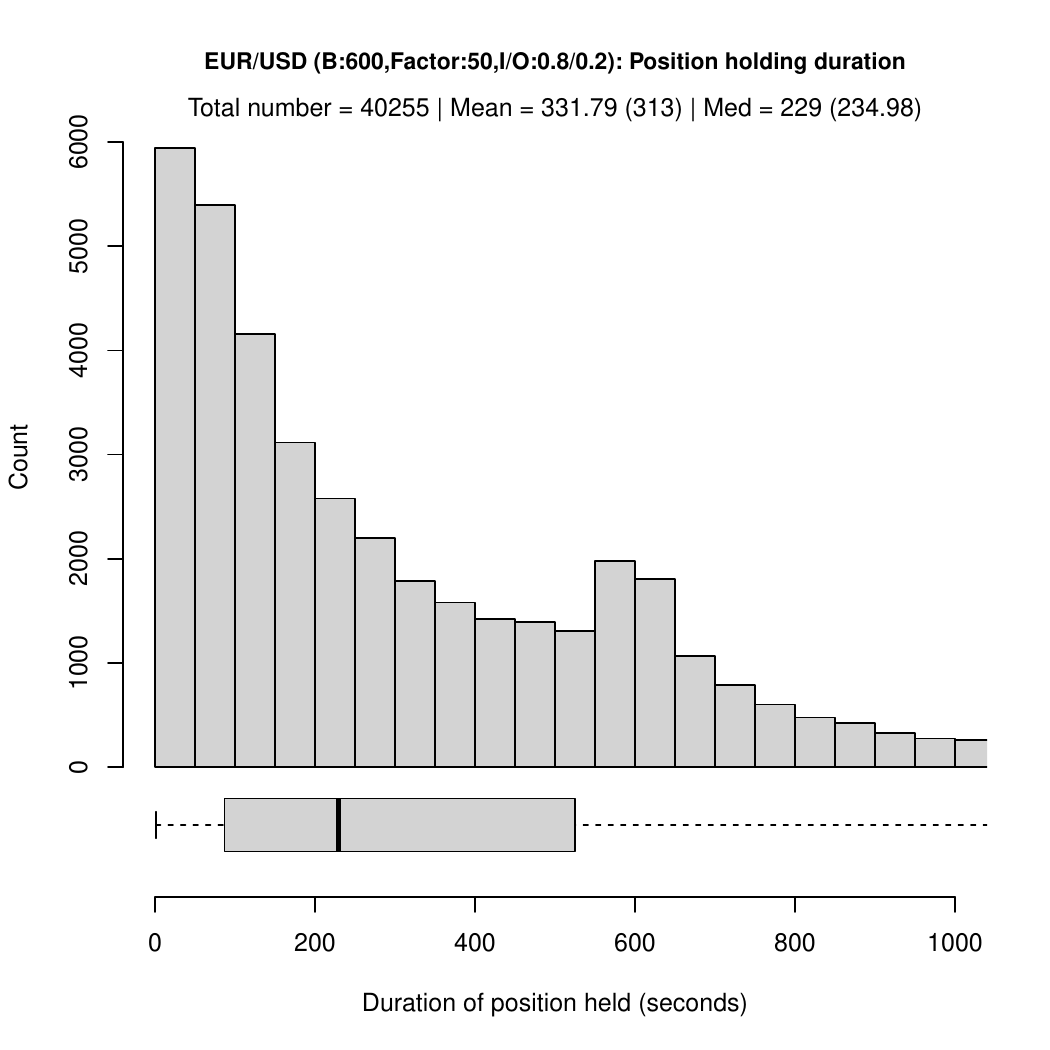} & \includegraphics[width=6.5cm]{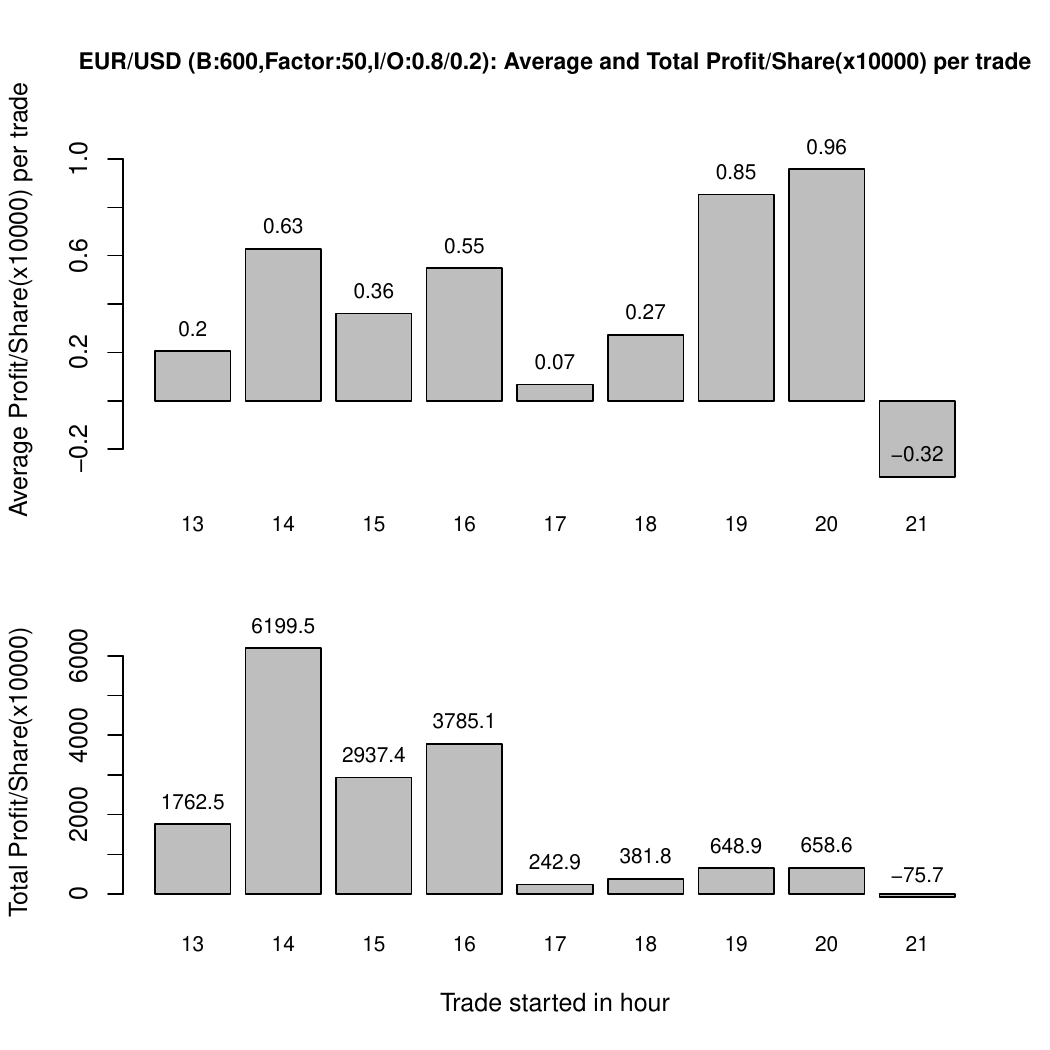}\\
		\includegraphics[width=6.5cm]{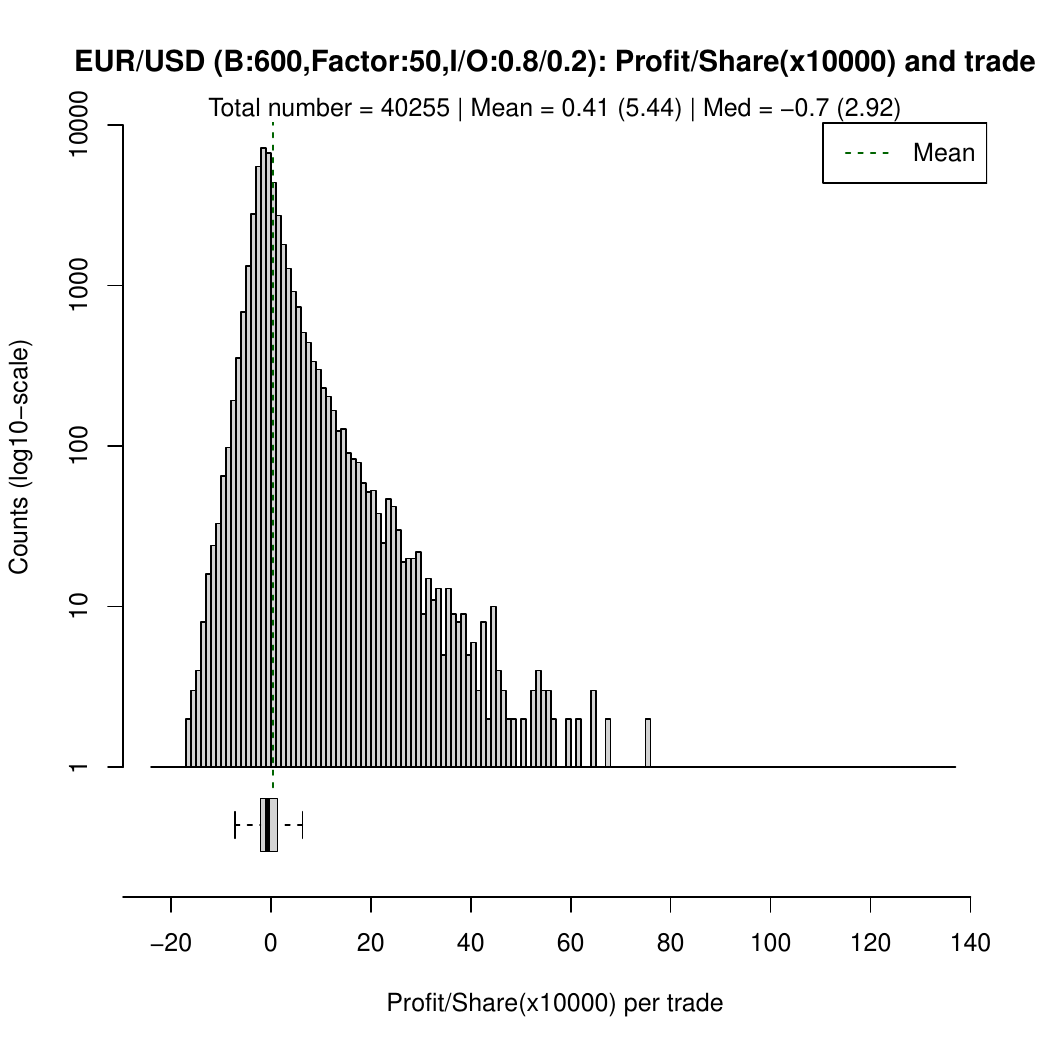} & \includegraphics[width=6.5cm]{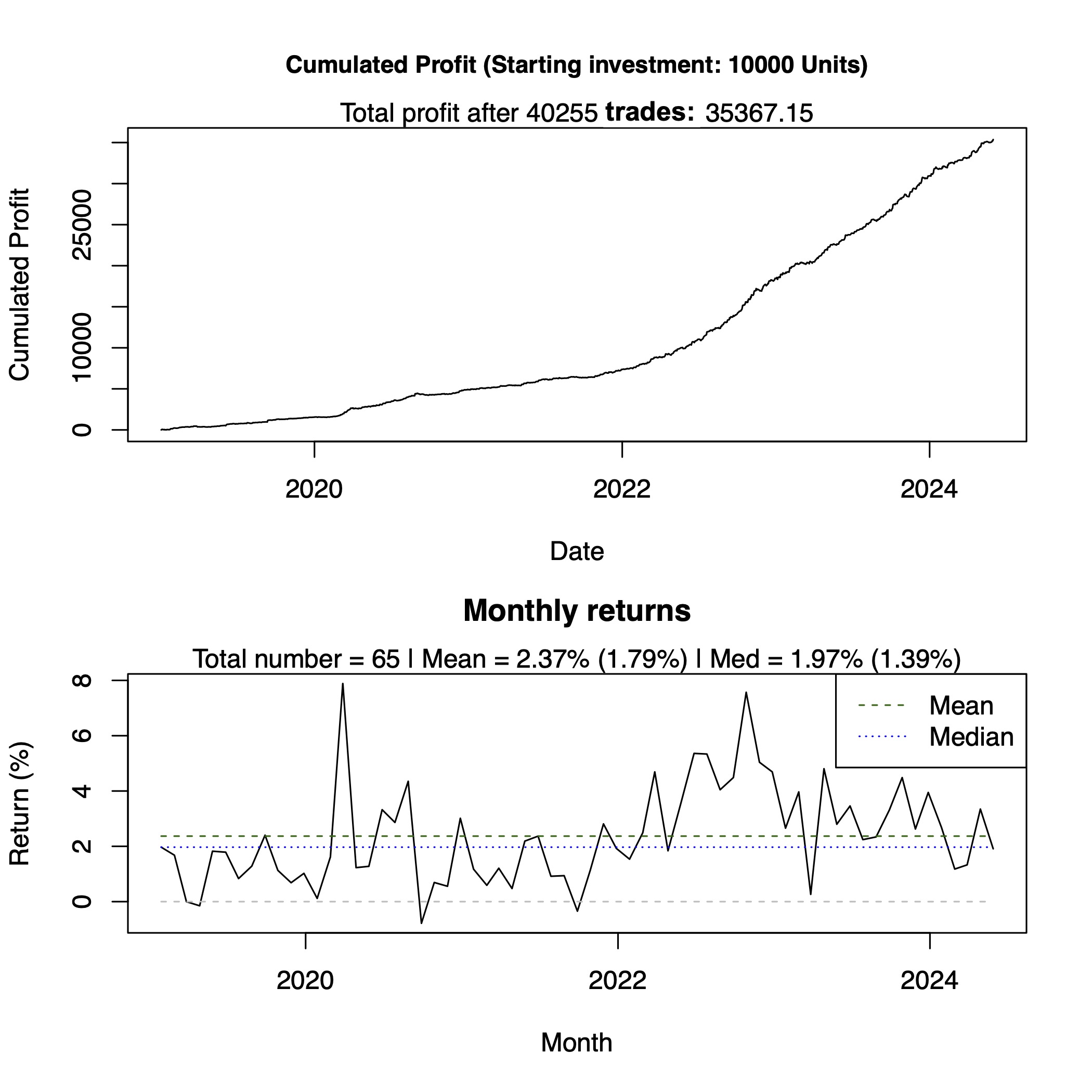}\\
	\end{tabular}
	\label{fig_EURUSD_600_OSC50_IN08_OUT02}
\end{figure}

\begin{figure}[h!]
	\centering
	\caption{EUR/USD, Bandwidth 300s, In/Out: 0.8/0.2: Characteristics of the trading strategy from 2019/01 to 2024/05.}
	\begin{tabular}{cc}
		\includegraphics[width=6.5cm]{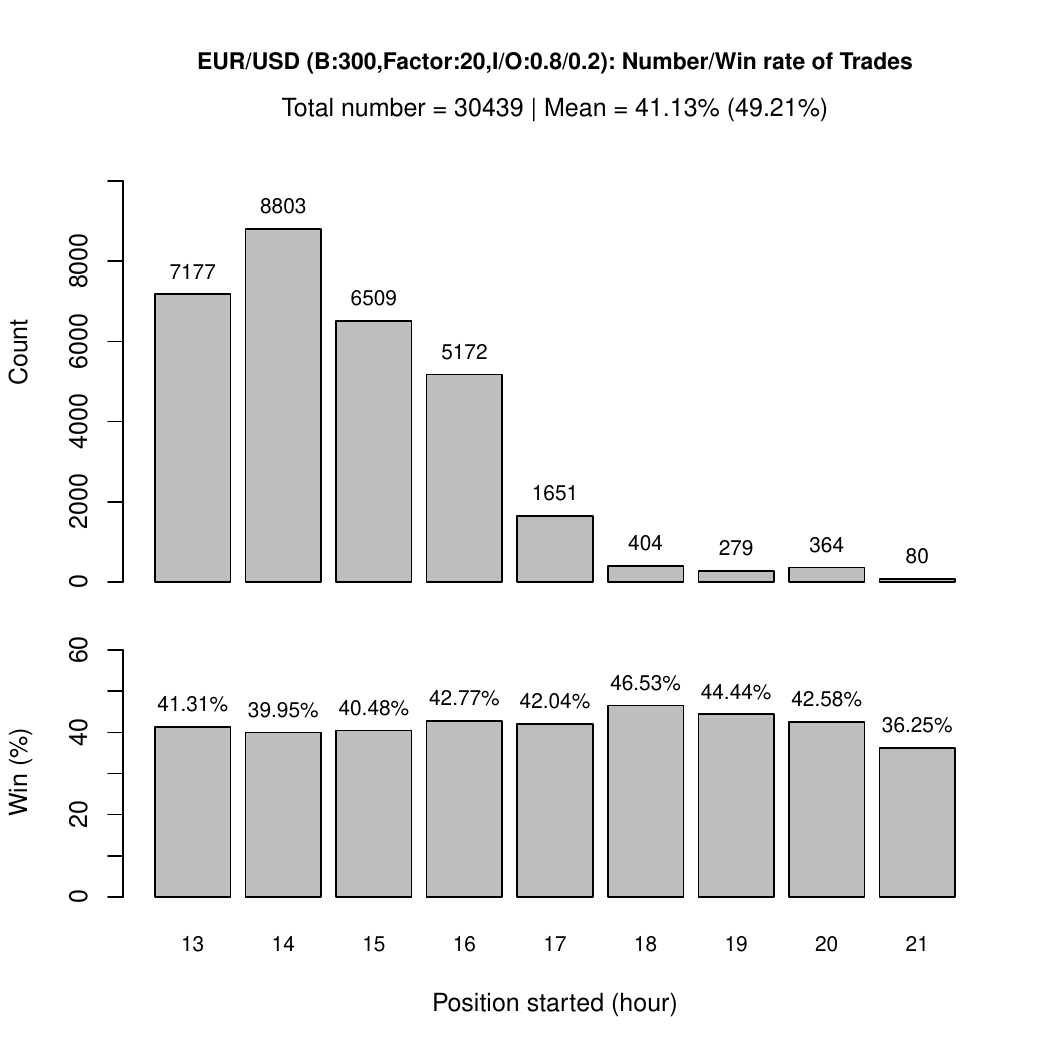} & \includegraphics[width=6.5cm]{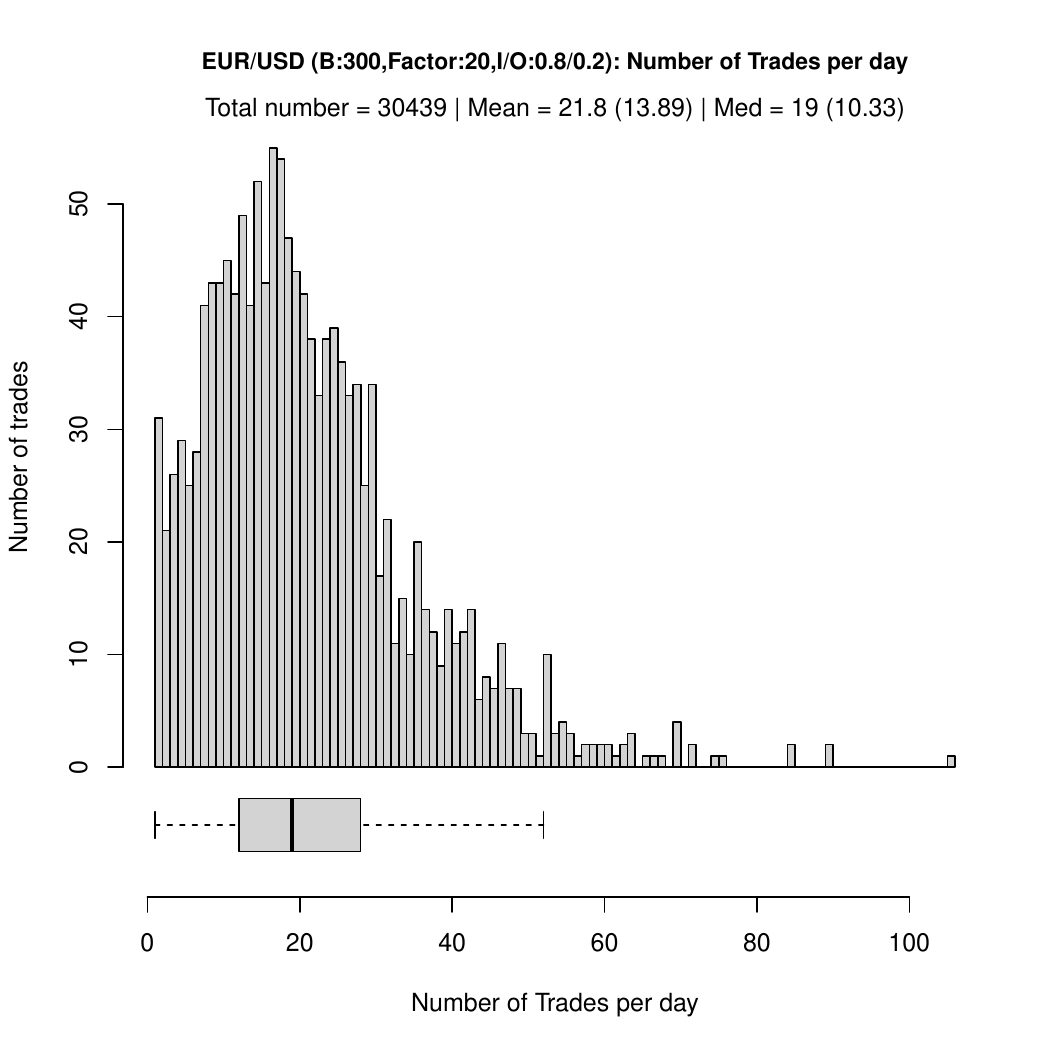}\\
		\includegraphics[width=6.5cm]{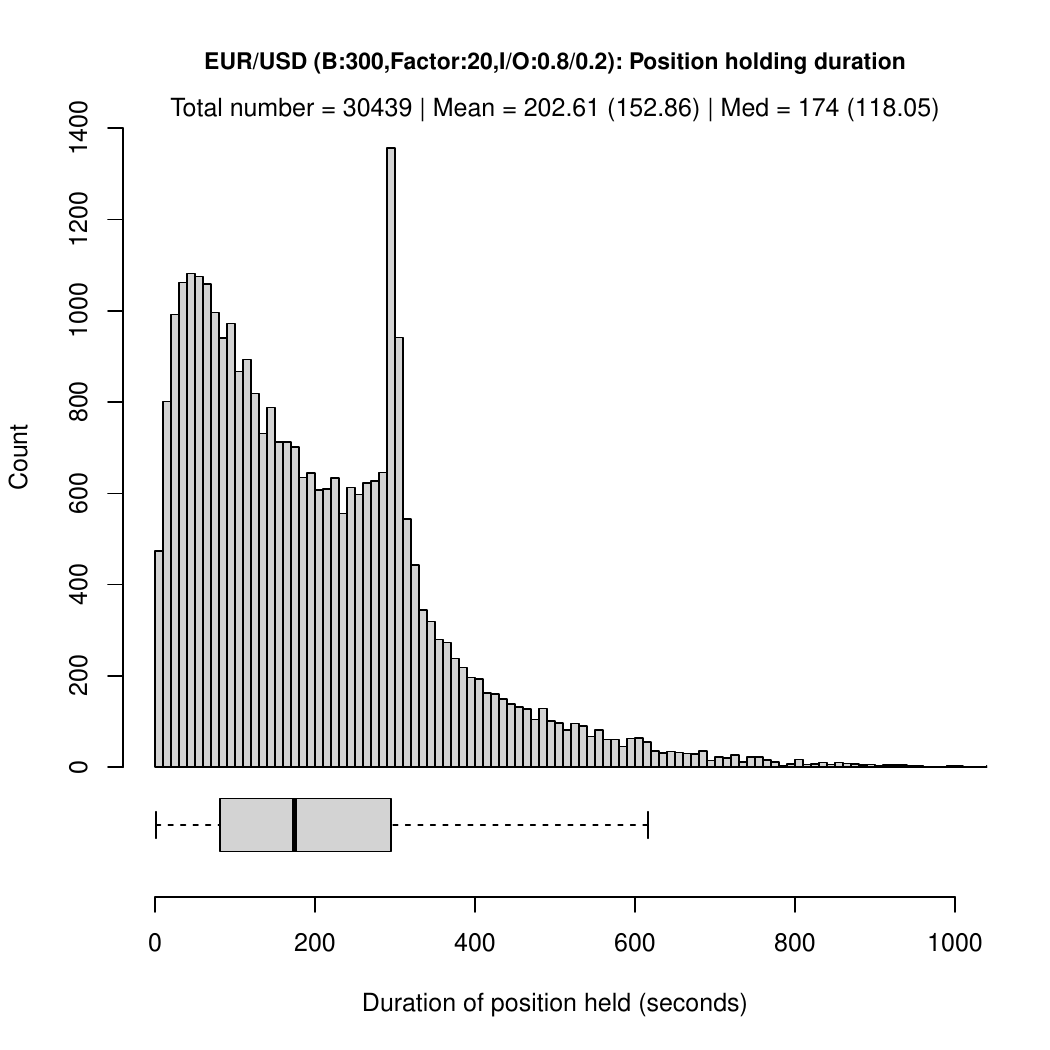} & \includegraphics[width=6.5cm]{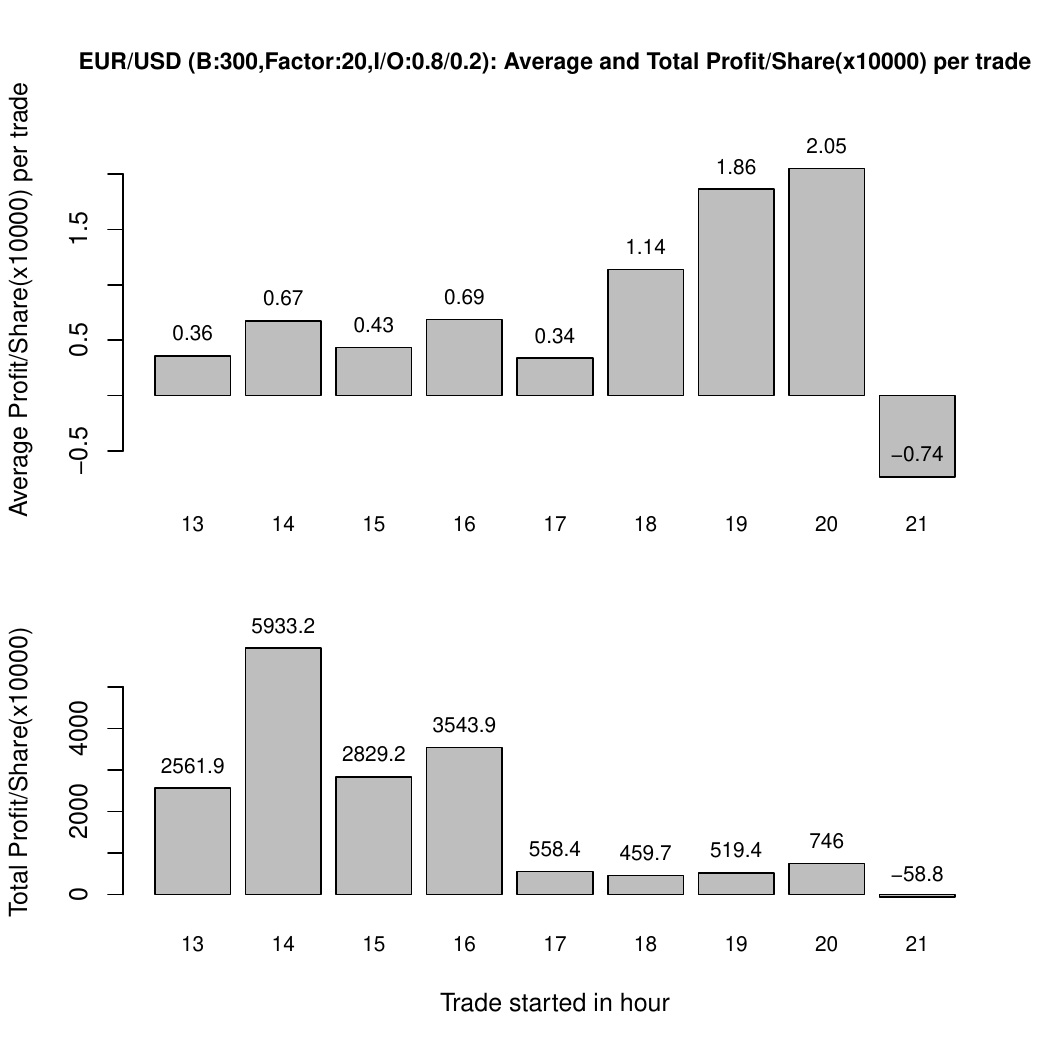}\\
		\includegraphics[width=6.5cm]{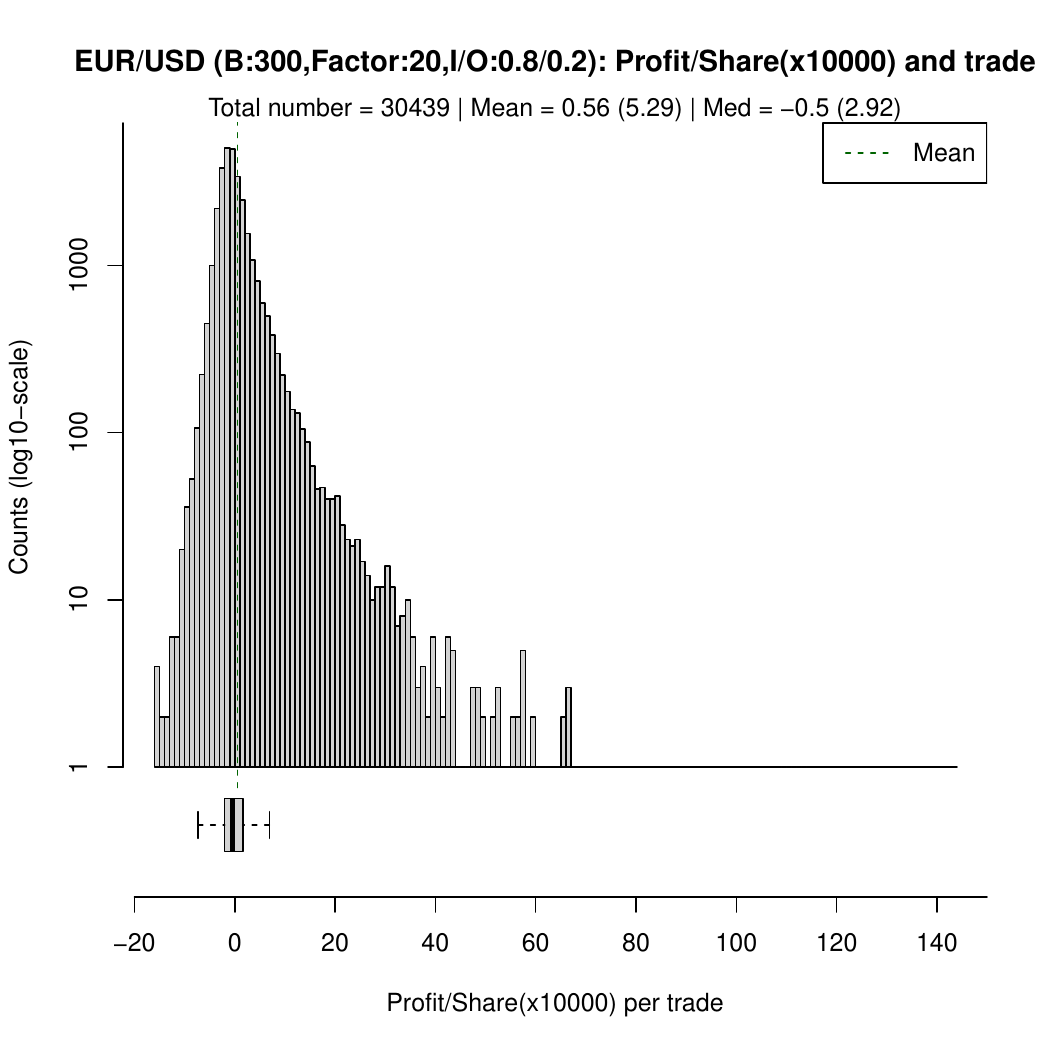} & \includegraphics[width=6.5cm]{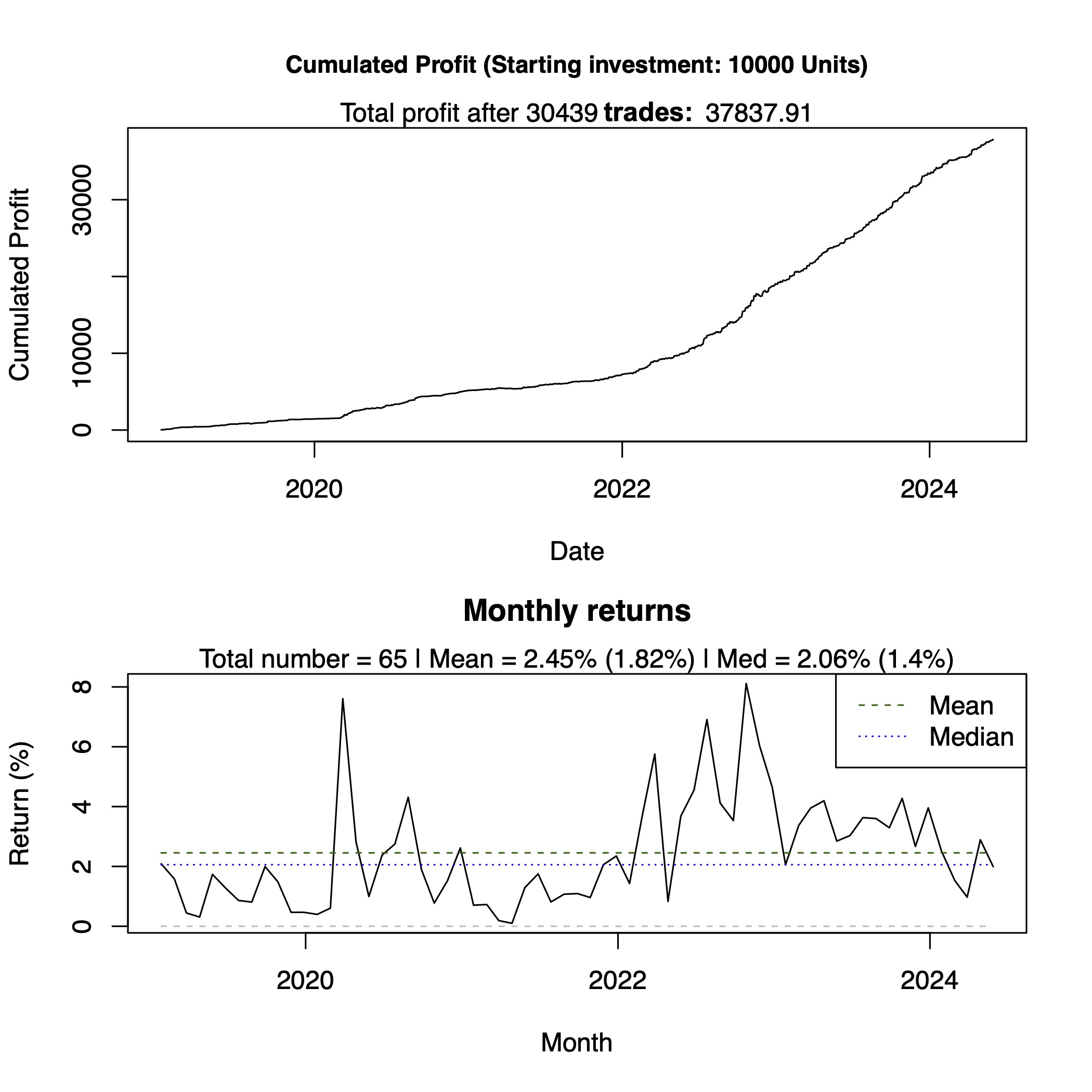}\\
	\end{tabular}
	\label{fig_EURUSD_300_OSC20_IN08_OUT02}
\end{figure}

\FloatBarrier

\FloatBarrier
\subsection{German DAX 40: Trading strategy behavior for other configurations}
\label{DAX_examples}

\begin{figure}[b!]
	\centering
	\caption{DAX, Bandwidth 600s, In/Out: 0.4/0.1: Behavior of the trading strategy from 2019/01 to 2024/05. For a detailed explanation of the plots, see Figure \ref{fig_EURUSD_300_OSC20_IN04_OUT01_example}.}
	\begin{tabular}{c}
		\includegraphics[width=13cm]{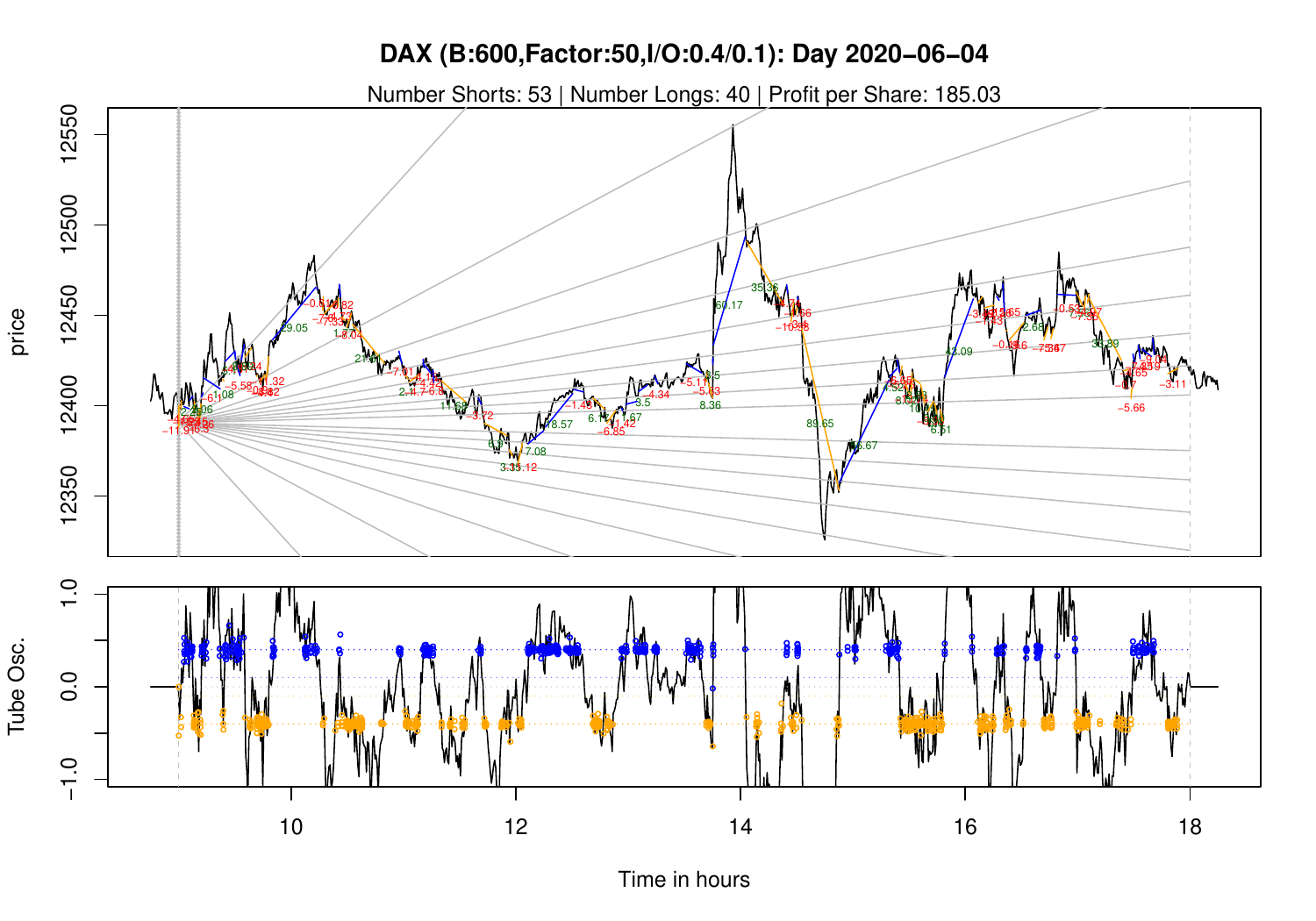}\\
		\includegraphics[width=13cm]{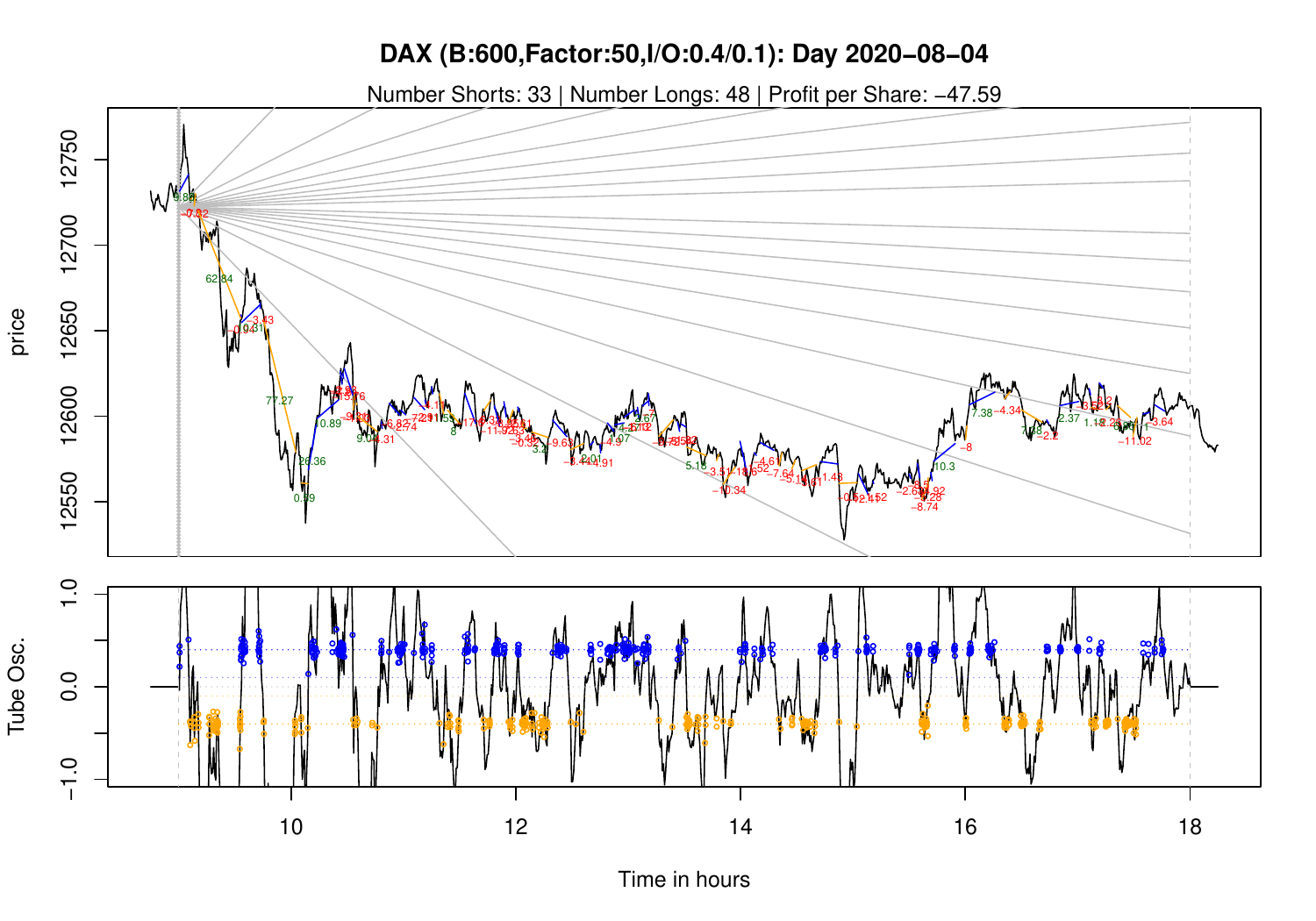}\\
	\end{tabular}
	\label{fig_DAX_600_OSC50_IN04_OUT01_example}
\end{figure}

\begin{figure}[h!]
	\centering
	\caption{DAX, Bandwidth 300s, In/Out: 0.4/0.1: Behavior of the trading strategy from 2019/01 to 2024/05. For a detailed explanation of the plots, see Figure \ref{fig_EURUSD_300_OSC20_IN04_OUT01_example}.}
	\begin{tabular}{c}
		\includegraphics[width=13cm]{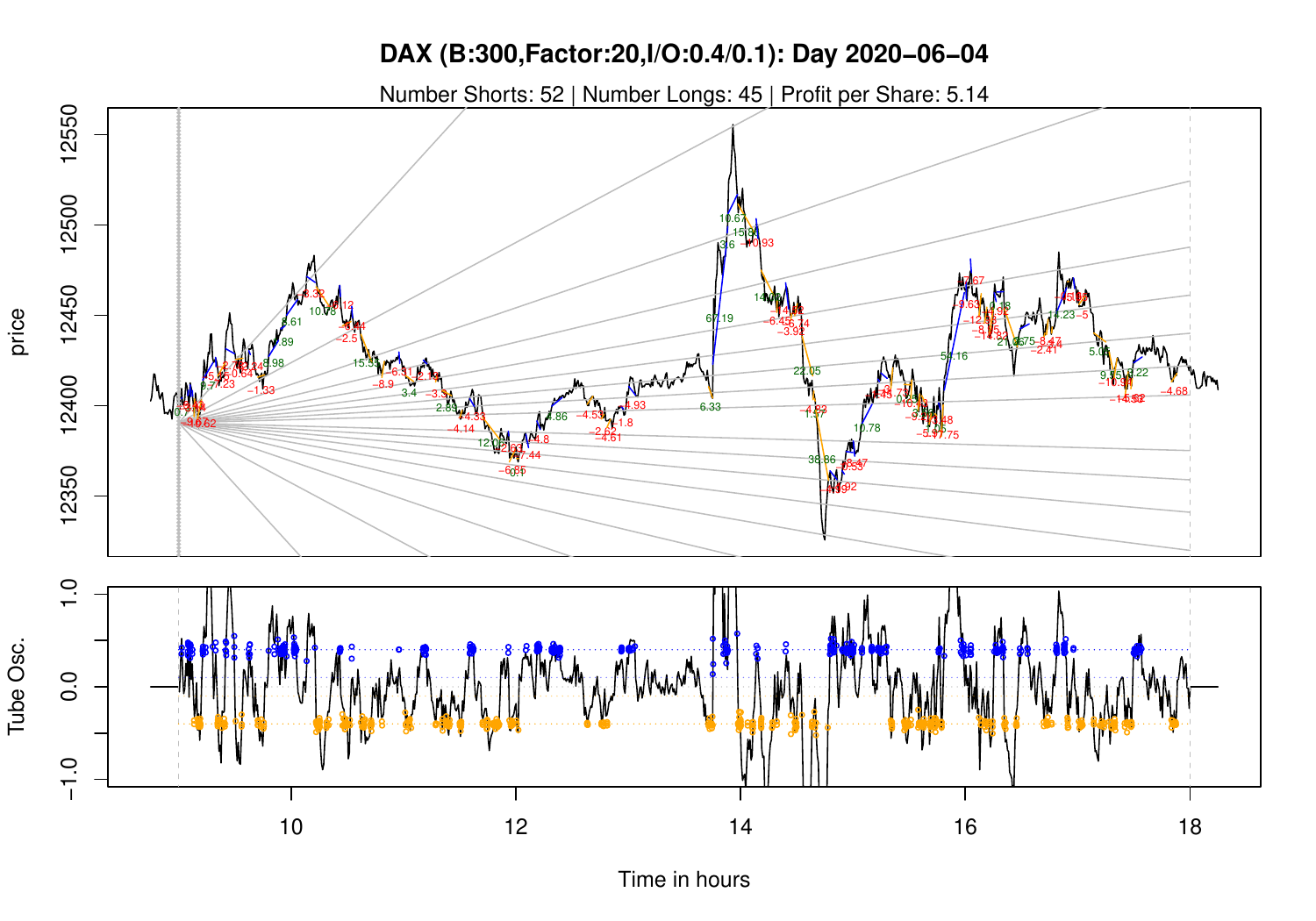}\\
		\includegraphics[width=13cm]{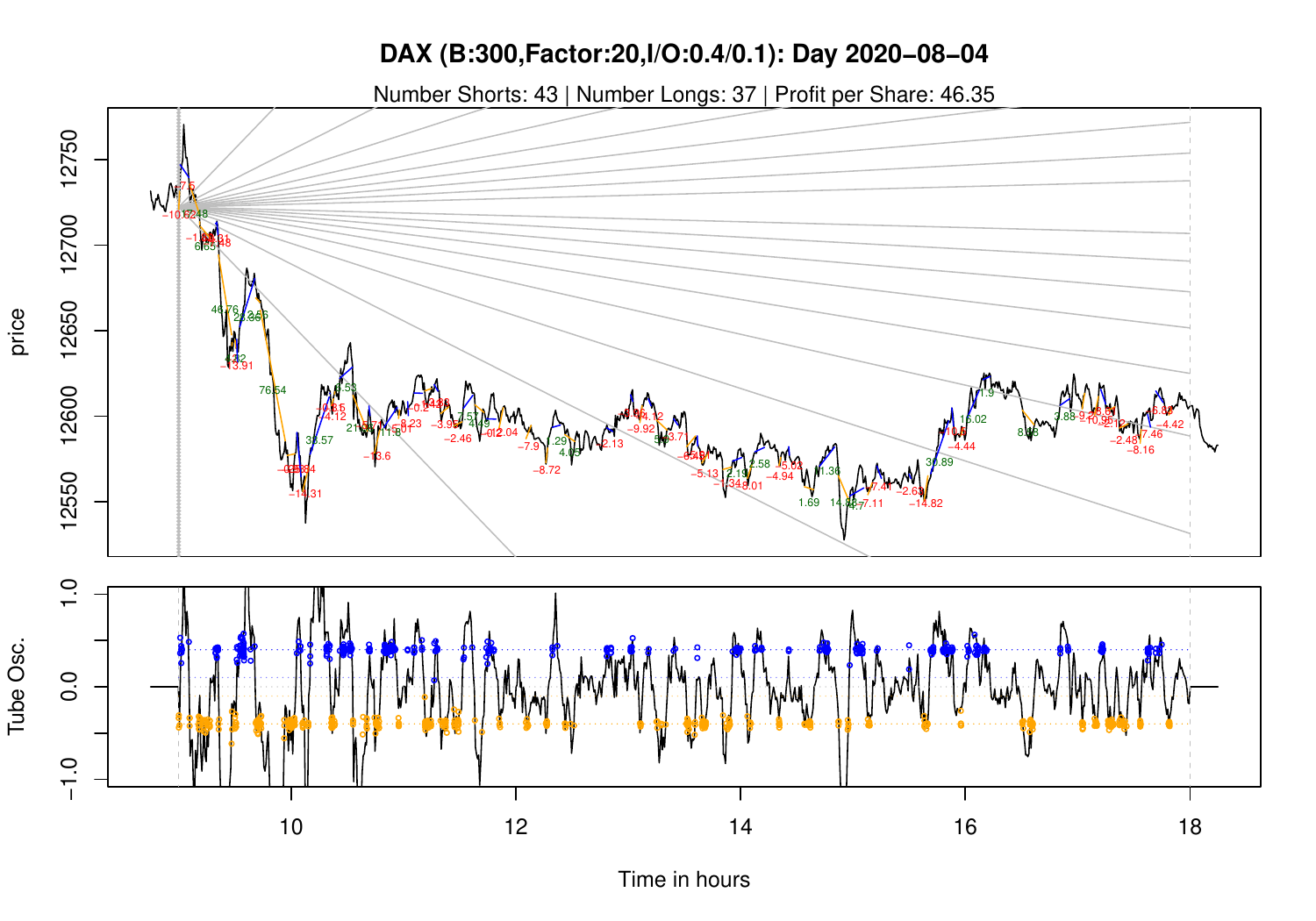}\\
	\end{tabular}
	\label{fig_DAX_300_OSC20_IN04_OUT01_example}
\end{figure}

\begin{figure}[h!]
	\centering
	\caption{DAX, Bandwidth 300s, In/Out: 0.8/0.2: Behavior of the trading strategy from 2019/01 to 2024/05. For a detailed explanation of the plots, see Figure \ref{fig_EURUSD_300_OSC20_IN04_OUT01_example}.}
	\begin{tabular}{c}
		\includegraphics[width=13cm]{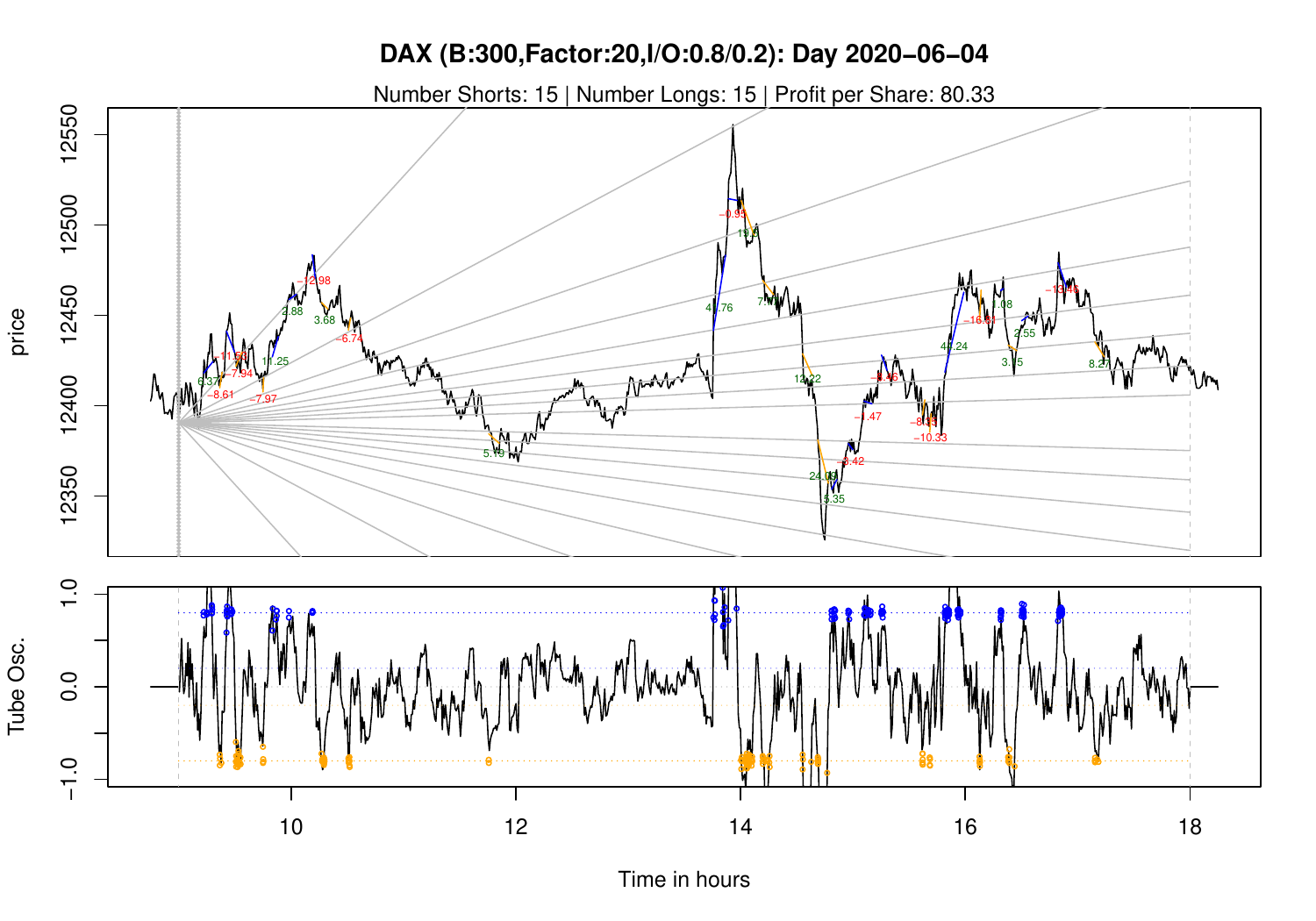}\\
		\includegraphics[width=13cm]{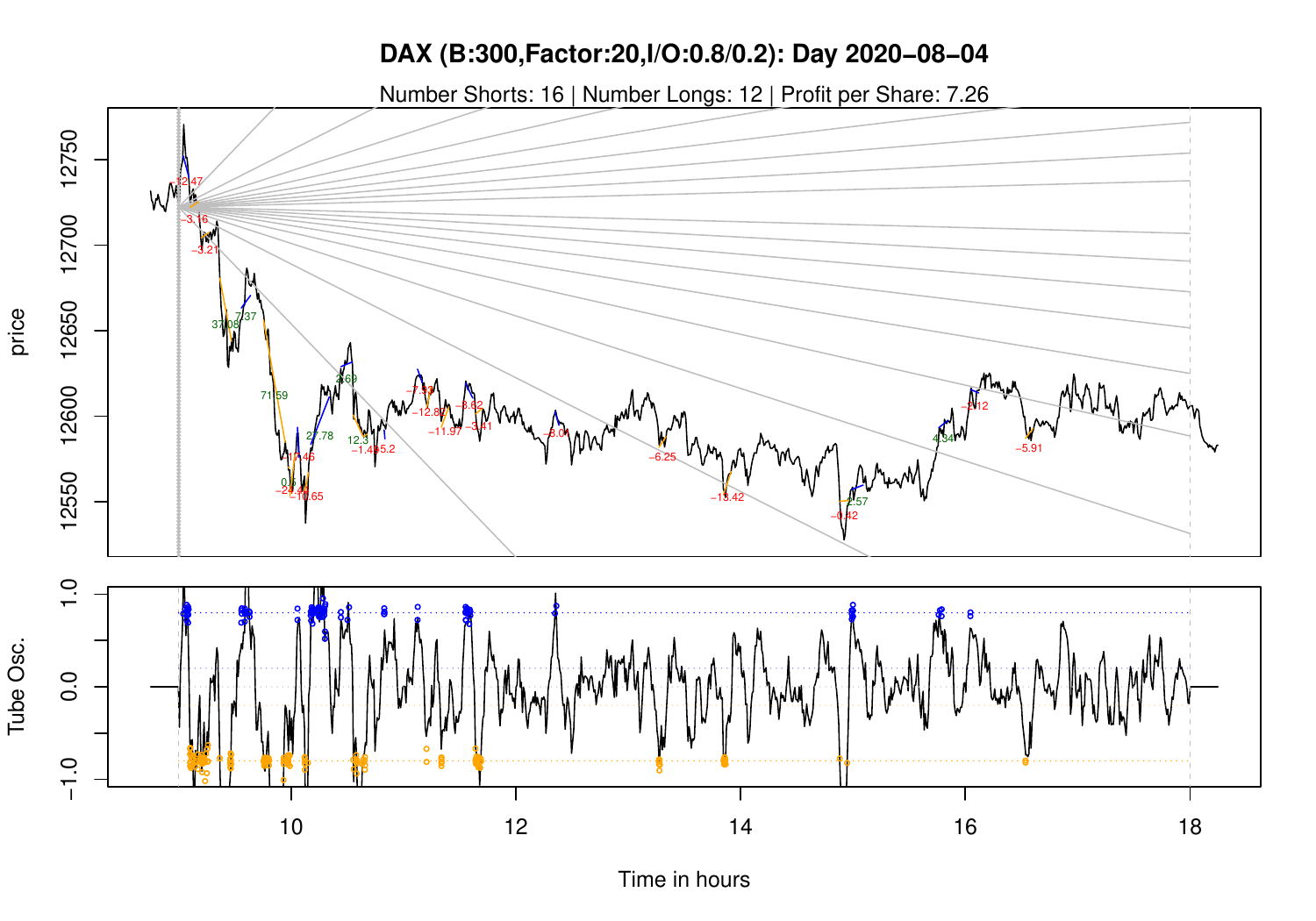}\\
	\end{tabular}
	\label{fig_DAX_300_OSC20_IN08_OUT02_example}
\end{figure}

\FloatBarrier

\subsection{German DAX 40: Trading strategy characteristics for other configurations}
\label{DAX_characteristics}

\begin{figure}[b!]
	\centering
	\caption{DAX, Bandwidth 600s, In/Out: 0.4/0.1: Characteristics of the trading strategy from 2019/01 to 2024/05.}
	\begin{tabular}{cc}
		\includegraphics[width=6.5cm]{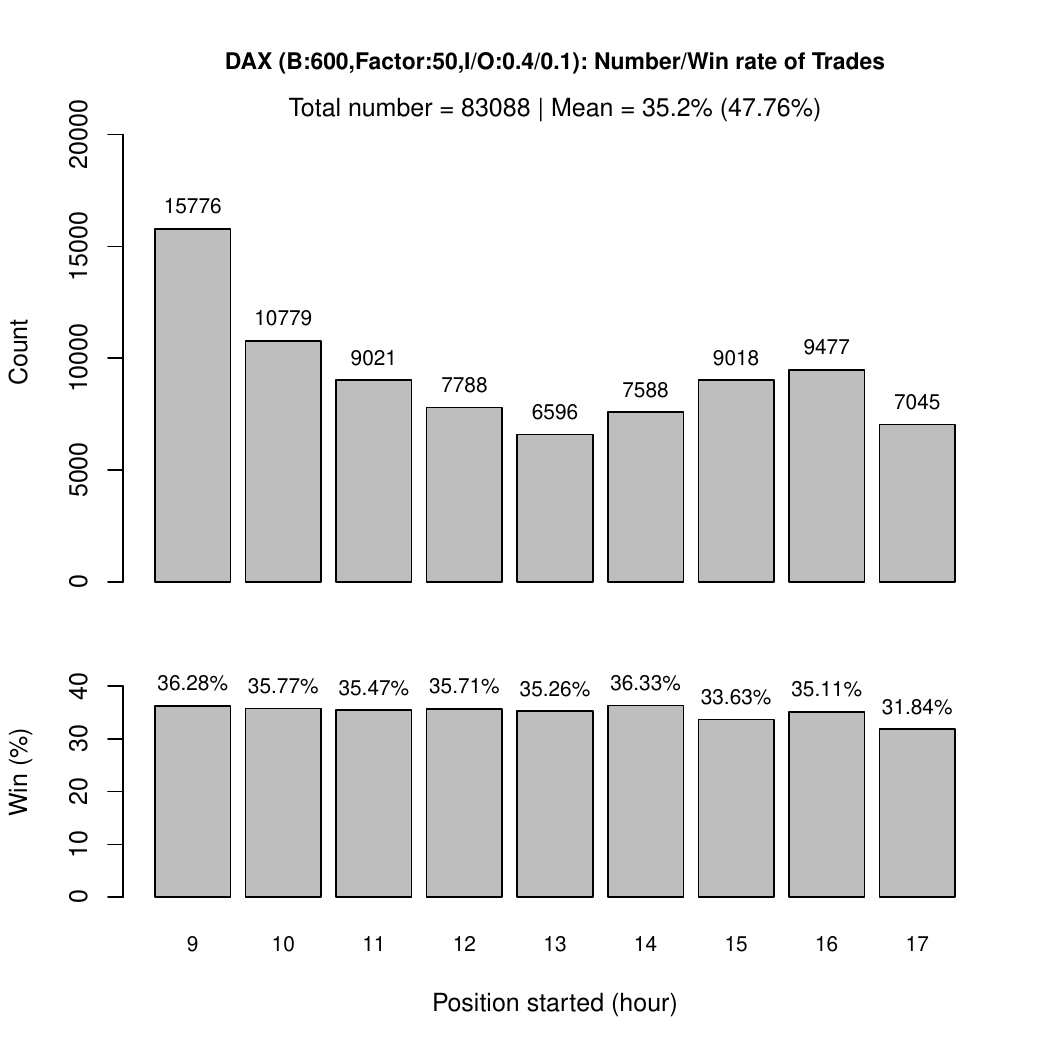} & \includegraphics[width=6.5cm]{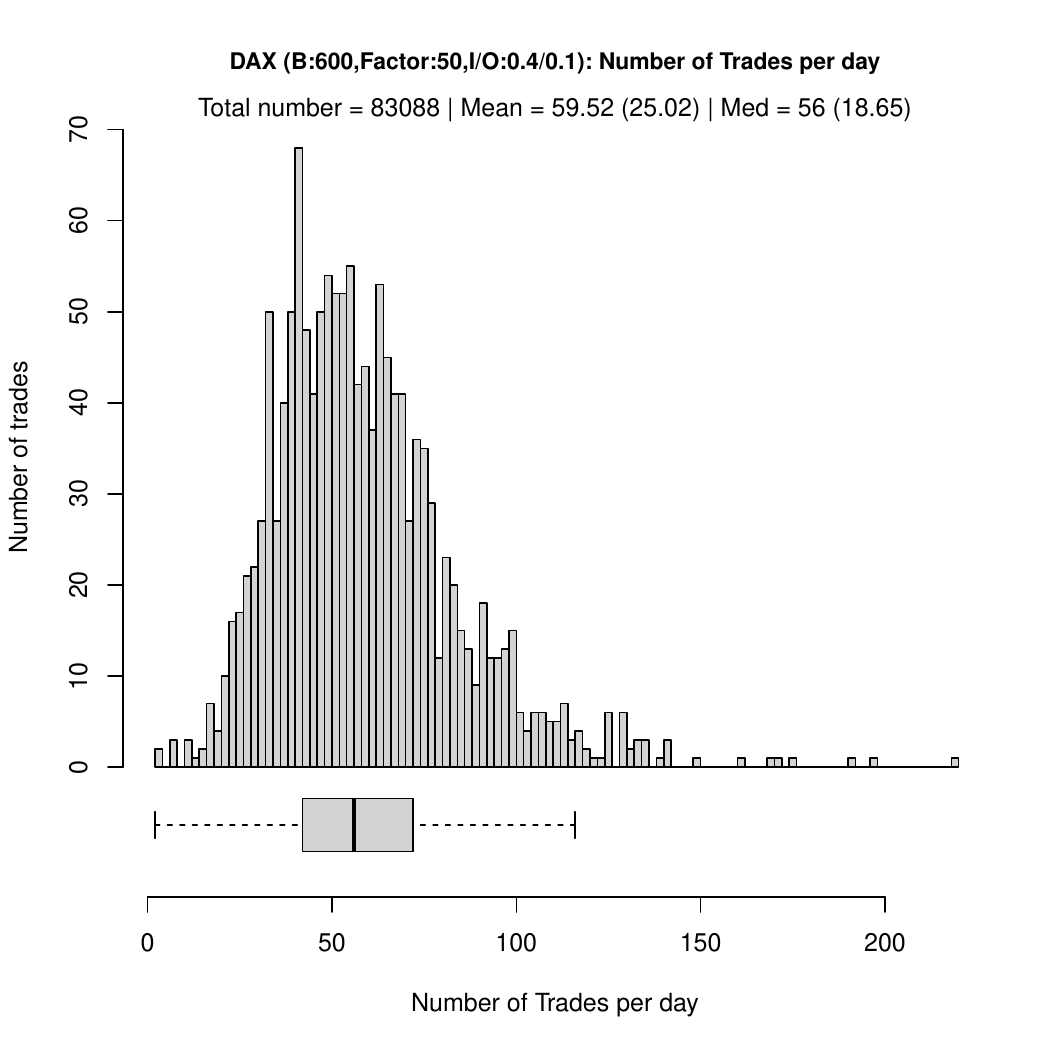}\\
		\includegraphics[width=6.5cm]{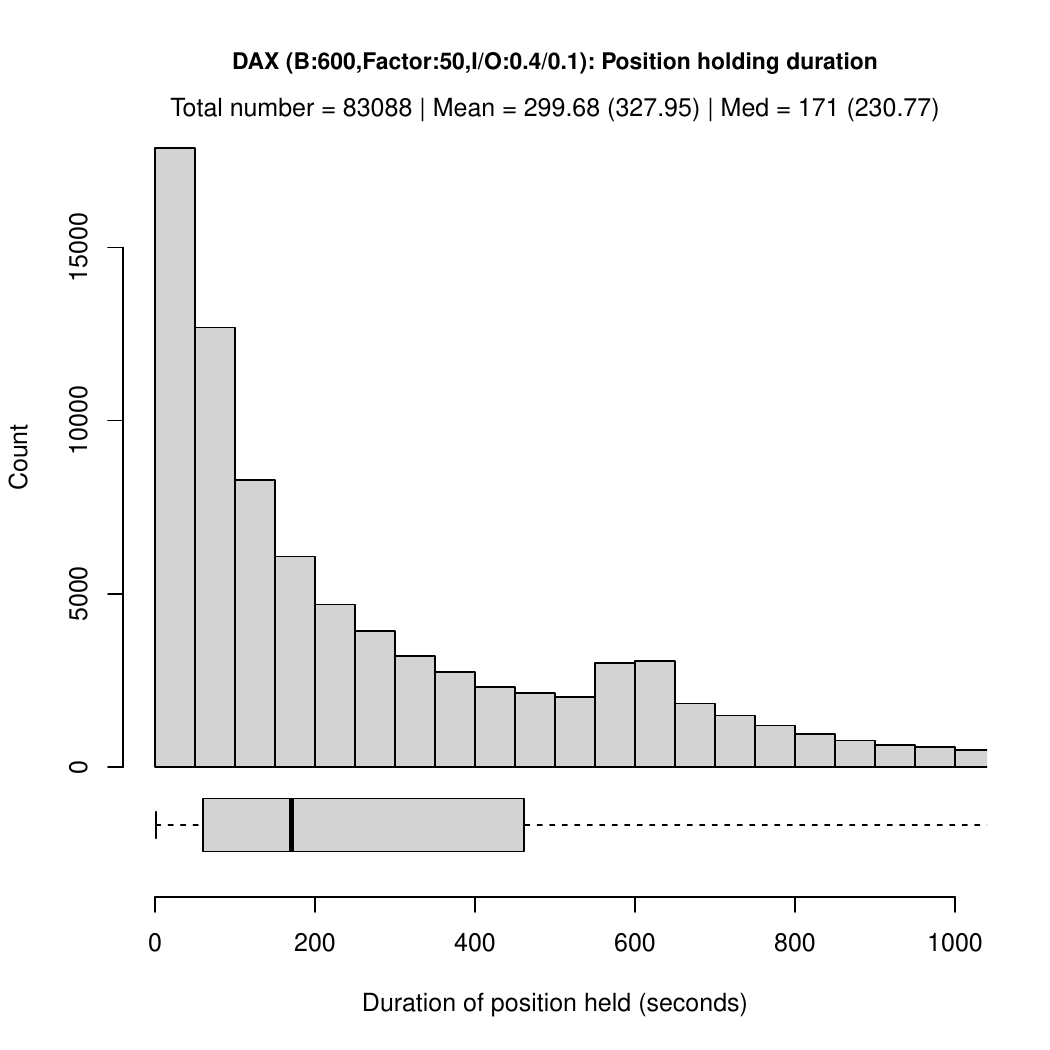} & \includegraphics[width=6.5cm]{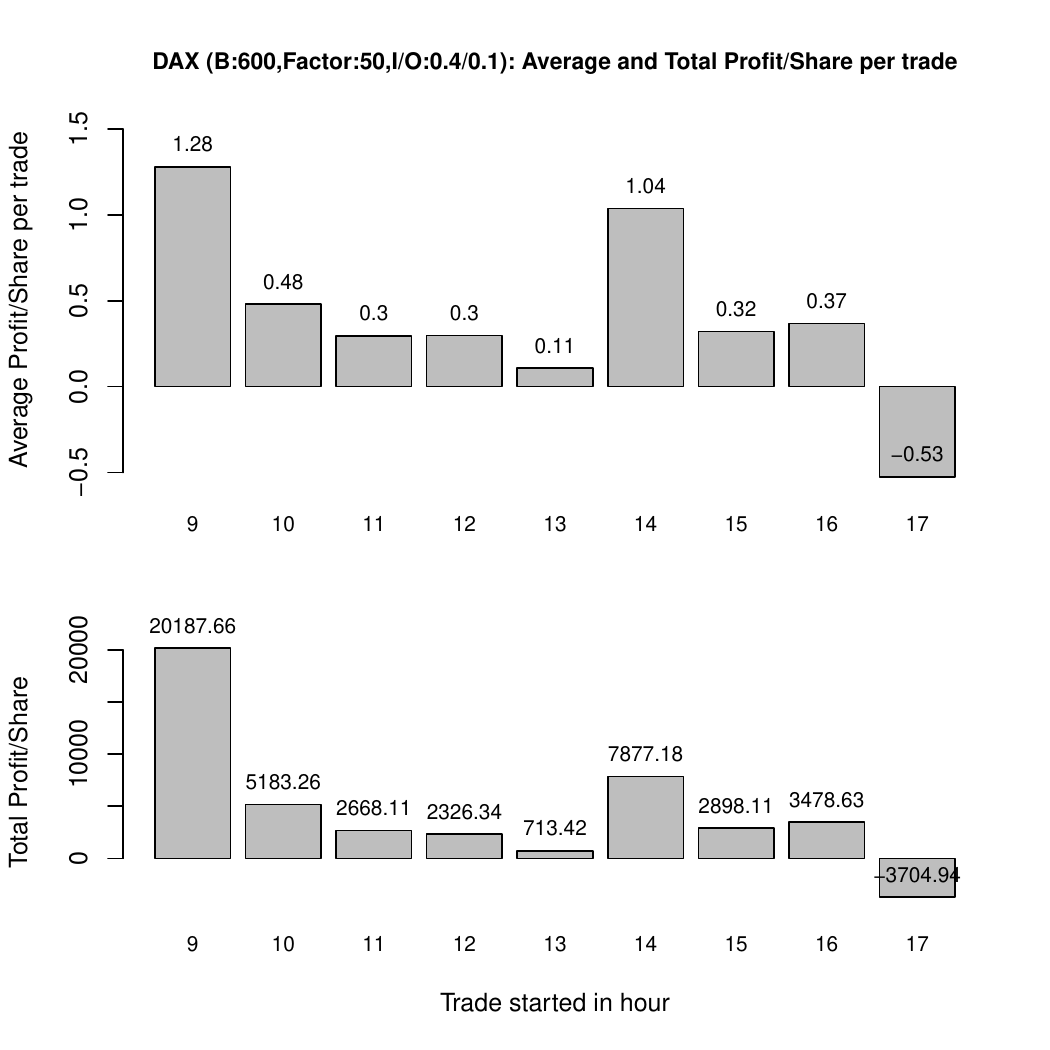}\\
		\includegraphics[width=6.5cm]{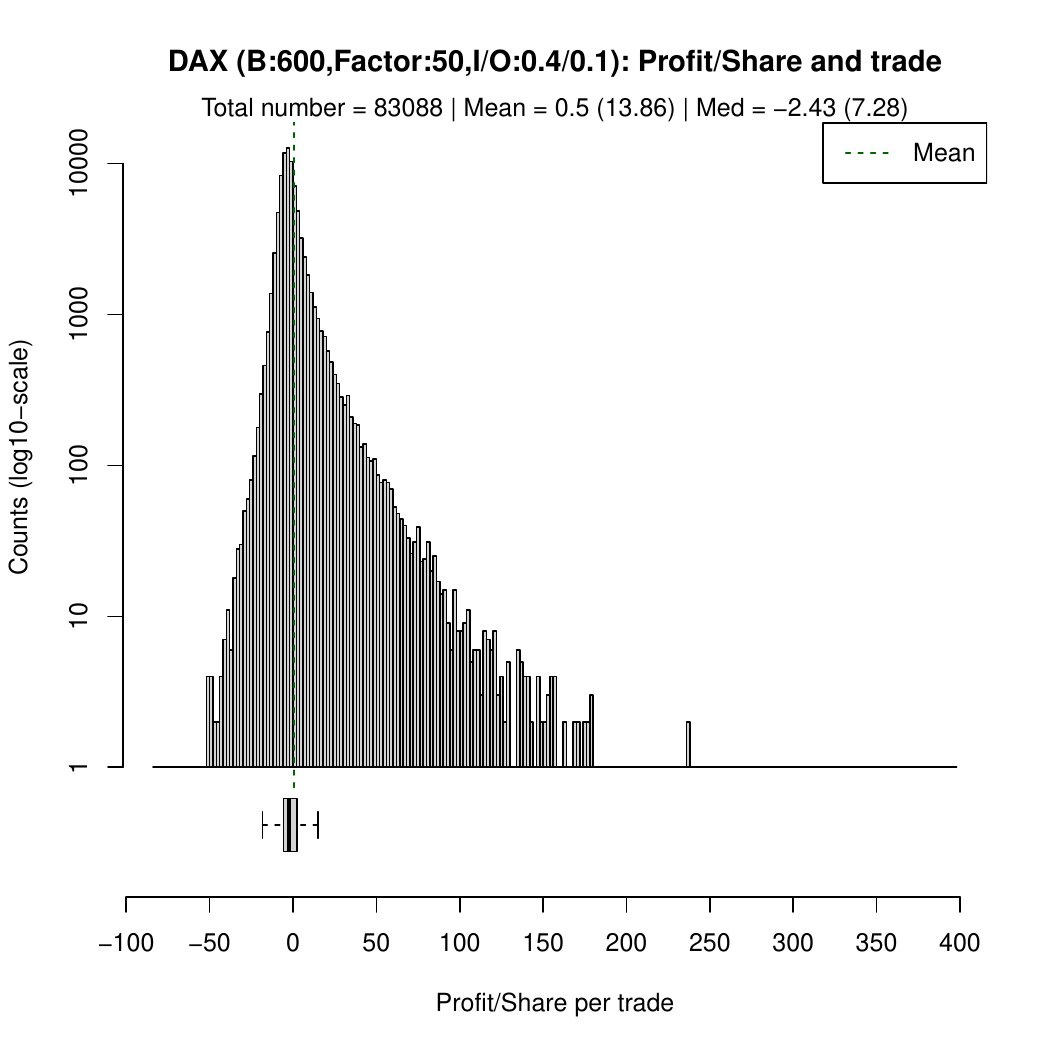} & \includegraphics[width=6.5cm]{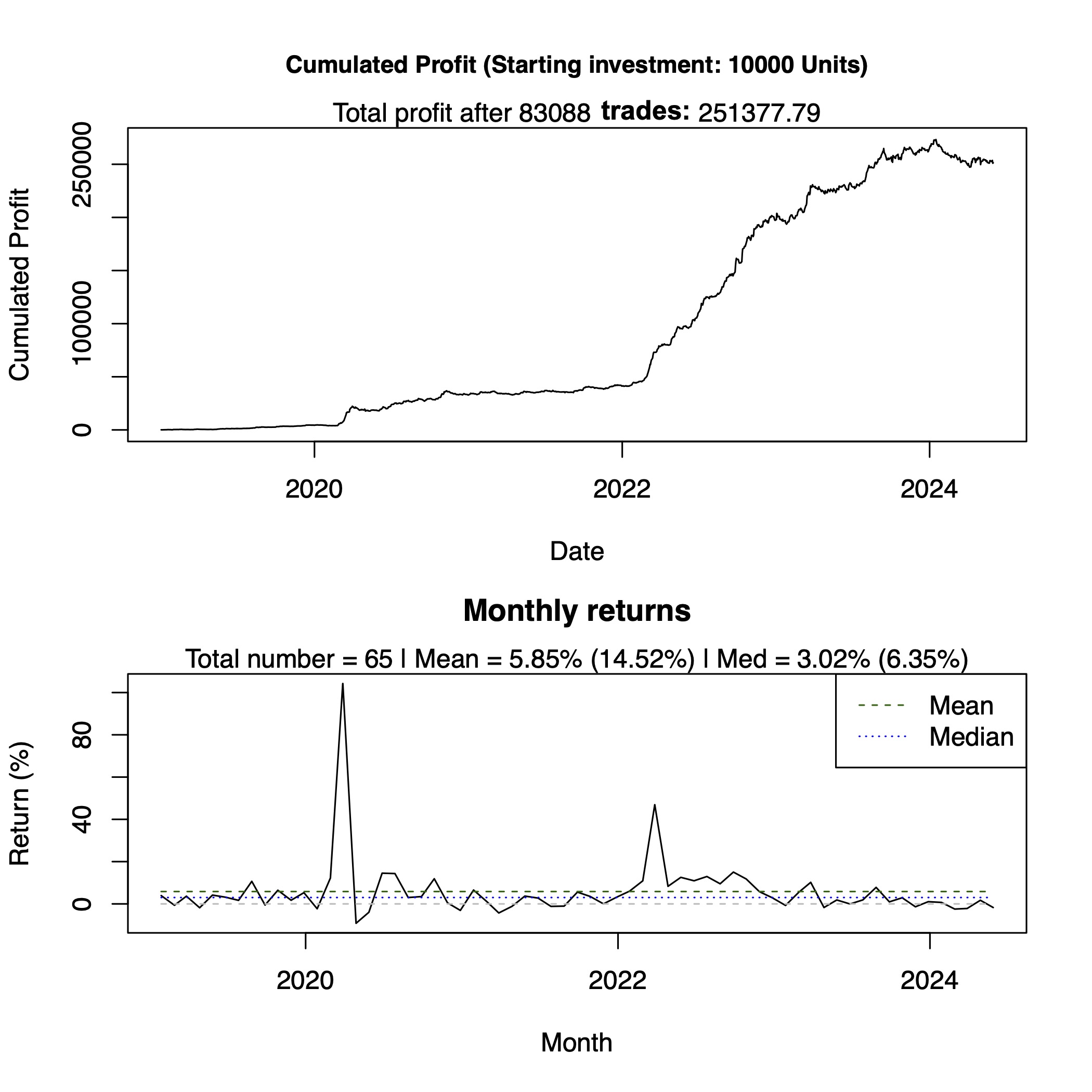}\\
	\end{tabular}
	\label{fig_DAX_600_OSC50_IN04_OUT01}
\end{figure}

\begin{figure}[h!]
	\centering
	\caption{DAX, Bandwidth 300s, In/Out: 0.4/0.1: Characteristics of the trading strategy from 2019/01 to 2024/05.}
	\begin{tabular}{cc}
		\includegraphics[width=6.5cm]{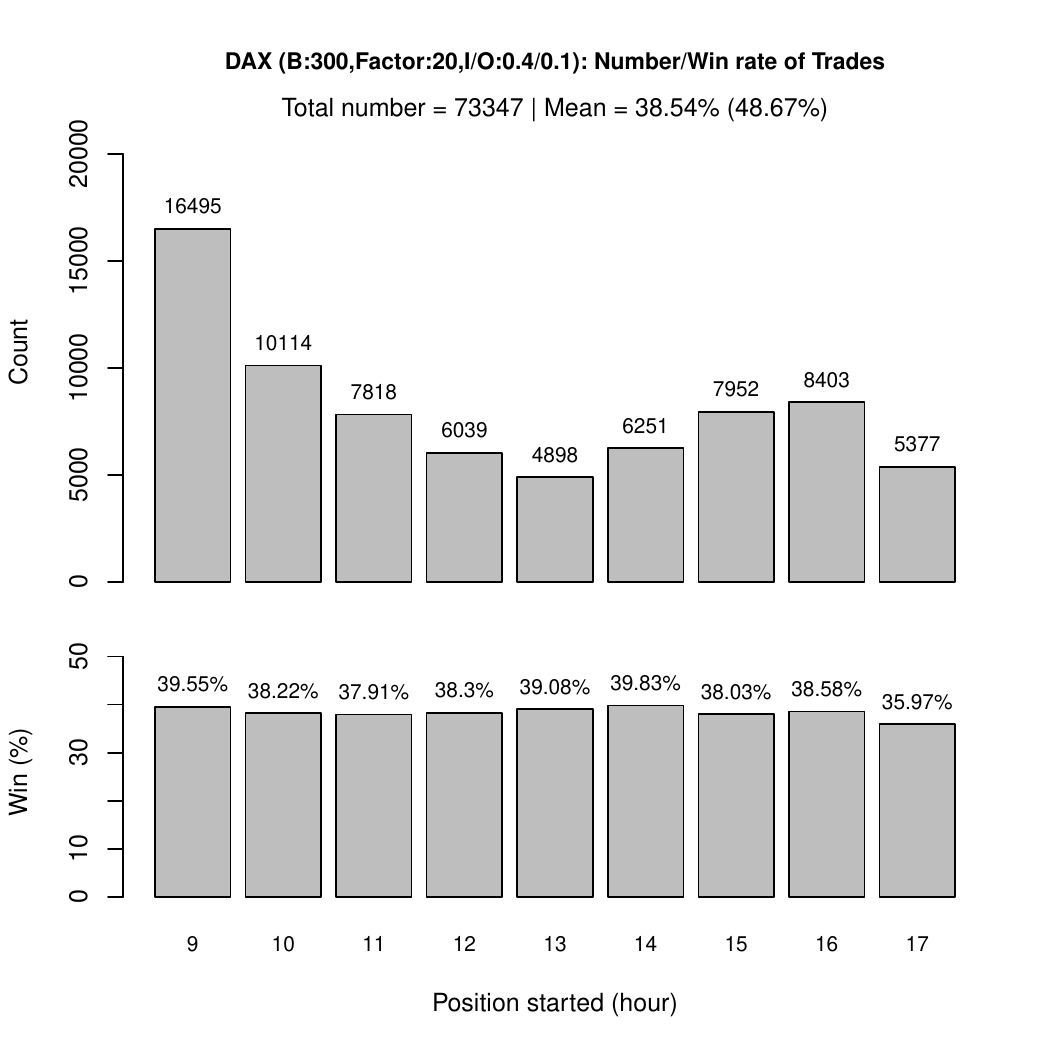} & \includegraphics[width=6.5cm]{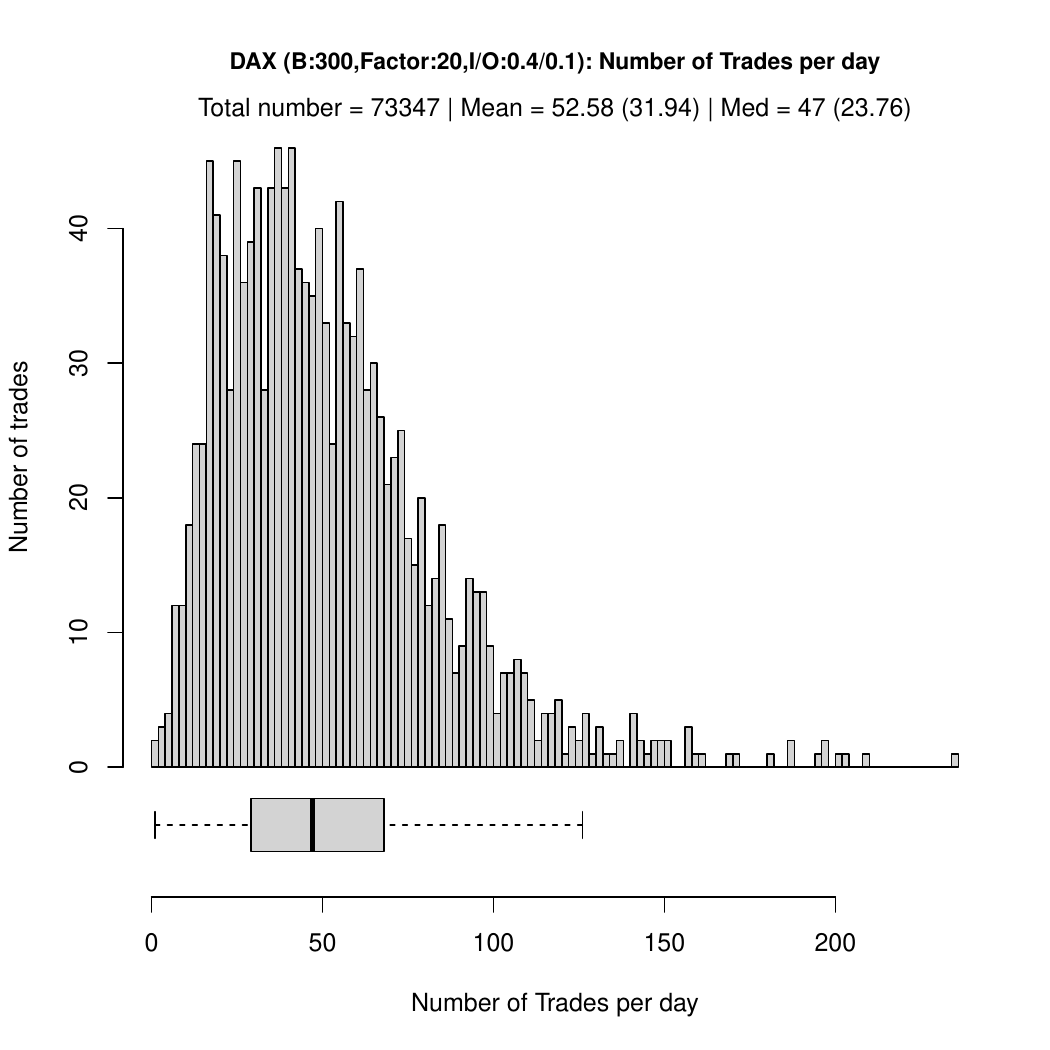}\\
		\includegraphics[width=6.5cm]{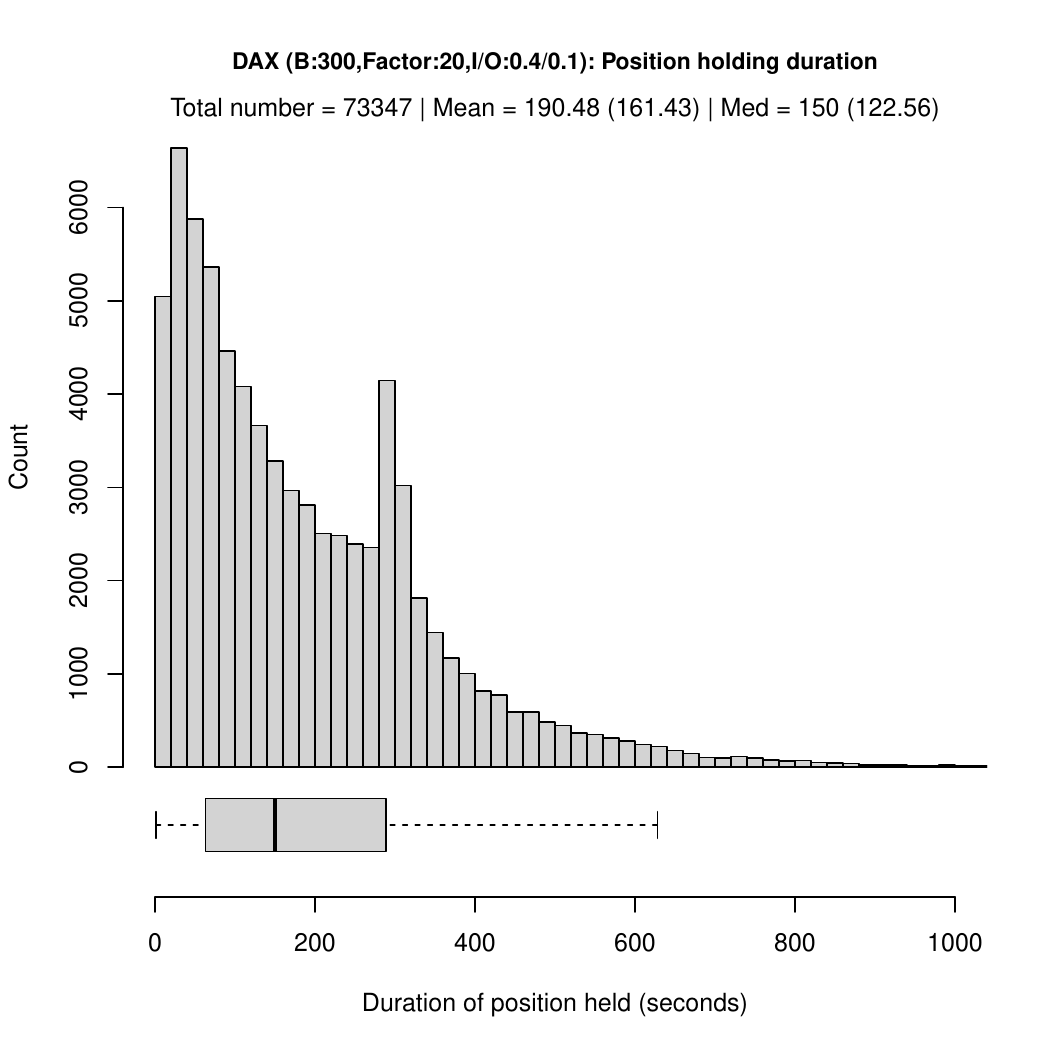} & \includegraphics[width=6.5cm]{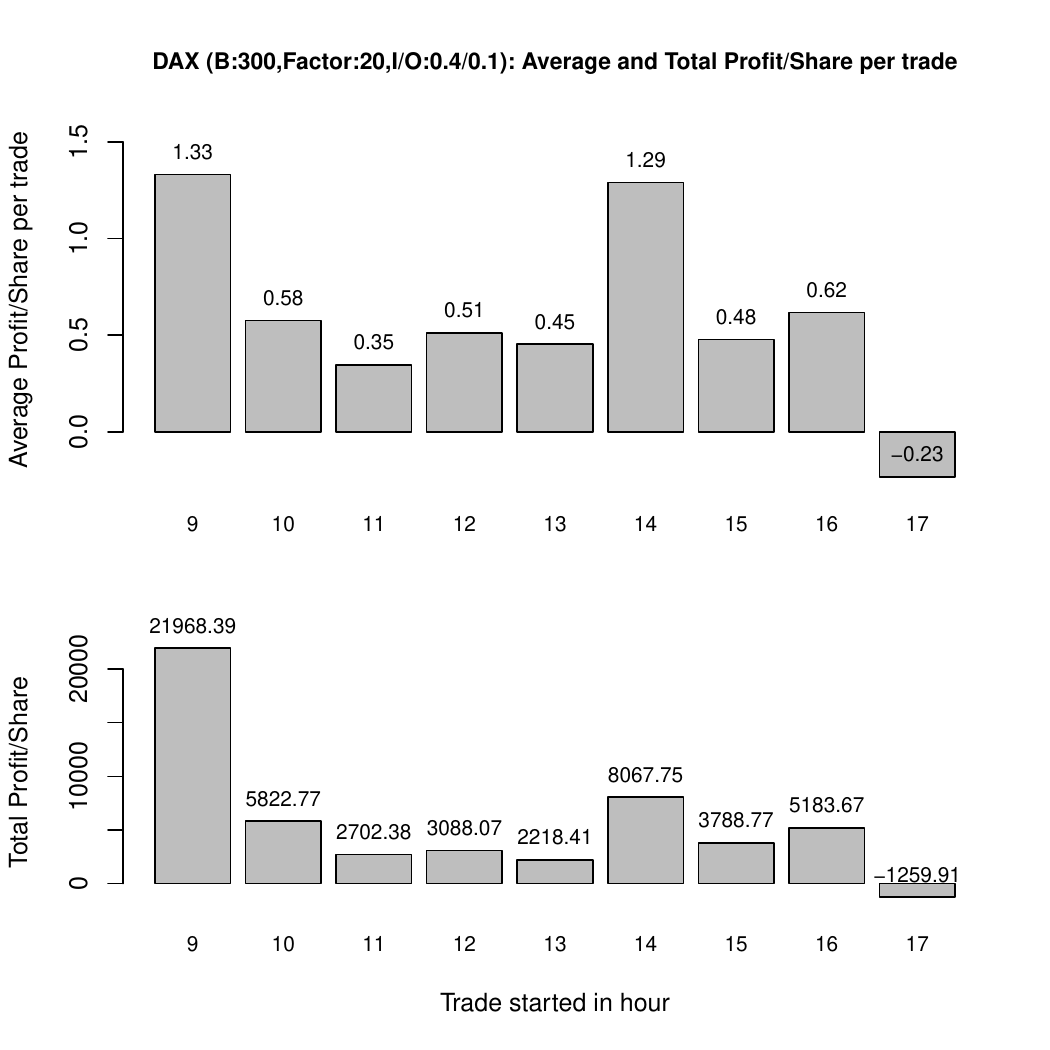}\\
		\includegraphics[width=6.5cm]{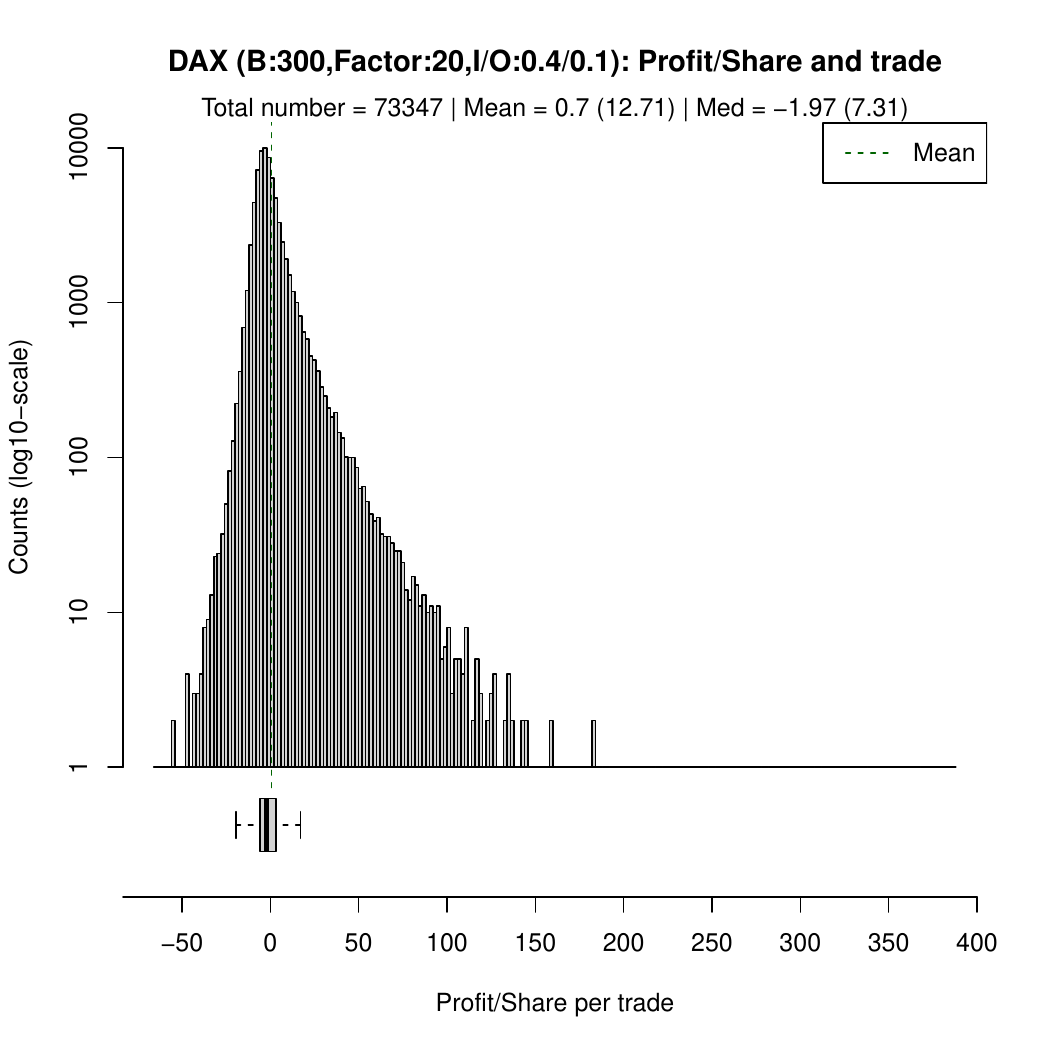} & \includegraphics[width=6.5cm]{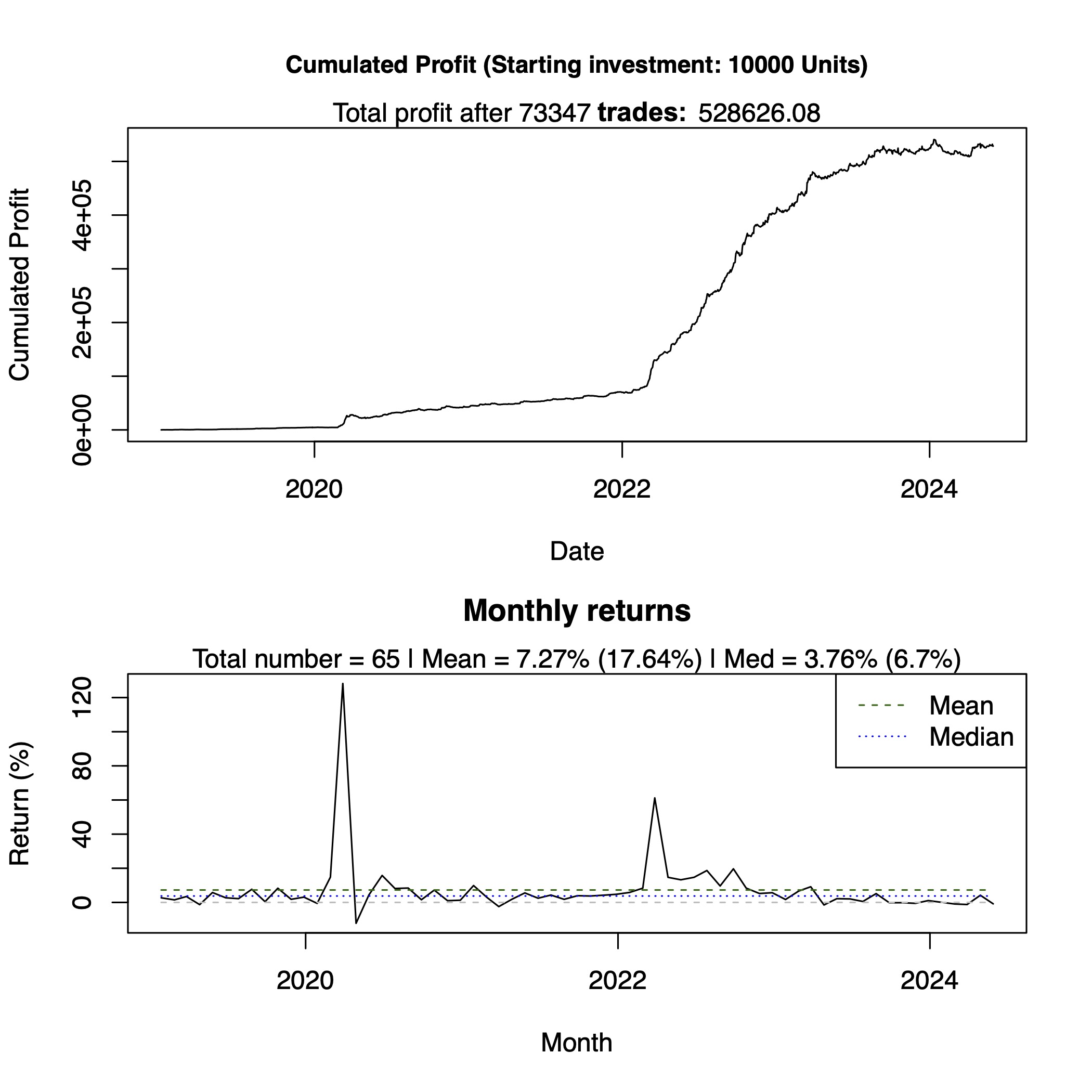}\\
	\end{tabular}
	\label{fig_DAX_300_OSC20_IN04_OUT01}
\end{figure}

\begin{figure}[h!]
	\centering
	\caption{DAX, Bandwidth 300s, In/Out: 0.8/0.2: Characteristics of the trading strategy from 2019/01 to 2024/05.}
	\begin{tabular}{cc}
		\includegraphics[width=6.5cm]{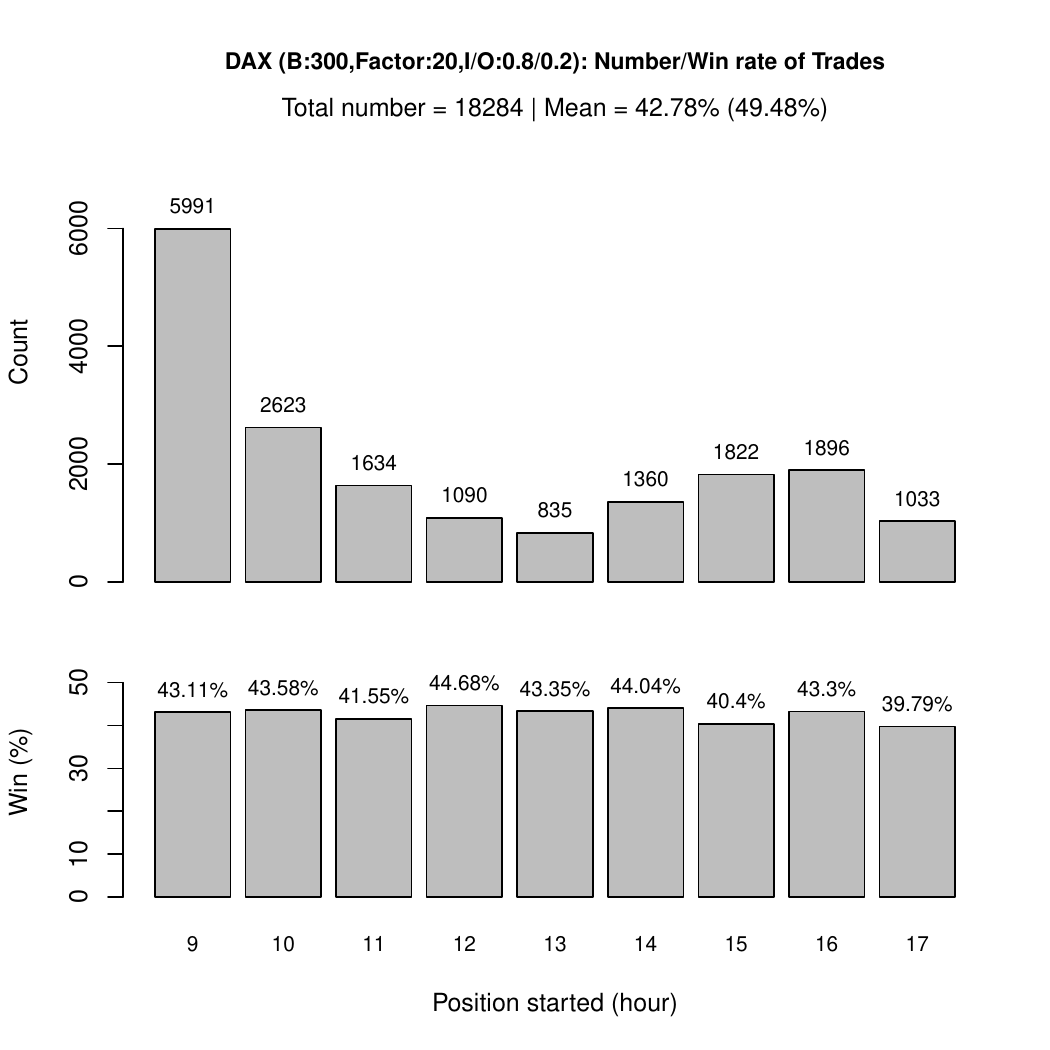} & \includegraphics[width=6.5cm]{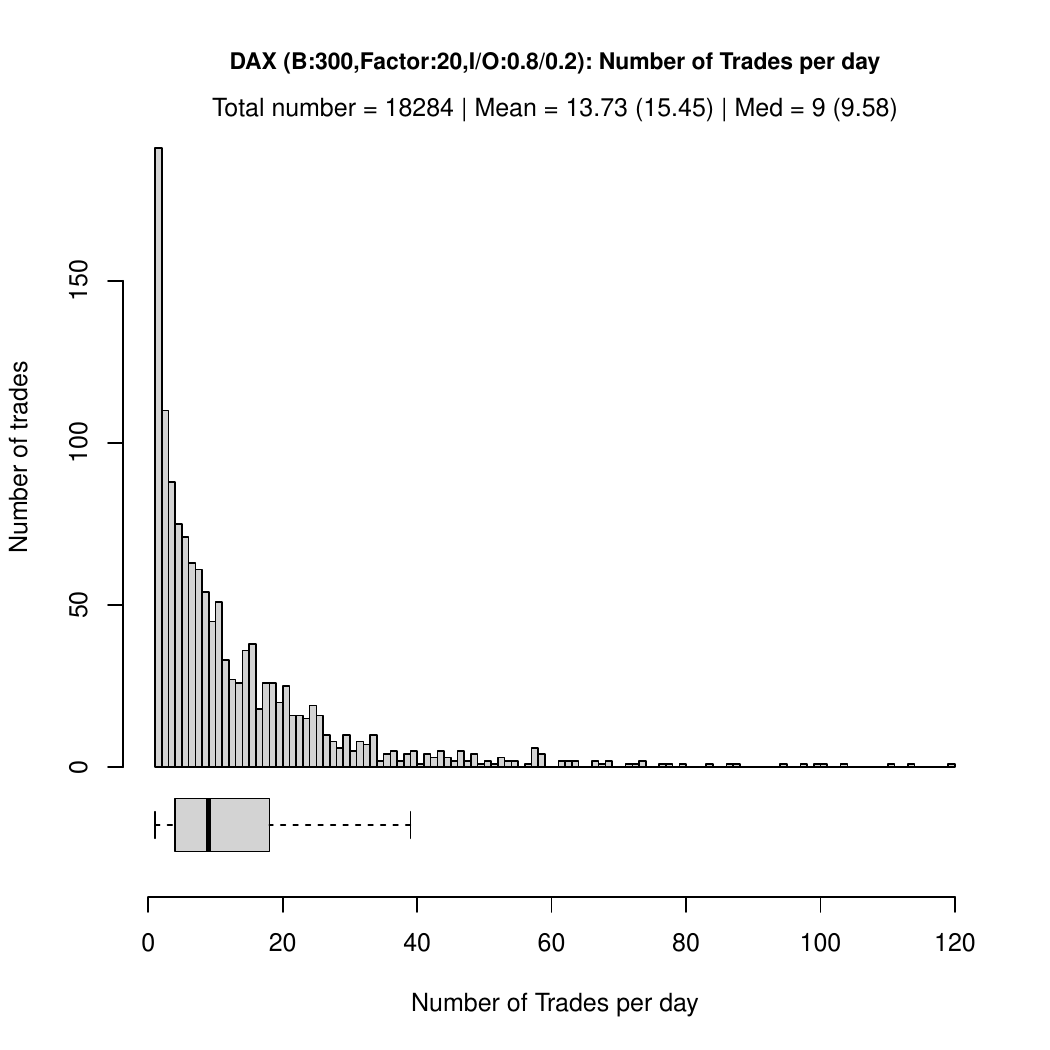}\\
		\includegraphics[width=6.5cm]{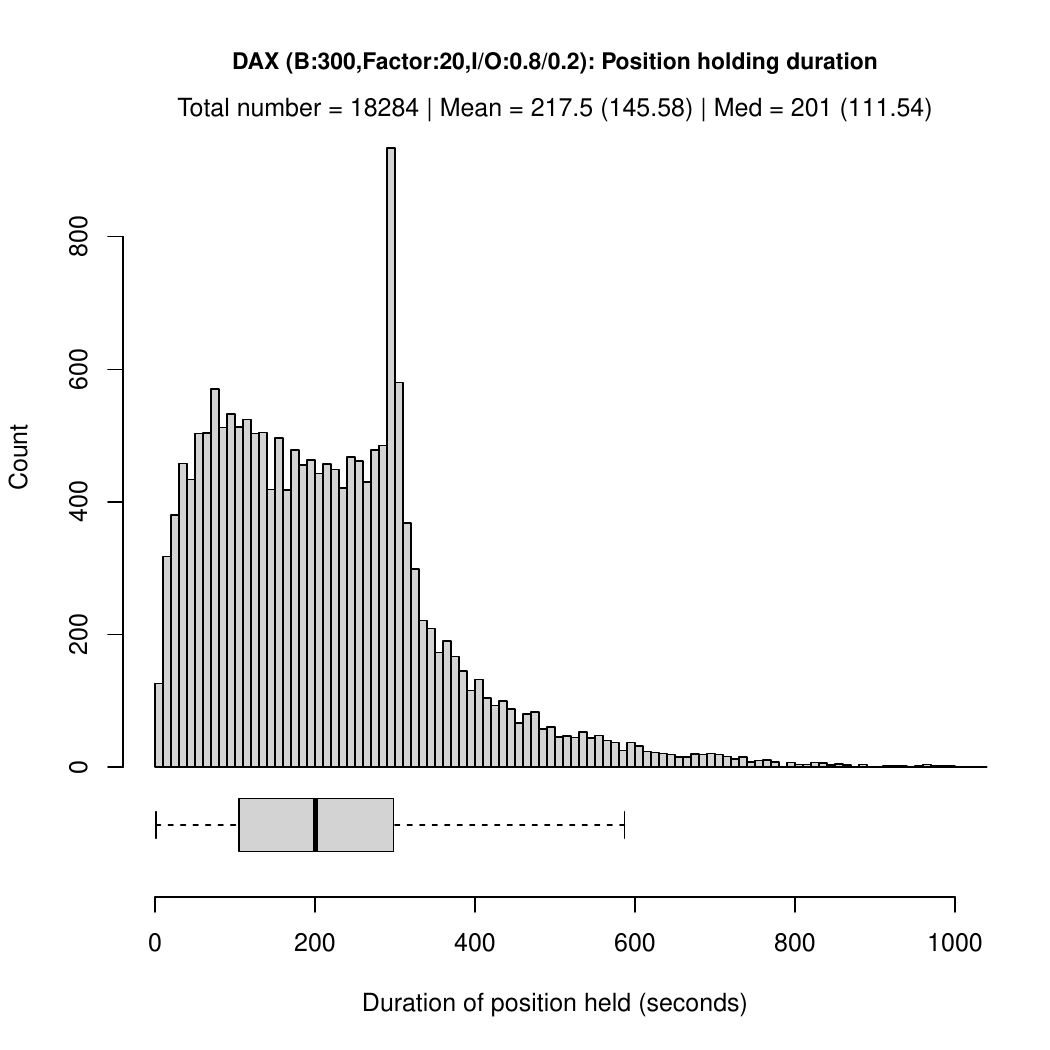} & \includegraphics[width=6.5cm]{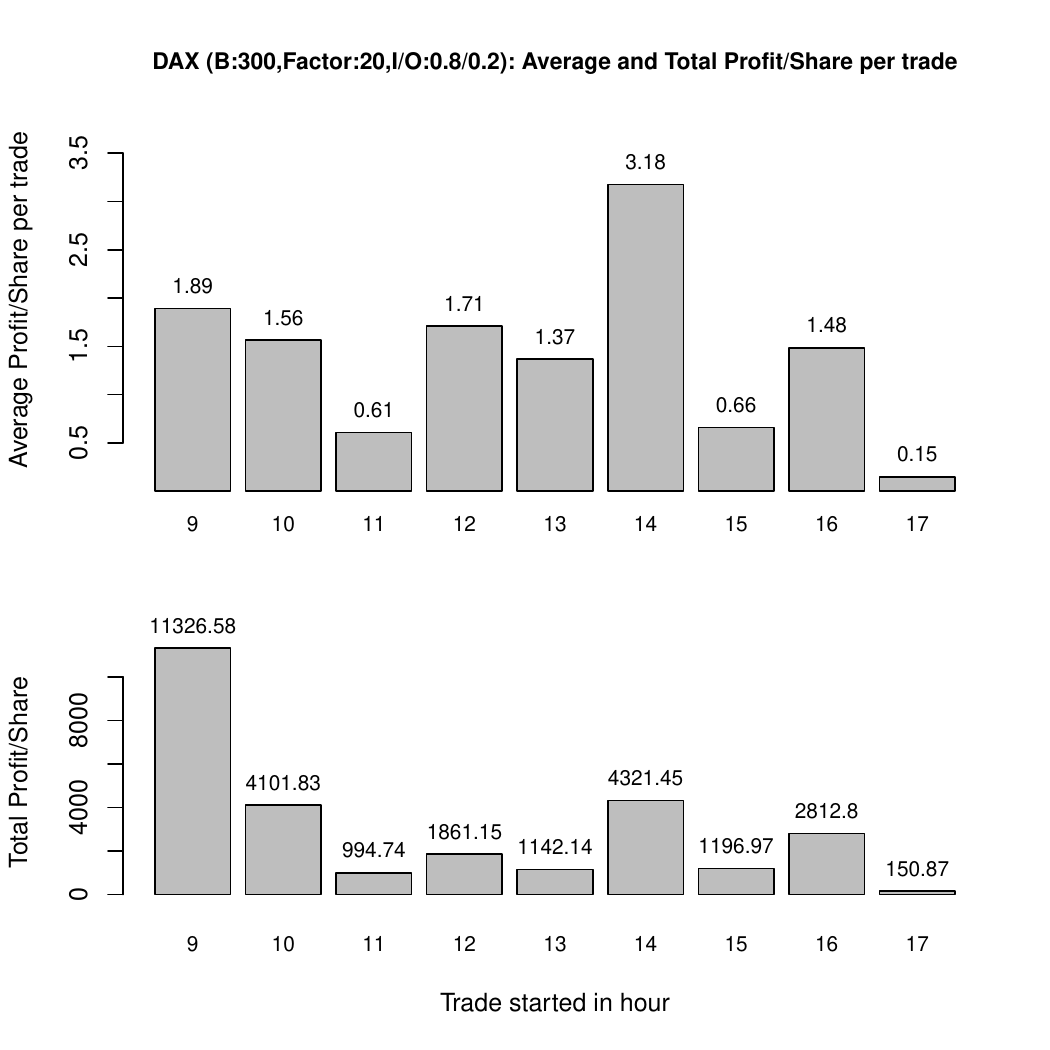}\\
		\includegraphics[width=6.5cm]{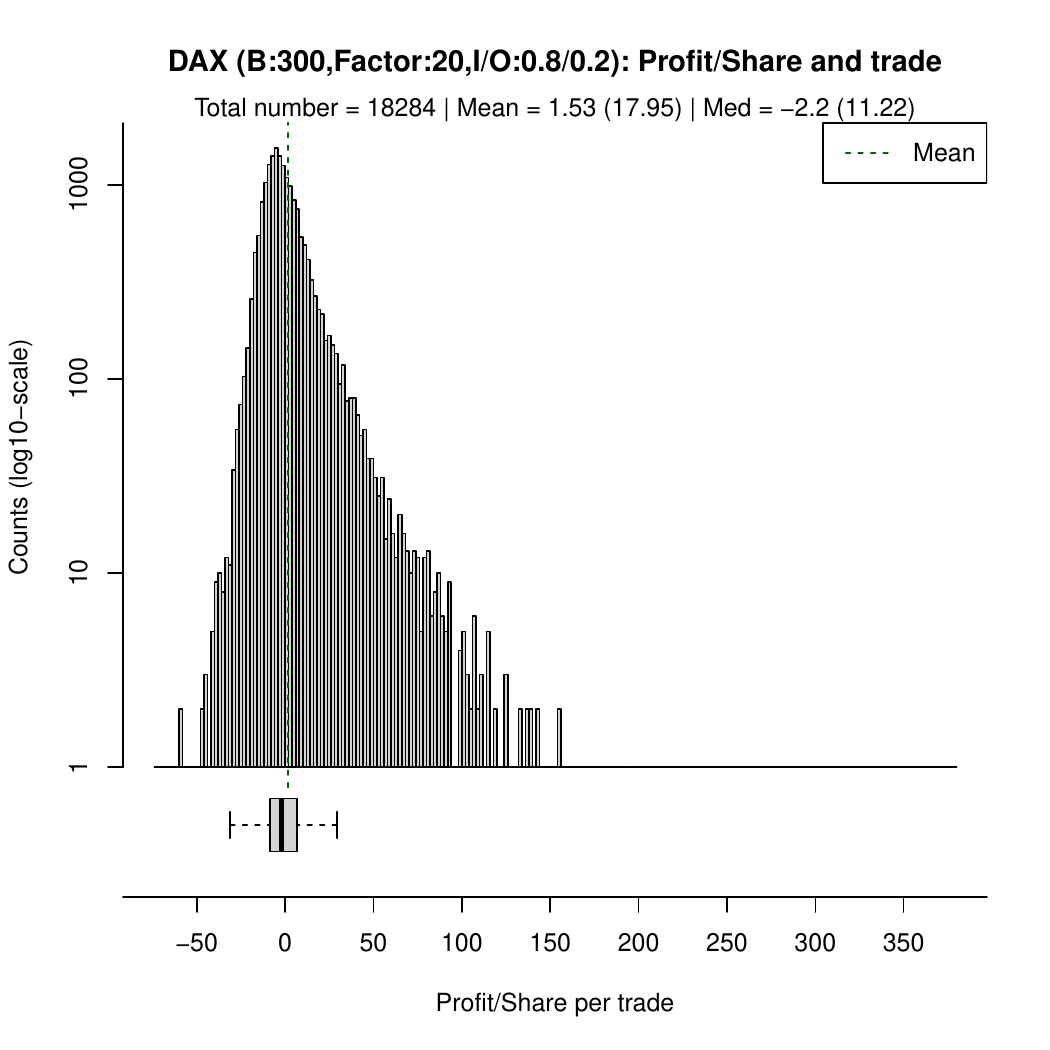} & \includegraphics[width=6.5cm]{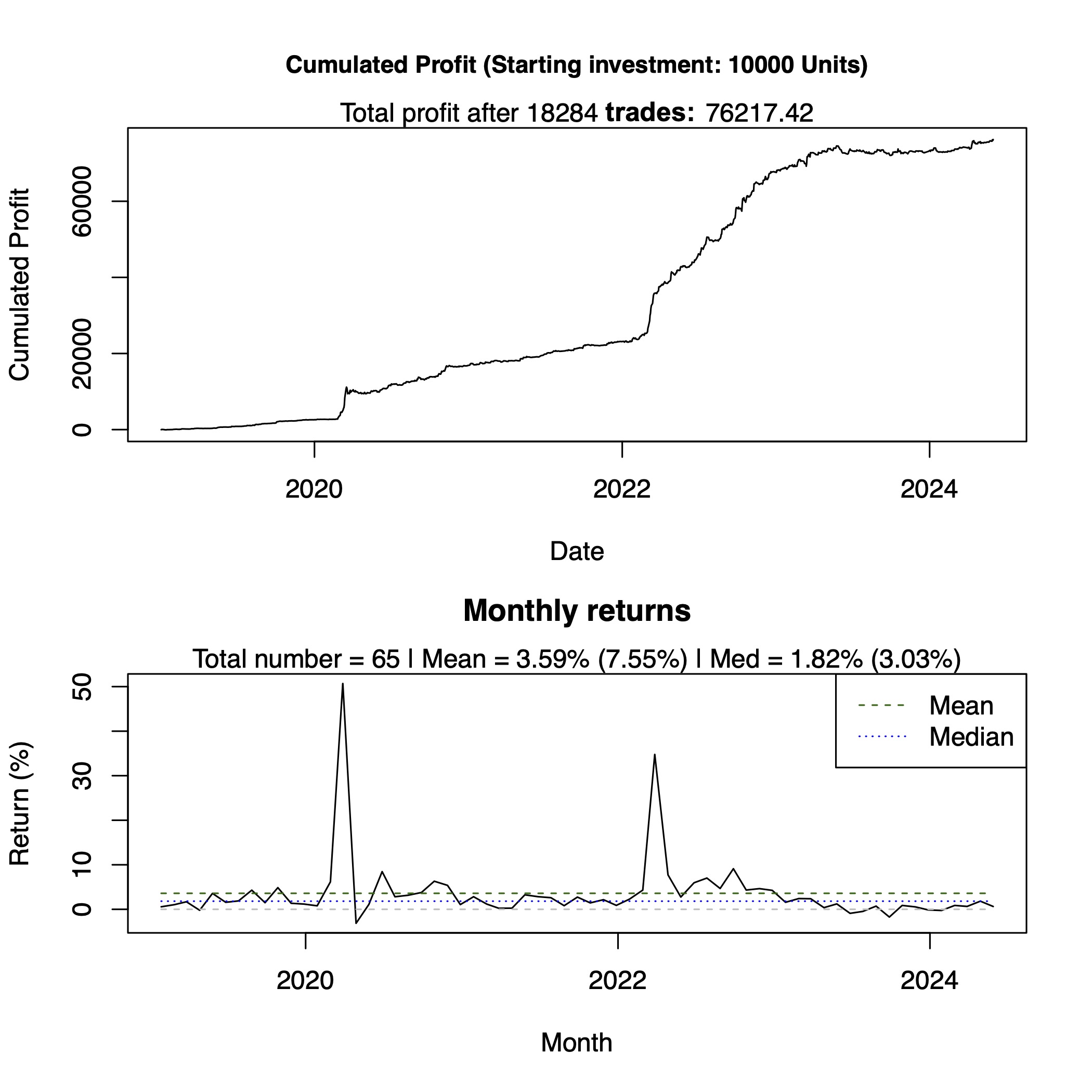}\\
	\end{tabular}
	\label{fig_DAX_300_OSC20_IN08_OUT02}
\end{figure}

\FloatBarrier

\end{document}